\documentclass[12pt]{iopart}

\usepackage{iopams}  
\usepackage{graphicx}
\usepackage{bm}
\usepackage[nosort]{cite}
\usepackage[hidelinks]{hyperref}
\usepackage{etoolbox}
\usepackage{tikz}
\usetikzlibrary{graphs,graphs.standard}
\usepackage{caption}
\DeclareMathAlphabet\mathbfcal{OMS}{cmsy}{b}{n}

\DeclareMathAlphabet{\mathpzc}{OT1}{pzc}{m}{it}

\expandafter\let\csname equation*\endcsname\relax
\expandafter\let\csname endequation*\endcsname\relax
\usepackage{amsmath}

\newcommand \vev [1] {\langle{#1}\rangle}
\def\bx{{\bf x}}
\def\ca{\mathpzc{a}}
\def\cA{\mathcal{A}}
\def\cB{\mathcal{B}}
\def\cC{\mathcal{C}}
\def\cb{\mathpzc{b}}
\def\cB{\mathcal{B}}
\def\cE{\mathcal{E}}
\def \cN{{\mathcal N}}
\def \cO{{\mathcal O}}
\def\cbO{{\mathbfcal O}}
\def \cI{{\mathcal I}}
\def \com{\mathbb{C}}
\def \Alpha{\mathrm{A}}
\def \Beta{\mathrm{B}}
\let\GGamma\Gamma
\renewcommand{\Gamma}{\mathrm{C}}
\let\DDelta\Delta
\renewcommand{\Delta}{\mathrm{D}}
\def\a{\alpha}
\def\b{\beta}

\DeclareMathOperator{\Gr}{Gr}
\DeclareMathOperator{\sdet}{sdet}
\DeclareMathOperator{\str}{str}

\makeatletter
\def\@makefntext#1{\parindent 1em\noindent 
	\makebox[1em][l]{\footnotesize\rm${}^{\arabic{footnote}}$}\footnotesize\rm  \hskip-.1em  #1}%
\def\@makefnmark{\hbox{${}^{\arabic{footnote}}\m@th$}}%
\def\footnoterule{\kern-3\p@
	\hrule \@width 2in \kern 2.6\p@} % the \hrule is .4pt high
\makeatother

\begin{document}
	\bibliographystyle{iopart-num}

	%%%%XXXX extra command to fix hyperlinks:
	%
	\newcommand{\eprint}[2][]{\href{https://arxiv.org/abs/#2}{#2}}
	%
	%%%%XXXX

	\begin{flushright}
		SAGEX-22-09
	\end{flushright}
	
	\title[Half BPS correlators]{The SAGEX Review on Scattering Amplitudes \\ 	
		Chapter 8: Half BPS correlators}
	
	\author{Paul Heslop}
	
	\address{Department of Mathematical Sciences, Durham University,\\ Upper Mountjoy Campus, Stockton Road, Durham  DH1 3LE, United Kingdom}
	\ead{paul.heslop@durham.ac.uk}
	\vspace{10pt}
	\begin{indented}
		\item[]January 2022
	\end{indented}
	
	\begin{abstract}		
		Half BPS correlators in $\cN=4$ super Yang-Mills theory are key quantities both in the AdS/CFT correspondence as well as in scattering amplitudes research. They are dual at strong  coupling to quantum gravity amplitudes. At weak coupling on the other hand they contain all  $\cN=4$ SYM amplitudes. 
		They have been found to possess  a number of hidden symmetries, for example  non-trivial permutation symmetry  of perturbative four-point integrands and a higher dimensional conformal symmetry. Their study has enjoyed continuous progress  since the discovery of AdS/CFT and they are now some of the best understood quantities of any four-dimensional quantum field theory.
		
		In this  review we outline the  current knowledge of half BPS correlators,	emphasising these two co-existing relations  to  scattering amplitudes at strong and weak coupling.

	\end{abstract}
	
	%\maketitle

	\newpage

	\section*{Contents}
	\makeatletter\@starttoc{toc}\makeatother

	\section{Introduction}
	\label{sec:intro}
	
	$\cN=4$ super Yang-Mills is the   most symmetric of all  four-dimensional quantum field theories. The simplest local operators in this theory are the half BPS operators. Despite this simplicity, the family of correlators of half BPS operators provides a remarkably  rich breeding ground of ideas  as well as being some of the most accurately known quantities in any 4d QFT. 
	They  impact  on many key areas in current theoretical physics, scattering amplitudes, integrability, positive geometry/amplituhedron, conformal bootstrap etc. 
	
	The  relation of half BPS correlators to scattering amplitudes is of particular interest and importance and has provided the thrust for most of the progress in our understanding of these correlators. This relation with amplitudes appears in two completely different ways, to amplitudes in two different theories, via AdS/CFT on the one hand and the correlator/amplitude duality on the other. 
	Half BPS correlators first became significant objects of interest immediately after the discovery in 1997 of the AdS/CFT correspondence which  tells us that we can interpret them as  IIB supergraviton amplitudes in string theory on an AdS${}_5\times S^5$ background. 
	Indeed the most accurate quantum gravity amplitude in curved space has recently been obtained by bootstrapping a half BPS correlator at strong coupling  and this provides a crucial arena for exploring quantum gravity.
	But then over a decade after AdS/CFT in 2010 another relation to amplitudes was discovered, the correlator/amplitude duality.
	The correlator/amplitude duality states that the correlators become $\cN=4$ SYM amplitudes on taking a certain polygonal lightlike limit. This duality  gives insight in both directions and indeed the most accurately known amplitude integrand  has been obtained via the correlator using this duality.

	In this review we will attempt to describe as much as possible of what is currently known of the family of half BPS correlators in $\cN=4$ SYM
	\begin{align}
		\langle \cO_{p_1}\dots \cO_{p_n}\rangle\ .
	\end{align}
	Here one can vary the {\em number} of operators in the correlator, $n$, the {\em nature}  of the operators in the correlators (ie their  charges $p_i$ - we will focus on single trace/ single particle operators  $\cO_p=\Tr(\phi^p)+..$), as well as  the Yang-Mills coupling, $\lambda$, and the rank of the $SU(N_c)$ gauge group, $N_c$.
	We have  arranged this review as illustrated here:
		\begin{center}
		\begin{tikzpicture}
			\node[anchor=west] (s2) at (0,0) {Section 2: $\langle\cO_2 \cO_2\cO_2\cO_2 \rangle$};
			\node[anchor=west] (s20) at (8,1){Intro: superconformal kinematics};		
			\node[anchor=west] (s21) at (8,0){2.1-2.4: Weak coupling};	
			\node[anchor=west] (s25) at (8,-1){2.5: Strong coupling};	
			\draw[-latex] (s2) -- (s20);
			\draw[-latex] (s2) -- (s21);
			\draw[-latex] (s2) -- (s25);
			\node[anchor=west] (s3) at (0,-3) {Section 3: $\langle\cO_{p_1} \cO_{p_2}\cO_{p_3}\cO_{p_4} \rangle$};
			\node[anchor=west] (s30) at (8,-2){Intro: superconformal kinematics};		
			\node[anchor=west] (s31) at (8,-3){3.1-3.2: Weak coupling};	
			\node[anchor=west] (s33) at (8,-4){3.3: Strong coupling};	
			\draw[-latex] (s3) -- (s30);
			\draw[-latex] (s3) -- (s31);
			\draw[-latex] (s3) -- (s33);
			\node[anchor=west] (s4) at (0,-6) {Section 4: $\langle\cO_{p_1} \dots\cO_{p_n} \rangle$};
			\node[anchor=west] (s41) at (8,-5.5){4.1-4.6: superconformal kinematics};		
			\node[anchor=west] (s47) at (8,-6.5){4.7-4.10: Weak coupling};	
			\draw[-latex] (s4) -- (s41);
			\draw[-latex] (s4) -- (s47);
		\end{tikzpicture}
	\end{center}
		Thus we focus first, in section~\ref{sec:simplest}, on the simplest (non-trivial) correlator, the four-point correlator of the lowest charge stress-tensor multiplets, $n{=}4$, $p_i{=}2$. We begin with its structure and the hidden symmetry of its perturbative integrands (leaving the derivations to later sections) and then where feasible giving, or at least outlining, explicit results, 
	first at weak coupling -- where we highlight the relation to $\cN=4$ SYM amplitudes -- and then at strong coupling.
	Then in section~\ref{sec:highercharges} we consider four-point functions of higher charge half BPS operators ($n=4$, arbitrary $p_i$) where a recently discovered hidden 10d conformal symmetry plays a  key role: when present all higher charge correlators can be obtained from the simplest, lowest charge, correlator.
	Finally in section~\ref{sec:higherpoints} we move to higher point correlators (arbitrary $n$,  focussing mostly on $p_i=2$ but also making some new speculations about all charge generalisations based on the ten dimensional conformal symmetry) and introduce the analytic  superspace formalism, perfectly suited for half BPS correlators. Pulling together material from a few different places we hope to explain in detail this formalism which in the process will  allow us to both prove many of the structural results stated in earlier sections as well as discuss the relatively little that is known explicitly at higher points.

	\section{Four-point stress-tensor correlators}
	\label{sec:simplest}
	
	%\subsection{Structure}

	We begin with the simplest non-trivial half BPS correlator in planar   $SU(N_c)$ $\cN=4$ SYM:
	\begin{align}\label{corbasic}
		&\langle \cO(x_1)\bar \cO(x_2)   \cO(x_3) \bar \cO(x_4)\rangle \notag \\&=  \frac{4c^2}{(4\pi^2)^4}\left(\frac1{x_{12}^4x_{34}^4} + \frac1{x_{14}^4x_{23}^4} + \frac1c \frac1{x_{12}^2 x_{23}^2x_{34}^2x_{14}^2} \left(1+ 2  F(x_i;\lambda,c)\right)\right)\ .
	\end{align}
	Here $\lambda= g_{YM}^2 N_c$ is the 't Hooft coupling and $c=(N_c^2{-}1)/4$ is the central charge. The operator $\cO= \Tr(\phi^2)$ where $\phi(x)$ is one of the three complex scalar fields in the adjoint of the gauge group.
	The operator $\cO$   is part of the simplest  half BPS supermultiplet, called the stress-tensor multiplet since it also contains the stress-tensor $T_{\mu \nu}$ as well as the on-shell Lagrangian. As was discovered soon after the AdS/CFT correspondence,  two- and three-point functions involving $\cO$ are all independent of the coupling constant, taking on their free theory values~\cite{hep-th/9807098,Intriligator:1998ig,hep-th/9808162,hep-th/9903094,hep-th/9905020,Eden:1999gh,hep-th/9907088,hep-th/9910197}. These can thus be obtained by simple Wick contractions, for example the  two-point function is $\langle\cO(x_1)\bar  \cO(x_2)\rangle =2c/(x_{12}^4(4\pi^2)^2)$. The four-point function is then the simplest non-trivial correlator involving $\cO$.  Superconformal symmetry is enough to completely fix the four point function of any four operators in the stress tensor multiplet  $\cO$ in terms of this correlator~\cite{hep-th/0009106} (see section~\ref{sec:an} for a derivation). 
	The first three terms in~\eqref{corbasic} give the free theory ($\lambda=0$) correlator which can be obtained by  Wick contractions. Thus $F$, which is a conformally invariant function,  vanishes at zero coupling, $F(x_i;0,c)=0$.

	\subsection{Loop integrands: Hidden symmetry}
	\label{sec:hidden}
	
	We first consider the correlator in perturbation theory at weak coupling.
	It was computed at one loop $O(\lambda)$ and two loops $O(\lambda^2)$ by direct (Feynman diagram) computation in  a series of papers soon after the discovery of AdS/CFT~\cite{Gonzalez-Rey:1998wyj,Eden:1998hh,Eden:1999kh,Eden:2000mv,Bianchi:2000hn}.
	The one-loop integrand was given in terms of the one loop box integral and the two-loop integrand in terms of  double box integrals together with the product of box functions in various orientations.  Over a decade later, following renewed interest arising from the correlator/amplitude duality, it was observed that these combinations of integrands, if placed over a common denominator have a suggestive, permutation symmetric form~\cite{Eden:2011we}.
	If we define the loop level correlators $F^{(l)}$ via the expansion in $\lambda$ and the  corresponding integrands $f^{(l)}$ as
	\begin{align}\label{Fl}
		F(x_1,..,x_4;\lambda,c)&= \sum_{l=1}^\infty{\left(\frac{\lambda}{4\pi^2}\right)^l}F^{(l)}(x_1,.,x_{4};c)\\
		F^{(l)}(x_1,..,x_{4};c)	&= \frac{\xi^{(4)}}{l!}\int \frac{d^4x_5}{(-4\pi^2)}..\frac{d^4x_{4+l}}{(-4\pi^2)}f^{(l)}(x_1,.,x_{4+l};c)\ \label{Finf}
	\end{align}
	where $\xi^{(4)}=x_{13}^4x_{24}^4x_{12}^2 x_{23}^2x_{34}^2x_{14}^2$ (all integrals are wick rotated to Euclidean signature). 
	Then the 1- and 2-loop correlators can be written compactly as:
	\begin{align}\label{1loop}
		f^{(1)}& = {1 \over \prod_{1\leq i<j \leq 5} x_{ij}^2}\,, 
		\\
		f^{(2)}& =
		{\frac1{48}\sum_{\sigma \in S_6} x_{\sigma_1
				\sigma_2}^2 x_{\sigma_3 \sigma_4}^2x_{\sigma_5 \sigma_6}^2
			\over \prod_{1\leq i<j \leq 6} x_{ij}^2}\,.
		\label{2loop}
	\end{align}
	The factor of $1/48$ in~\eqref{2loop} just factors out the over-counting in the sum over $S_6$ permutations so that  each term appears with coefficient 1: there are 15 independent terms in the sum over $6!=15\times 48$. 
	Now in this form we see that both denominator and numerator are fully $S_{4+l}$ symmetric in~\eqref{1loop},\eqref{2loop} and this suggests the presence of such  a symmetry also at higher loops. The subgroup $S_4\times S_l\subset S_{4+l}$ arises simply from crossing symmetry together with symmetry of the integration variables: the highly non-trivial aspect of this hidden $S_{4+l}$ symmetry occurs in the interchanging of external $x_1,..,x_4$ variables and integration $x_5,..,x_{4+l}$ variables.
	This higher loop permutation symmetry indeed turns out to be present~\cite{Eden:2011we,Eden:2012tu}. The proof involves relating the integrand to higher point correlators with Lagrangian operators which are related by SUSY to $\cO$.  We give  this proof later in section~\ref{sec:proofhidden} but for now we simply assume this symmetry and explore its consequences.

	The functions  $f^{(1)}$ and $f^{(2)}$ are uniquely defined (up to an overall coefficient) by this hidden $S_{4+l}$ permutation symmetry, together with conformal symmetry which says that they are  functions of the Lorentz invariants $x_{ij}^2$ only and have conformal weight 4 at each point as well as the assumption that they only have simple poles in $x_{ij}^2$. 
	
	Going further it is useful to consider graphs representing the integrands in the standard way - associating vertices with the space-time points and edges between vertices $i$ and $j$ whenever there is a propagator term $1/x_{ij}^2$. The permutation symmetry means we needn't worry about a specific labelling since we have to sum over all labellings. We call such graphs $f$ graphs. The $f$ graphs corresponding to the 1- and 2-loop correlators~\eqref{1loop},\eqref{2loop} are: 
	\begin{align}\label{fgraph12}
		f^{(1)}=
		\raisebox{-.5\height}{		\includegraphics[width=2.2cm]{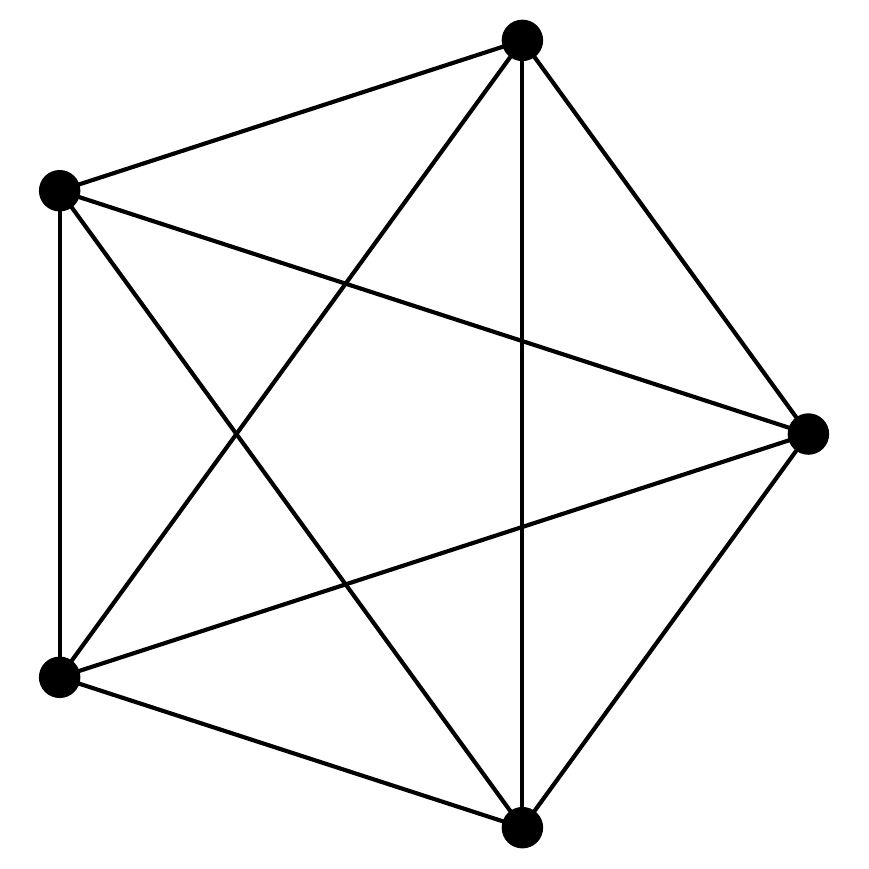}}
		\qquad \qquad \qquad 
		f^{(2)}=
		\raisebox{-.5\height}{		\includegraphics[width=2.2cm]{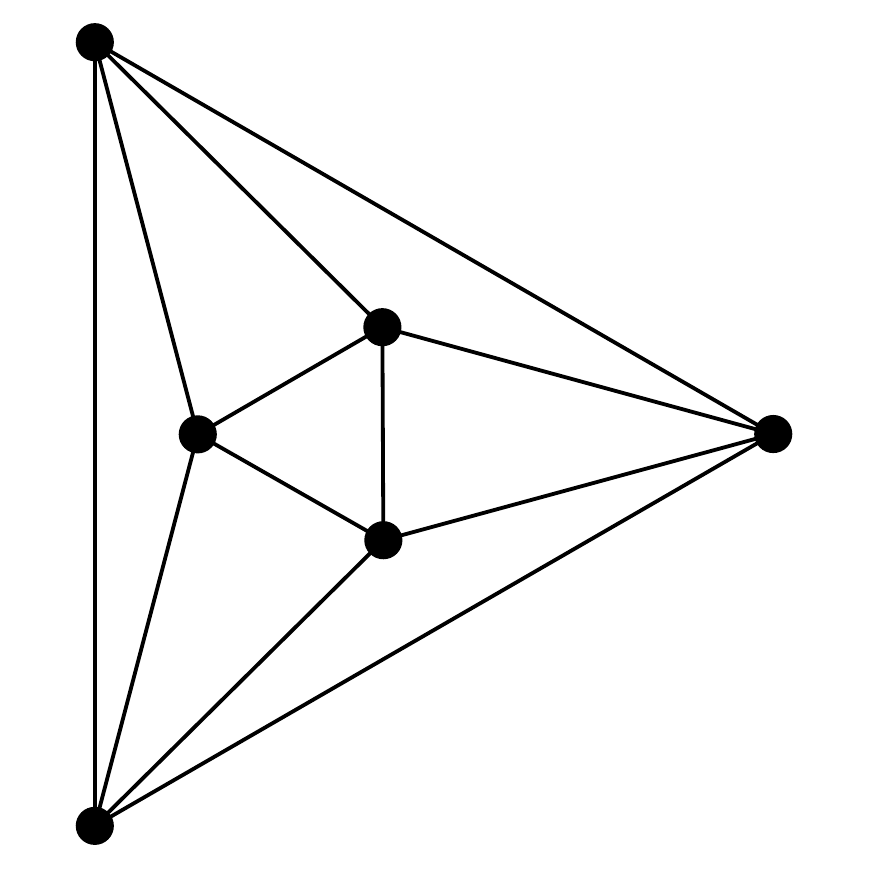}}
	\end{align}
	More generally  a possible contributing  $l$-loop $f$-graph is an $l+4$ point graph of (net) vertex degree 4. We may also have non-cancelled numerator terms $x_{ij}^2$ which we represent by dashed lines. Since the conformal weight at each point is 4, this means the number of edges minus the number of dashed edges (which we call the net vertex degree) equals 4 at each point. 
	A simple algorithm for obtaining all possible $f$ graphs is then to list all $(4{+}l)$- point graphs of vertex degree 4 or higher at each point, and then consecutively try to add numerator lines to any vertices of degree higher than 4 to bring the net vertex degree down to 4 at each point.
	
	This is true for any value of $N_c$. If we consider the large $N_c$ limit, there is a further powerful constraint on the allowed $f$ graphs, namely we expect all component correlator graphs to be {\em planar}. This suggests the $f$-graphs themselves should be planar also (excluding the numerator, dashed edges). We notice that in~\eqref{fgraph12} the graph $f^{(2)}$ is indeed planar.%
	\footnote{A little awkwardly $f^{(1)}$ is non-planar however. The reason for this is that the function $f$ itself is not the correlator, but appears multiplied by $x_{12}^4x_{34}^4$ (see~\eqref{corbasic}) which deletes some edges. When these edges are deleted the resulting correlator is represented by planar graphs. Moreover this is true for the four-point function of any other operators related by SUSY to $\cO(x)$.  However at higher loops the only way to achieve  planar graphs for all component correlators is for the $f$ graph itself to be planar.} 
	So at 3 loops we now simply look for all planar $f$ graphs, and we  we find that there is only 1 possibility!
	\begin{align}\label{3loop}
		f^{(3)}=	\raisebox{-.5\height}{	\includegraphics[height=2.5cm]{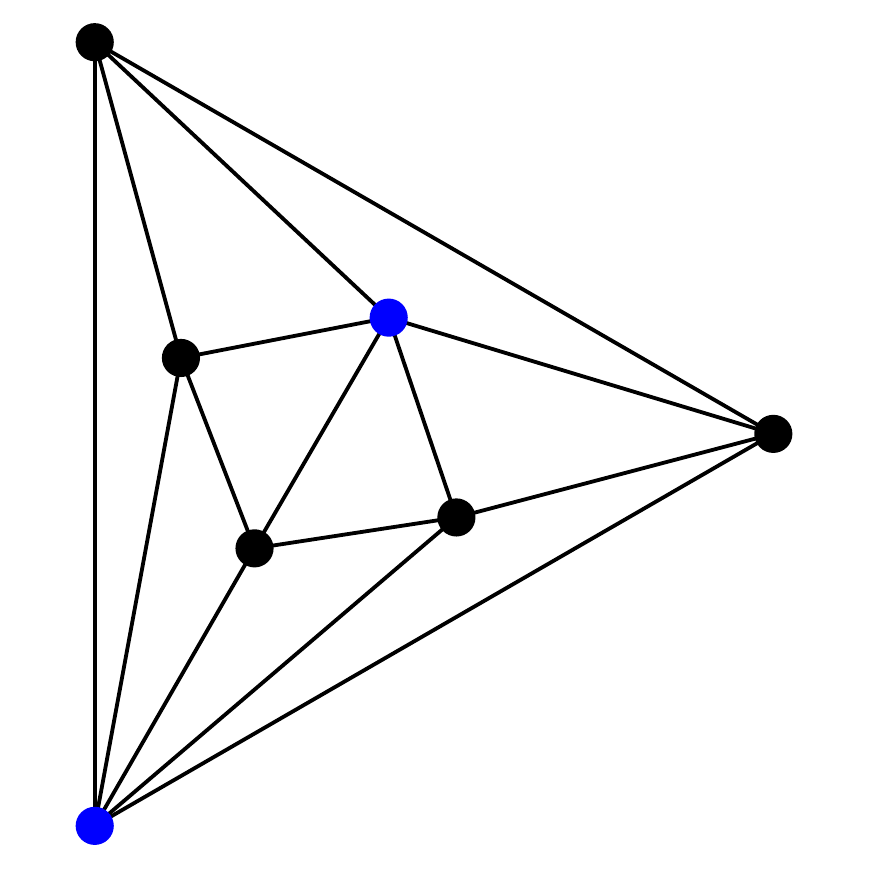}} =  {{1 \over 20} \sum_{\sigma \in S_7} x_{\sigma_1\sigma_2}^4 x_{\sigma_3\sigma_4}^2 x_{\sigma_4\sigma_5}^2 x_{\sigma_5\sigma_6}^2 x_{\sigma_6\sigma_7}^2
			x_{\sigma_7\sigma_3}^2 \over \prod_{1\leq i<j \leq 7}
			x_{ij}^2}
	\end{align}
	where we have suppressed a numerator line between the two blue nodes (where the colour blue indicates that the vertex has degree 5 and thus needs a numerator line to bring it back to net degree 4).
	Its overall coefficient (which is 1) can be fixed by various considerations the most direct being the correlator/amplitude duality.
	
	Continuing this process we find three planar 4-loop $f$-graphs and seven planar 5-loop $f$-graphs. Again various considerations (to be discussed shortly) fix all their coefficients to be $\pm 1$.
	We thus obtain the following remarkably compact expressions for the correlator to 5 loops
	\begin{align}
		f^{(4)}=&	\raisebox{-.5\height}{\includegraphics[height=2.5cm]{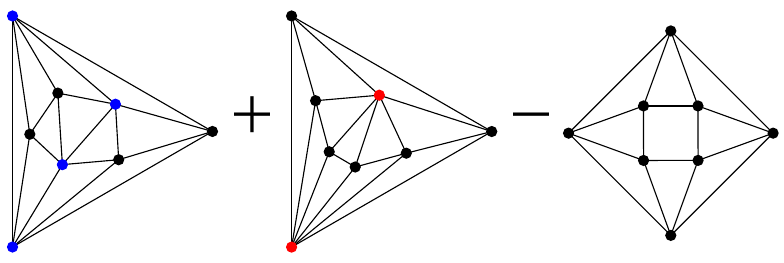}}\\
		f^{(5)}=&	\raisebox{-.5\height}{\includegraphics[height=2cm]{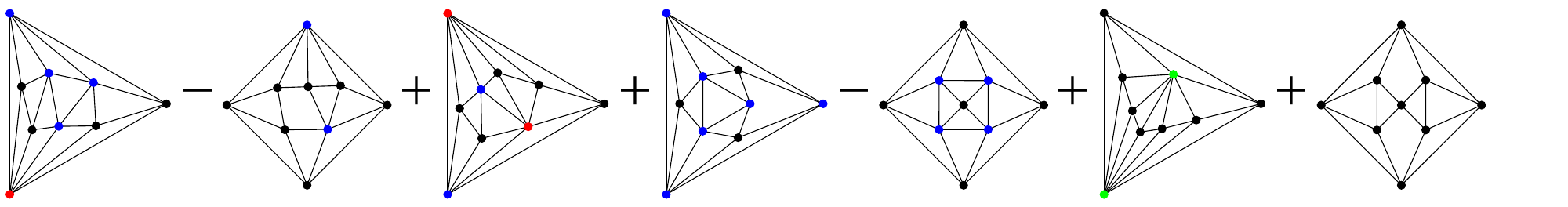}}\label{f5}
	\end{align}
	where again the numerator lines are suppressed in the figures but can be deduced by ensuring the net degree is 4 at each vertex.%
	\footnote{As above,  a blue vertex is degree 5 and needs one numerator line ending there, whereas a red vertex is degree 6 and needs two numerator lines and a green numerator line is degree 7 and needs 3 numerator lines ending there.}
	
	At 6 loops~\cite{Eden:2012tu} there are 36 planar $f$ graphs, and for the first time not  all of them contribute (some have vanishing coefficient). Furthermore, for the first time a non unit norm coefficient appears, namely the  pentagram graph comes with coefficient 2
	\begin{align}\label{6loops}
		f^{(6)}= \dots  + \ 2
		\raisebox{-.5\height}{	\includegraphics[width=2cm]{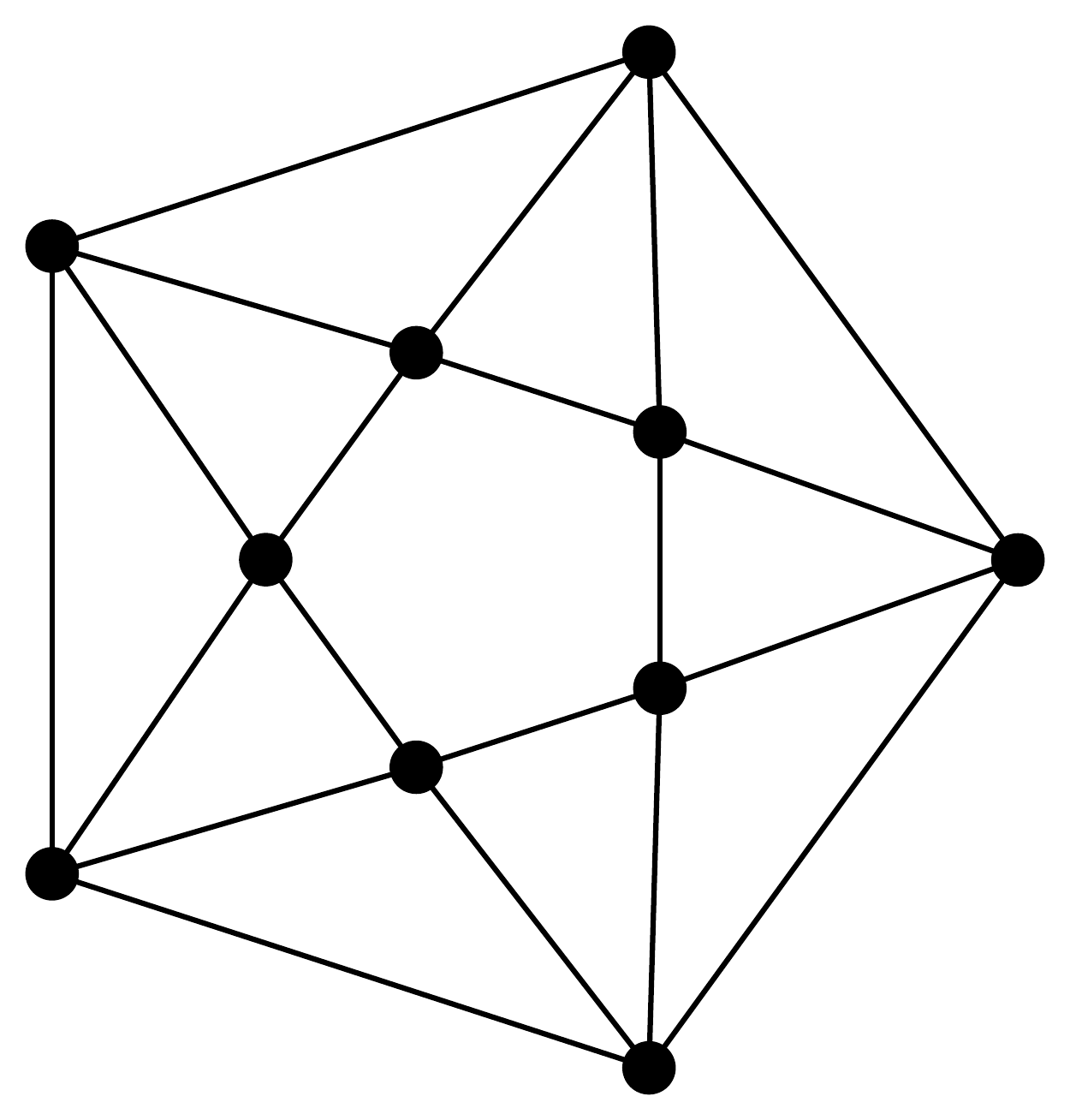}}
	\end{align}
	This program of listing all the planar $f$ graphs, and then fixing their  coefficients has been pursued up to ten loop order~\cite{Eden:2012tu,Bourjaily:2015bpz,Bourjaily:2016evz}. But how do we fix the coefficients?
	By far the most efficient method   is to deduce  equations which act  purely on the individual graphs themselves without having to evaluate their actual algebraic expressions (which in particular involves summing over $(4+l)!$ terms which for example is $8.7\times 10^{10}$ at ten loops!) 
	
	In~\cite{Bourjaily:2016evz} three graphical rules were implemented, called the triangle rule, the square rule and the pentagon rule. All three were needed to fix all the coefficients at ten loops. 
	The triangle rule is straightforward to state. It says that the result obtained by shrinking all triangles of all $f$ graphs at $l$ loops is twice the result of shrinking all edges at $(l-1)$ loops. Each side of the equation is a sum of graphs.
	% (but not $f$ graphs since they have the wrong net vertex degree after the shrinking). 
	The origin of this rule arises from considering the coincident limit $x_{i} \rightarrow x_j$ using the OPE, realising that the log of the correlator at $l$ loops only has the divergence of the 1 loop correlator, and understanding the implications of this reduced divergence at the level of the integrand~\cite{Eden:2012tu,Bourjaily:2015bpz}.
	
	The other rules, the square and pentagon rules arise from structural considerations of the correlator/amplitude duality.
	
	\subsection{Correlator/Amplitude duality for the four-point correlator}
	\label{sec:ampcor}
	
	In~\cite{Alday:2010zy} it was shown very generally that if you take   operator insertion points to be consecutively lightlike separated the resulting correlator  becomes proportional to a Wilson loop in the adjoint representation of the gauge group on the lightlike polygonal contour whose vertices are these insertion points. But the amplitude/ Wilson loop duality~\cite{Alday:2007hr,Drummond:2007aua,Brandhuber:2007yx,Bern:2008ap,Drummond:2008aq,Mason:2010yk,Caron-Huot:2010ryg} 
	relates the large $N_c$, $\cN=4$ SYM polygonal Wilson loop in the fundamental representation to scattering amplitudes. Furthermore, since in the planar limit a Wilson loop in the adjoint rep equates to the square of Wilson loops in the fundamental rep these relations imply that correlators should reduce to squares of amplitudes in the light-like limit. This is indeed the case as has been shown and proven in a number of works~\cite{Eden:2010zz,Alday:2010zy,Mason:2010yk,Eden:2010ce,Caron-Huot:2010ryg,Eden:2011yp,Adamo:2011dq,Eden:2011ku}. An important point is that this correlator/amplitude duality works directly at the level of the  {\em integrands},  and thus becomes a direct relation between rational functions, avoiding regularisation issues.
	
	The duality can be applied to correlators at any loop order and for any number of points. We will consider the light-like limit of general higher-point correlators  later, in section~\ref{sec:superampcor} but in this section we will consider applying $n$-gon light-like limits to the four-point correlator. Even this case has remarkably powerful  implications. 
	
	\subsubsection*{The four-point light-like limit}
	
	First we consider applying the four point polygonal light-like limit $x_{12}^2,x_{23}^2,x_{34}^2,x_{41}^2 \rightarrow 0$ to the connected ($O(c)$) part  of the correlator~\eqref{corbasic}. The duality states that this (divided by tree-level) yields the square of the four-point amplitude. More precisely
	\begin{align}\label{ampcor4pnt}
		\lim_{x_{12}^2,..,x_{41}^2 \rightarrow 0} \frac{\langle \cO \bar \cO \cO \bar \cO \rangle|_{c\phantom{,\lambda=0}}\!\!\!\!\!\!\!\!\!\!\!  } {\langle \cO \cO \bar \cO \bar \cO \rangle|_{c,\lambda=0}\!\!\!\!\!\!\!\!\!\!\!\! }\qquad \quad  =\quad \cA_4(x_i;\lambda)^2\ 
	\end{align}
	where $\cA_4(x_i;\lambda)$ is the planar four-point amplitude, divided by the tree-level amplitude,%
	%
	%\footnote{Or said more precisely,  the planar four-point amplitude is equal to the tree-level amplitude multiplied by $\cA_4(x_i;\lambda)$.}
	%
	and written in terms of region momenta so $p_i=x_{i\,i+1}$.
	Defining the amplitude integrand in the obvious way
	\begin{align}\label{coramp4}
		\cA_4(x_1,..,x_{4};\lambda)	&= \sum_{l=0}^\infty \frac{\lambda^l}{(4\pi^2)^ll!}\int \frac{d^4x_5}{(-4\pi^2)}..\frac{d^4x_{4+l}}{(-4\pi^2)}\cA_4^{(l)}(x_1,.,x_{4+l})\ ,
	\end{align}
	the duality~\eqref{ampcor4pnt} becomes 
	\begin{align}\label{ampcorf}
		\lim_{x_{12}^2,..,x_{41}^2 \rightarrow 0} \left(2 \xi^{(4)} \, f^{(l)}(x_i;\lambda)\right) = \sum_{l'=0}^l \binom{l}{l'}
		\cA_4^{(l')} \cA_4^{(l-l')}\ 
	\end{align}
	where $\xi^{(4)}=x_{13}^4x_{24}^4x_{12}^2 x_{23}^2x_{34}^2x_{14}^2$ and on the rhs the binomial coefficient is really shorthand for   a sum over all inequivalent  ways of distributing the $l$ loop variables between  $\cA_4^{(l')}$ and  $\cA_4^{(l-l')}$.
	
	This duality can be beautifully represented graphically on the $f$ graphs. Only terms in $f$ which contain all four poles $x_{12}^2 ..x_{41}^2$ survive the limit on the lhs of~\eqref{ampcorf}. Graphically this corresponds to a four-cycle. This breaks the planar graph into two halves which can be thought of as an `inside' and an `outside'. The number of vertices inside is $l'$ in~\eqref{ampcorf} and the number of vertices outside is then $l-l'$. Furthermore the graph inside the four cycle will be a graph contributing to the amplitudes $\cA_4^{(l')}$. So graphically from any $f$ graph we read off the corresponding amplitude graphs contributing to the product by taking the inside and outside of the 4-cycle 
	\begin{align}
		\begin{tikzpicture}
			\node[inner sep=0pt] (cor2amp) at (0,0)	{	\includegraphics[width=3.5cm]{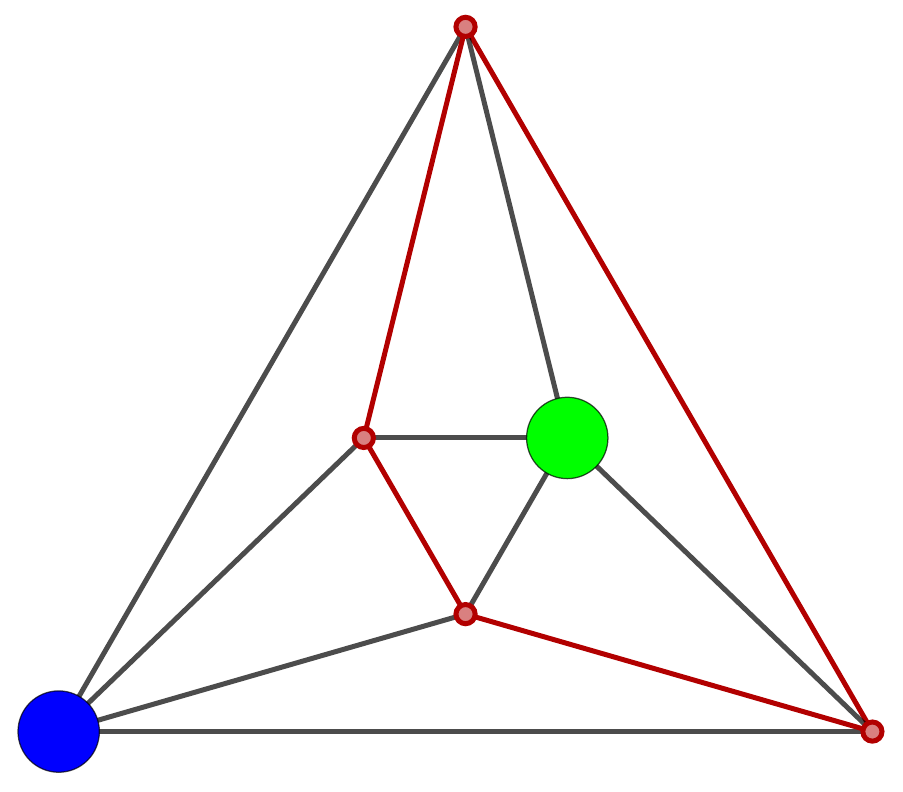}};
			\node[inner sep=0pt] (arrow) at (3,0)	{	$\rightarrow$ };
			\node[inner sep=0pt] (amp1) at (6,0)	{	\includegraphics[width=2cm]{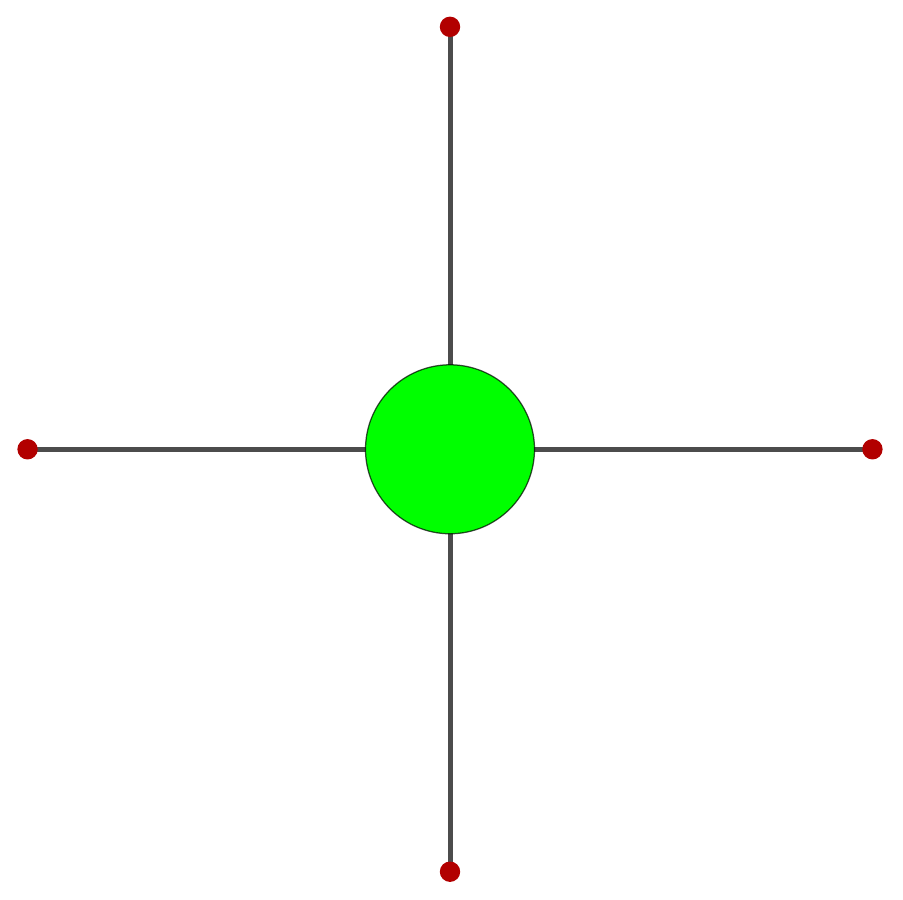}};  
			\node[inner sep=0pt] (times) at (8,0)	{	$\times$ };
			\node[inner sep=0pt] (amp2) at (10,0)	{	\includegraphics[width=2cm]{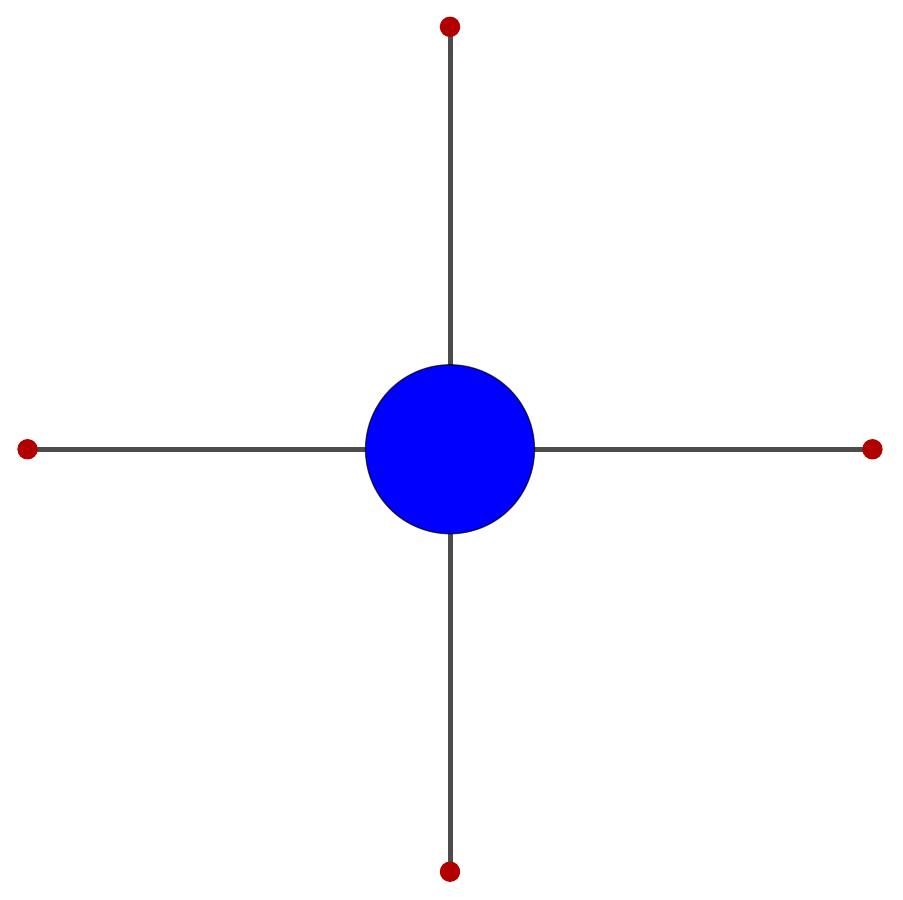}};
			\node[inner sep=0pt] (lp) at (.6,.2)	{	\scriptsize{$l'$}};
			\node[inner sep=0pt] (lp2) at (-1.7,-1)	{	\scriptsize{$l{-}l'$}};
			\node[inner sep=0pt] (lp) at (6.3,.3)	{	\scriptsize{$l'$}};
			\node[inner sep=0pt] (lp2) at (10.5,.3)	{	\scriptsize{$l{-}l'$}};
			\node[inner sep=0pt] (lp2) at (0,-2)	{$f^{(l)}$-graph contribution};
			\node[inner sep=0pt] (lp2) at (8,-2)	{ 	$\cA_4^{(l')} \times \cA_4^{(l-l')}$ amplitude graph contributions};
		\end{tikzpicture}
	\end{align}
	This is true for any value of $l'$ including $l'=0$ (or equivalently $l'=l$) in which case the inside or outside is empty. 
	This has two immediate implications. 
	
	First, extracting the amplitude $\cA_4^{(l)}$ from $f^{(l)}$ is simply a case of taking all 4 cycles with no inside (or alternatively no outside) from all $f$ graphs i.e. taking all quadrilateral faces (where by quadrilateral face we include two adjacent triangular faces). Importantly the resulting amplitude graphs obtained inherit the coefficient of the $f$ graph. In this way the  four-point amplitude of $\cN=4$ SYM to ten loops has been  obtained directly from the correlator~\cite{Bourjaily:2016evz}.

	But second, this gives a recursive rule for building higher loop $f$ graphs, with their coefficients, from lower loop ones. Take any two  quadrilateral faces of two  $f$-graphs and glue them together along this face. The result will be an $f$-graph at  higher loops  contributing with coefficient given by the product of the coefficients of the two original $f$ graphs.
	\begin{align}
		\begin{tikzpicture}
			\node[inner sep=0pt] (cor2amp) at (0,0)	{	\includegraphics[width=3cm]{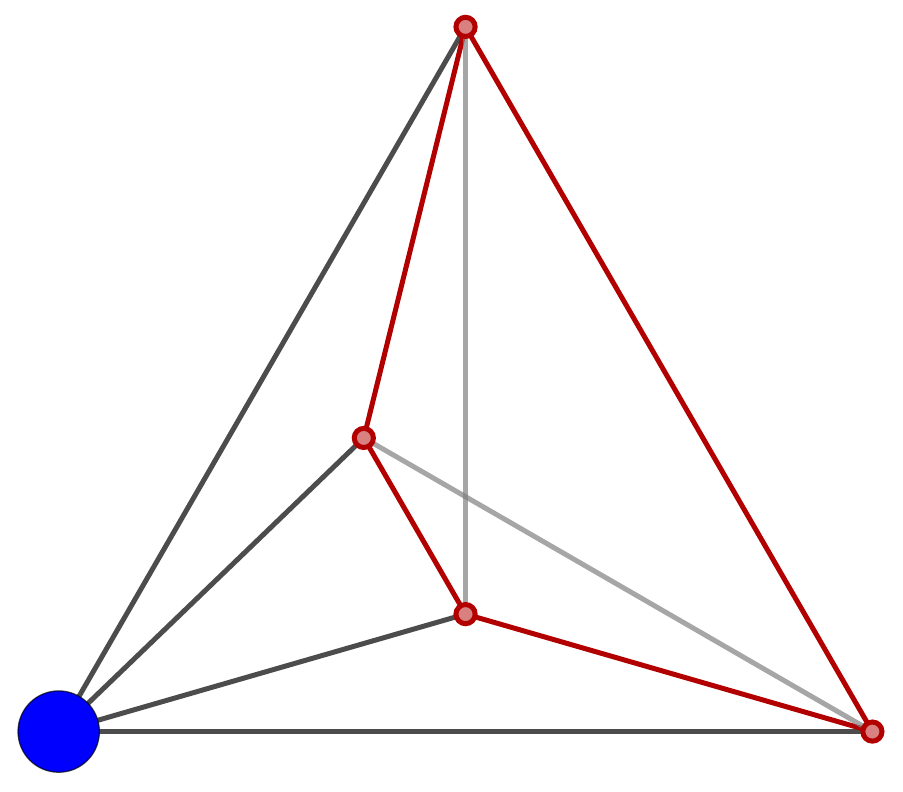}};
			\node[inner sep=0pt] (amp1) at (4,0)	{	\includegraphics[width=2cm]{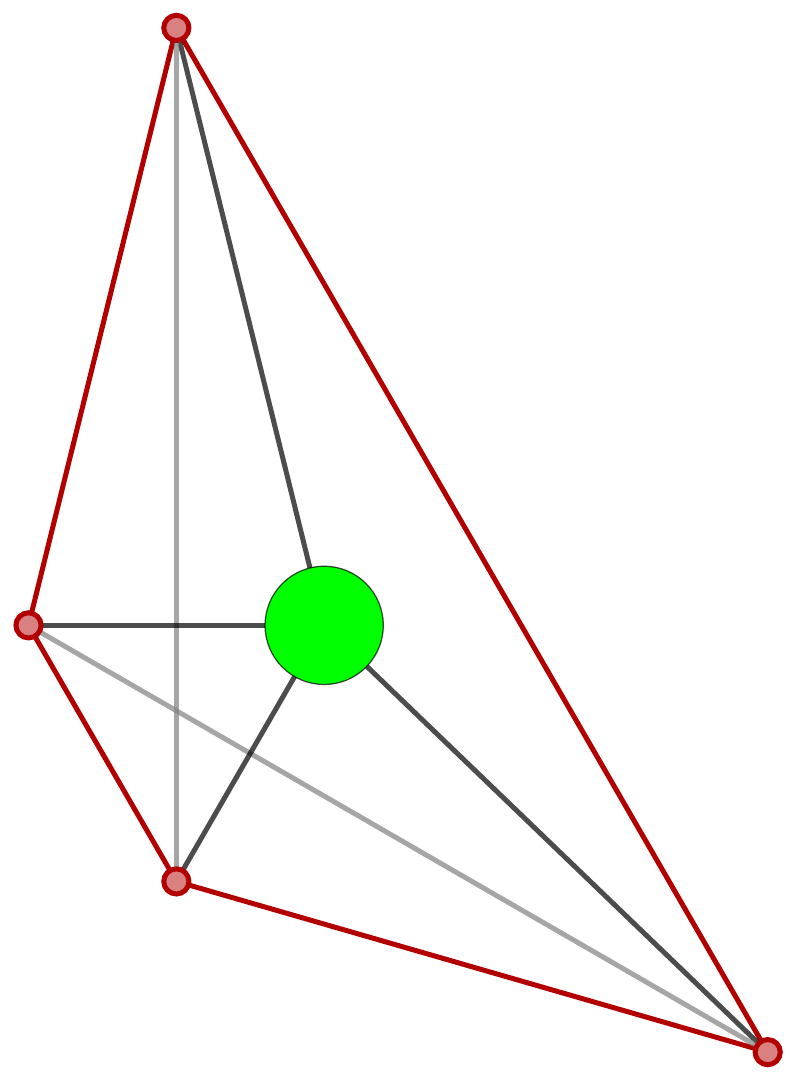}};  
			\node[inner sep=0pt] (times) at (6.5,0)	{	$\rightarrow$ };
			\node[inner sep=0pt] (times) at (6.5,-2)	{	$\rightarrow$ };
			\node[inner sep=0pt] (amp2) at (10,0)	{	\includegraphics[width=3.5cm]{cor2ampgraph}};
			\node[inner sep=0pt] (lp) at (10.6,.2)	{	\scriptsize{$l'$}};
			\node[inner sep=0pt] (lp2) at (8.3,-1)	{	\scriptsize{$l{-}l'$}};
			\node[inner sep=0pt] (lp) at (3.9,.2)	{	\scriptsize{$l'$}};
			\node[inner sep=0pt] (lp2) at (-1.7,-1)	{	\scriptsize{$l{-}l'$}};
			\node[inner sep=0pt,align=center]  at (0,-2)	{$f^{(l-l')}$-graph has \\ coefficient $a$ };
			\node[inner sep=0pt,align=center]  at (4,-2)	{$f^{(l')}$-graph has \\ coefficient $b$ };
			\node[inner sep=0pt,align=center] (lp2) at (10,-2)	{$f^{(l)}$-graph  has \\ coefficient $ab$ };
		\end{tikzpicture}
	\end{align}
	Note that the two $f$-graphs on the lhs look non-planar due to the grey lines. However,  at most one of these grey lines is actually present due to the existence  of numerators $x_{13}^2$ or $x_{24}^2$ which will cancel the grey  lines. The numerator lines are implicit in the blue or green blobs, representing amplitude graphs. Indeed the only case where no grey line are cancelled by numerators is the 1 loop case where we know that the $f$ graph is in fact non-planar.

	The above  directly leads to the `square rule' when we consider $l'=1$, and corresponds to  gluing pyramids on the quadrilateral face~\cite{Eden:2011yp,Eden:2012tu,Bourjaily:2016evz}. 
	This was one of the constraints used to fix the correlator to 10 loops~\cite{Bourjaily:2016evz} as mentioned at the end of section~\ref{sec:hidden}.%
	\footnote{Although the case $l'>1$ can  give more general predictions that were not used in~\cite{Bourjaily:2016evz}. I thank Gabriele Dian for discussions and for performing explicit checks of this point.}

	\subsubsection*{The higher-point light-like limit}
	
	The higher-point amplitude/ correlator duality has implications even for the  four-point correlator. This is due to the fact that  the  four-point loop level correlator is itself a higher point correlator and then takes part in the higher point amplitude duality. We will derive this in detail in section~\ref{sec:superampcor}. For now we just consider the implications which are that the $n$-point polygonal light-like limit $x_{12}^2,x_{23}^2,...,x_{n1}^2 \rightarrow 0$ of the {\em four}-point correlator integrand gives a sum of products of  $n$-point N${}^k$MHV $l$ loop superamplitude integrands divided by the tree-level MHV amplitude $\mathcal{A}_{n;k}^{(l)}$ as follows:
	\begin{align}\label{npll}
		\lim_{x_{12}^2,..,x_{n1}^2 \rightarrow 0} \Big(2  \xi^{(n)}f^{(l+n-4)}\Big)
		=\sum_{l'=0}^{l}\sum_{k=0}^{n-4}\binom{l}{l'} \frac{\mathcal{A}_{n;\hspace{0.5pt}k}^{(l')}\,\mathcal{A}_{n;\hspace{0.5pt}n-4-k}^{(l-l')}}{\mathcal{A}_{n;\hspace{0.5pt}n-4}^{(0)}}\ .
	\end{align}
	where 
	\begin{align}
		\xi^{(n)}\equiv\prod_{i=1}^n x^2_{i\,i+1}x^2_{i\,i+2}\ .\label{xi_general_definition}
	\end{align}
	
	Similarly to the four-point case above,  this relation can be used for $n=5$ to extract the full $5$-point amplitude from the 4-point correlator~\cite{Ambrosio:2013pba} graphically. 
	The limit~\eqref{npll} becomes
	\begin{align}\label{5pll}
		\lim_{x_{12}^2,..,x_{51}^2\rightarrow 0} \Big(  \xi^{(5)}f^{(l+1)}\Big)
		=\sum_{l'=0}^{l}\binom{l}{l'} \mathcal{A}_{5}^{(l')}\, \overline{\mathcal{A}}_{5}^{(l-l')}\ .
	\end{align}
	The five-point amplitude is a little more complicated than the 4-point case. One only has MHV and $\overline{\text{MHV}}$ amplitudes, both of which factorise into tree-level times integrals depending on $x$ only. But the amplitude integrands are no longer functions of $x_{ij}^2$ only. Rather they  split into a parity even $\cA + \overline \cA$ and a parity odd $\cA - \overline \cA$ part. While the parity even part only involves $x_{ij}^2$, the parity odd part contains another object, most neatly represented in the 6d embedding space formulation of 4d Minkowski space, using a 6d epsilon-tensor contracted with 6 $x_i$s,  $\epsilon_{i_1i_2 i_3 i_4 i_5 i_6}$, which maintains manifest dual conformal invariance. The  parity even part of the 5-point amplitude at $l$ loops can be extracted graphically directly from $f^{(l+1)}$ in a completely analogous way to the extraction of the four-point amplitude. Namely taking all pentagonal faces (by which  we include three adjacent triangular faces and adjacent triangle/quadrilateral faces), the graphs remaining after removing the pentagon gives the amplitude graphs (with coefficient) of the parity even part of the five point amplitude.
	To obtain the parity odd part we need to go one loop higher, to $f^{(l+2)}$,   and extract all pentawheels (pentagons with a single vertex inside and 5 spokes from this vertex to the pentagon). A pentawheel necessarily has a numerator line from its central vertex (to produce net degree 4) to some other vertex in the remaining part of the $f$ graph. We now remove the pentawheel and mark the vertex which the numerator line ends on. The resulting marked graph (with accompanying coefficient)  contributes directly to the parity odd part of the $l$-loop  5-point amplitude.
	
	Consistency of this 5-point lightlike limit is quite non-trivial. For example the product  of two parity odd amplitudes involves the product of epsilons, which leads to a number of terms, some of which output non-planar $f$-graphs which must cancel in the sum. This then leads to  consistency conditions relating $f$ graphs at the same loop order. One such consistency condition was extracted and lead to the `pentagon rule' which is used to fix the correlator to 10 loops~\cite{Bourjaily:2016evz} as mentioned at the end of section~\ref{sec:hidden}.

	At higher points, $n>5$,  the duality with the four-point correlator~\eqref{npll}  becomes even more non-trivial, although  the consequences are harder to extract at least from a purely graphical approach. Here the rhs of~\eqref{npll} inevitably involves non trivial N${}^k$MHV amplitudes which have complicated particle dependent rational contributions. Nevertheless there is strong evidence that by assuming Yangian invariance and dual conformal invariance,   {\em all $n$-point, loop-level, any $k$ amplitudes can be extracted from the four-point correlator~\cite{Heslop:2018zut}}. More precisely, writing the  scattering amplitudes as an arbitrary sum of N${}^k$MHV Yangian invariant rational terms $R_{k;\hspace{0.5pt}i}$ times $l$-loop dual conformal invariant integrands $\cI^{(\ell)}_j $
	\begin{align}
		\mathcal{A}_{n;\hspace{0.5pt}k}^{(\ell)} = \sum_{ij}\alpha_{ij}  R_{k;\hspace{0.5pt}i} \hspace{0.5pt} \cI^{(\ell)}_j  ,\label{ansatz}
	\end{align}
	and inserting this into the duality~\eqref{npll}, then the 4-point correlator completely fixes $ \mathcal{A}_{n;\hspace{0.5pt}k}^{(\ell)}$ (see~\cite{Heslop:2018zut}).
	It would be fascinating to explore this further and see if there is a  more direct, graphical way to extract higher point amplitudes from the four-point correlator and furthermore to understand more systematically what constraints the existence of this duality imposes on the correlator.
	
	There are also other graphical possibilities for constructing the higher loop  correlator, arising from the amplituhedron. In particular the deep cuts of~\cite{Arkani-Hamed:2018rsk,Langer:2019iuo} can be interpreted as highly non-trivial, even constructive, graphical rules  for the correlator~\cite{dhs}.

	\subsection{The Non-planar Correlator}
	\label{sec:np}
	Having reviewed the story for perturbative planar correlators  let us now consider the non-planar theory perturbatively. 
	From equation~\eqref{3loop} onwards we derived the $f$-graph basis for the large $N_c$ 4-point correlator using conformal symmetry, hidden permutation symmetry, knowledge of the pole structure and {\em planarity}. We now ask what happens if we drop planarity. The listing of the basis of  $f$-graphs continues in the same way, but simply without  imposing that the $f$ graphs must be planar. It is then easier to consider numerator graphs rather than $f$-graphs (the advantage of $f$ graphs in the planar case is purely that we can impose planarity). A numerator graph is a graph obtained by multiplying the $f$ graph by the product of all possible poles $\prod_{i<j}x_{ij}^2$ to obtain a polynomial in $x_{ij}^2$. The polynomial will have weight  $l-1$ at each point. Thus associating $x_{ij}^2$ with an edge between vertex $i$ and $j$, then we obtain  a degree $l-1$ graph. However  we can have repeated  edges arising from  eg $x_{ij}^4$. Thus this numerator becomes equivalent to a degree $l-1$ multi-graph on $l+4$ vertices (the multigraph can also be disconnected unlike the $f$ graph).
	
	At 2-loops, the only degree 1 multigraph on 6 vertices is given by 3 disconnected edges. The numerator in~\eqref{2loop} is precisely of this form.   
	
	At 3-loops we can list all  degree 2 multigraphs on 7 vertices. There are four possibilities,  corresponding to a 7-cycle, a 5-cycle  $\times$ 2-cycle, 4-cycle $\times$ 3-cycle and 3-cycle$\times$2-cycle$\times$2-cycle. The only one that produces a planar $f$ graph is the 5-cycle  $\times$ 2-cycle which one can observe agrees with the numerator of~\eqref{3loop}. The other three produce non-planar $f$ graphs. To fix the coefficients one can use the triangle rule~\cite{Eden:2012tu}%
	\footnote{In~\cite{Eden:2012tu} the double coincidence limit was used but this is equivalent to the triangle rule.}
	which does not rely on planarity
	(unlike the square and pentagon rules which are intrinsically planar).
	This fixes the 3-loop non-planar result  up to a single free coefficient. However examining more carefully one sees that this free coefficient multiplies a combination of numerator graphs which becomes algebraically equal to the fully permutation invariant, conformally invariant,   vanishing Gram determinant  $\det(x_i.x_j)$ with $i,j=1,..,7$ (the $x$s are in the 6d embedding space formalism). 
	Thus the non-planar (all $N_c$ ) 3-loop correlator is completely fixed and is in fact equal to the large $N_c$ planar correlator~\eqref{3loop}.
	
	Moving to 4-loops a similar analysis can be performed. First listing all numerator graphs which here are degree 3 multi-graphs on 8 vertices, one obtains  32  possibilities. Then fix coefficients using the triangle rule relating it to lower loops.  In fact it turned out that a more powerful (algebraic rather than graphical) rule involving a lightlike limit and arising from the duality between correlators and Wilson loops fixes more coefficients. This fixed all but 7 coefficients. But as before there are again conformally covariant permutation invariant vanishing  Gram determinants. This time there are three independent combinations of the numerator polynomials which are vanishing in 4d. 
	Thus one is left with 4 remaining free unfixed coefficients for the 4 loop non-planar correlator. 
	These four remaining coefficients have themselves been fixed in  a remarkable computation~\cite{Fleury:2019ydf} using the formulation of correlators in twistor space in~\cite{Chicherin:2014uca} (reviewed briefly in section~\ref{sec:twistor}).

	\subsection{Integrals}
	
	The discussion so far has been focussed on the {\em integrand} of loop level correlators, which are themselves physical, Born-level, higher point correlators and as we have seen possess additional symmetries. But it is of course interesting to perform the integrals in~\eqref{Finf} not least because this then allows the extraction of non-trivial CFT data via a conformal partial wave decomposition.

	The integrals contributing to the correlator $F^{(l)}$ in~\eqref{Fl} are conformally invariant 
	and hence depend on the $x_i$ through the two independent cross-ratios
	$	F^{(l)}(x,\bar x)$
	where the cross-ratios $x,\bar x$ are defined as:
	\begin{align}\label{ftoFsimp}
		x \bar x &= u=\frac{x_{12}^2 x_{34}^2}{x_{13}^2 x_{24}^2} \qquad &(1{-}x) (1{-}\bar x) =v= \frac{x_{14}^2 x_{23}^2}{x_{13}^2 x_{24}^2}\ .
	\end{align}
	
	At 1-  and 2-loops the integrals one obtains are all ladder diagrams with known explicit  expressions in terms of polylogs~\cite{Usyukina:1993ch}. At 3-loops there are two new types of integrals, dubbed ``easy'' and ``hard'' in~\cite{Eden:2011we}. These integral were obtained in~\cite{Drummond:2013nda}. To obtain them an assumption about the functional form was used, namely that they had the form rational $\times$  (generalised) polylogarithms of total weight 6 (using the principle of uniform transcendentality which is  a property of many quantities in $\cN=4$ SYM). We assume that one can write the symbol~\cite{Goncharov:2010jf} of the polylogs in such a way that the only letter appearing  in each term is $x,\bar x, 1-x,1-\bar x$ and $x-\bar x$. Furthermore they are ``single-valued'', implying that the symbol can also be written such that first entry of every term is a $u$ or a $v$.
	Leading singularity methods were used to determine the rational prefactors. The asymptotic expansion  of the integrals as one of the cross ratios vanishes was obtained in~\cite{Eden:2012rr}. These asymptotics together with crossing symmetry are enough to uniquely fix the symbol which can then be integrated up to give the full result~\cite{Drummond:2013nda}. Interestingly, most of the integrals only have $x,\bar x, 1-x,1-\bar x$ in the symbol. Such polylogarithms are called harmonic polylogarithms and the single valued ones are known as single-valued harmonic polylogarithms and have nice properties, in particular there is 
	a straightforward linear basis for them~\cite{brown2004single}. The hard integral on the other hand requires the additional  $x-\bar x$ letter, and is written in terms of the more general Goncharov polylogarithms. Recently such single valued Goncharov polylogarithms have also been understood~\cite{Schnetz:2021ebf}.

	At 4-loops the correlator is still not known at the integrated level, although it is possible to evaluate the integrals with current technology, and one such non-trivial case was done in~\cite{Eden:2016dir}, and so the remaining correlator is presumably within reach. It is interesting that the same types of integral are also of interest from a more mathematical, number theoretic, perspective and are known in that context as graphical functions~\cite{Borinsky:2021gkd}. 
	
	Note that although the correlators themselves are not known fully beyond 3 loops, it has nevertheless still proven possible to extract certain  data at higher loop order directly from the integrands without fully integrating them. 
	In particular, in~\cite{Eden:2012fe} the anomalous dimension of the Konishi operator  (the lowest dimension operator with an anomalous dimension in the theory) was obtained up to 5 loops, by manipulating the integrands in~(\ref{fgraph12})-(\ref{f5}) and performing a simpler integral of one loop lower. This used the fact that this anomalous dimension is the coefficient of the leading single logarithmic singularity of the logarithm of the correlation
	function in short-distance limit in which two operator positions coincide (this is very closely related to the derivation of the triangle rule). Similar methods, but keeping more terms in the asymptotic expansion around the singular limit can also give OPE coefficients for the Konishi in the $\cO \bar \cO$ OPE. This has also now been achieved to 5 loop order in the planar theory~\cite{Eden:2012rr,1607.02195,1608.04222,1710.06419} and four-loops in the non-planar theory~\cite{Fleury:2019ydf}.
	
	Note that in this subsection we have discussed obtaining  analytic correlators by integrating integrands that have been obtained by bootstrap-type methods. For scattering amplitudes in ${\cal N}{=}4$ SYM there has been a very successful programme (see SAGEX review chapter 5~\cite{chapter5}) bootstrapping analytic amplitudes directly and thus bypassing integrands completely. It would be fascinating to attempt a similar approach on the correlator side, although the more complicated structure of the leading singularities suggests it would be an even more intricate story than for amplitudes.

	\subsection{Strong coupling AdS/CFT}
	\label{sec:strong}
	Having covered what is known about the simplest half BPS four point function in perturbation theory, we now turn to the other extreme, strong coupling $\lambda \rightarrow \infty$. Here there is a whole new and fascinating story which we will review  (see also~\cite{Bissi:2022mrs} for a recent review overlapping with this topic). 
	
	The AdS/CFT correspondence~\cite{Maldacena:1997re,Gubser:1998bc,Witten:1998qj} relates half BPS correlators to supergraviton scattering amplitudes in IIB string theory on AdS${}_5\times S^5$ space. In the current context, the simplest half BPS operator discussed here, $\cO$, relates to amplitudes of particles purely in AdS${}_5$, with no dependence on the sphere.
	In the string theory / quantum gravity  dual,  $1/c=G_N$ Newton's constant and $1/\sqrt{\lambda}=\alpha'$ the inverse string tension.  
	Thus now  we are expanding
	in small $G_N,\alpha'$ which means a $1/\lambda$, $1/c$ expansion  around infinite $\lambda, c$. 
	The results of this section are summarised in figure~\ref{fig1}.
	
	Consider  the expansion in $1/c$ of the correlator $F$ (recall its definition~\eqref{corbasic})
	\begin{align}
		F(x,\bar x;\lambda,c) = \sum_{g=0}^\infty   c^{-g} F^{SG;(g)}(x,\bar x;\lambda)\ .
	\end{align}
	
	\subsubsection*{Tree-level supergravity}
	
	The genus 0 term $F^{SG;(0)}(x,\bar x;\infty)$ at infinite $\lambda$ will correspond,  according to AdS/CFT,  to tree-level supergravity on AdS. More concretely this function can be  directly read off from the quartic terms in the supergravity action linearized around AdS space. It was obtained in this way a few years after the AdS/CFT correspondence was discovered in~\cite{Liu:1998ty, hep-th/9903196,hep-th/9911222,hep-th/0002170,Arutyunov:2002fh} and found to be given by the expression
	\begin{equation}\label{sugra}
		F^{SG;(0)}(x,\bar x;\infty) = - \tfrac12u v \partial_u \partial_v(1+u \partial_u + v \partial_v) \Phi^{(1)}(u,v) = -\tfrac12 u v\bar{D}_{2422}(u,v)\,,
	\end{equation}
	where 
	\begin{align}
		\Phi^{(1)}(u,v)= \tfrac1{x-\bar x} \left(\log(u)\log\left(\tfrac{1-x}{1-\bar x}\right)+ 2\text{Li}_2(x)-2\text{Li}_2(\bar x)     \right)
	\end{align}
	is the one-loop scalar box integral 
	and we also give the expression in terms of scalar contact AdS Witten diagrams (in the form of $\bar{D}$-functions introduced in \cite{hep-th/0112251} and  defined below in~\eqref{Dbar} which are derivatives of the one loop scalar box function).
	
	\subsubsection*{Tree-level  string corrections}
	
	The $1/\lambda$ corrections to this result, corresponding to tree-level string corrections,   have  polynomial Mellin amplitudes and finite spin support in their conformal partial wave decompositions~\cite{0907.0151,1011.1485,1410.4717}. The first correction, at $O(1/\lambda^{3/2})$, is  constant in Mellin space and this constant was obtained in~\cite{Goncalves:2014ffa} by comparing with the flat space limit.  Indeed certain contributions to all orders in $1/\lambda$ can be read off directly from the flat space amplitude (which is known to all orders in $\alpha'=1/\sqrt \lambda$, the Virasoro Shapiro amplitude). However lifting from flat space to curved space is not unique, as can be seen by considering the corresponding effective action. Derivative terms  in the effective action lift to covariant derivatives which no longer commute,  thus commutator terms will vanish in the flat space limit and be undetermined by it.
	Despite this, higher order corrections have been pinned down in~\cite{Binder:2019jwn,Chester:2020dja} up to order $1/\lambda^3$ using supersymmetric localisation in $\cN=4$ SYM. Indeed the coefficients are even known as full functions of the string coupling in terms of generalised Eisenstein series~\cite{Chester:2019jas,Chester:2020vyz}.

	The results are most simply quoted as Mellin amplitudes $M(s,t;\lambda)$ defined by
	\begin{align}
		F(x,\bar x;\lambda,c)
		= \frac12\int_{-i \infty}^{i \infty} \frac{ds\, dt}{(4 \pi i)^2} \big(x \bar x\big)^{\frac s2} \big((1{-}x)(1{-}\bar x)\big)^{\frac t2 - 2}
		\GGamma \Big[2 {-} \frac s2 \Big]^2 \GGamma \Big[2 {-} \frac t2 \Big]^2 \GGamma \Big[2 {-} \frac u2 \Big]^2
		M(s, t;\lambda,c)
	\end{align}
	with $u{=}4{-}s{-}t$.  Mellin amplitudes  can be viewed as analogues of flat space amplitudes and indeed in the large $s,t,u$ limit they reduce to flat space amplitudes with $s,t,u$ the Mandelstam invariants~\cite{1011.1485}.
	The Mellin amplitude for the supergravity solution~\eqref{sugra} has the very simple form $M^{SG;{(0)}}(\lambda{=}\infty)=8/((s{-}2)(t{-}2)(u{-}2))$. The string corrections are symmetric polynomials in $s,t,u$, closely mimicking the behaviour of flat space amplitudes in momentum space.
	The result for the next few $\alpha'=1/\sqrt{\lambda}$ corrections, obtained in~\cite{Binder:2019jwn,Chester:2020dja} is
	\begin{align}\label{m0}
		M^{(0)}(s, t;\lambda)=&	\frac{8}{(s{-}2)(t{-}2)(u{-}2)}{+}\frac{120\zeta(3)}{\lambda^{\frac32}}   {+}\frac{630\zeta(5)}{\lambda^{\frac52}}\left[s^2{+}t^2{+}u^2{-}3\right] \notag \\
		& {+}\frac{5040\zeta(3)^2}{\lambda^{3}}\left[stu{-}\frac14(s^2{+}t^2{+}u^2){-}4\right]{+}O(\lambda^{{-}3})\ .
	\end{align}
	These correspond to $R^4$ type corrections to the string effective action. The conversion of these expressions to position space is a straightforward procedure~\cite{1011.1485}. They correspond via AdS/CFT to   contact 4-point scalar Witten diagrams
	\begin{align}\label{adscont}
		D_{\DDelta_1\DDelta_2\DDelta_3\DDelta_4}(x_i) = \cN^{-1}\int_\text{AdS}  \frac{d^{d+1} z}{(z.x_1)^{\DDelta_1}(z.x_2)^{\DDelta_2}(z.x_3)^{\DDelta_3}(z.x_4)^{\DDelta_4}}\ ,
	\end{align} 
	where $z$ is a $(d{+}2)$-component   bulk coordinate, which  contracts with $d$ dimensional Minkowski space coordinates in the  $d+2$ component embedding space formalism,  $x_i$, using the $SO(2,d)$ metric, $z.x_i$. The integral in~\eqref{adscont} is taken over the $(d{+}1)$-dimensional  subspace  corresponding to AdS${}_{d+1}$. The normalisation is 
	\begin{align}\label{norm}
		\cN^{\text{AdS}_{d+1}}_{\Delta_i}=\frac{\tfrac12 \pi^{d/2} \GGamma(\Sigma_\DDelta{-}d/2)(-2)^{\Sigma_\DDelta}}{\prod_i\GGamma(\DDelta_i)}\ ,
	\end{align}
	where $\Sigma_\DDelta=(\DDelta_1{+}\DDelta_2{+}\DDelta_3{+}\DDelta_4)/2$ which makes the integral independent of the space-time dimension $d$ (as a function of invariants Lorentz invariants $x_{ij}^2$).
	All contact diagrams can be expressed in terms of derivatives of the scalar box function $\Phi^{(1)}$~\cite{Arutyunov:2002fh}. A prefactor is commonly pulled out of the contact diagrams to make them conformally invariant $\bar D$ functions
	\begin{align}\label{Dbar}
		\bar D({u,v}):= \frac{(x_{13}^{2})^{\Sigma- \DDelta_4}(x_{24}^{2})^{\DDelta_2}}{(x_{14}^{2})^{\Sigma-\DDelta_1-\DDelta_4}(x_{34}^{2})^{\Sigma-\DDelta_3-\DDelta_4}} D_{\DDelta_1\DDelta_2\DDelta_3\DDelta_4}(x_i)\ .
	\end{align}

	For example the first correction at $O(\lambda^{-3/2})$ is proportional to  $ D_{4444}$, the Witten contact diagram for the four-point function of four dimension 4 scalars. This is no coincidence in fact as it turns out that the entire quartic interaction sector of the string theory effective action on AdS${}_5\times S^5$  can be written in terms of a single, dimension 4 scalar action~\cite{Abl:2020dbx} (see section~\ref{sec:hpstrong}).

	\subsubsection*{Quantum gravity loop corrections}

	The first $1/c$ correction, $F^{SG;(1)}|_{\lambda^0}$   corresponding to quantum gravity loop corrections in AdS${}_5\times S^5$, is now known , computed by OPE bootstrap  techniques in $\cN=4$ SYM. More precisely, the $1/\lambda$ expansion of the one loop correction actually has a super-leading term at $O(\lambda^{1/2})$, which we will discuss shortly, and the expansion takes the following form
	\begin{align}\label{oneloop}
		F^{SG;(1)} =  {\lambda^{1/2}}F^{SG;(1)}|_{\lambda^{{\scriptscriptstyle{1/2}}}} + F^{SG;(1)}|_{\lambda^0} + {\lambda^{-1}}F^{SG;(1)}|_{\lambda^{{\scriptscriptstyle{-1}}}} + {\lambda^{-3/2}}F^{SG;(1)}|_{\lambda^{{\scriptscriptstyle{-3/2}}}} + \dots \ .
	\end{align}
	All four of the above terms in the expansion are known completely.
	
The one loop quantum gravity correction, $F^{SG;(1)}|_{\lambda^0}$, was obtained in~\cite{Aprile:2017bgs} as follows.
	First the full functional coefficient of the leading $\log^2 x_{12}^2$  divergence was obtained by extracting the information from  lower order correlators as follows. All terms in the  OPE of $\cO(x_1) \bar \cO(x_2)$ contain a factor $  C_{\cO\bar \cO}^{\hat \cO}(x_{12}^2)^{\gamma(c)/2}$ where $\gamma(c)=\gamma_1/c+O(1/c^2)$ is the anomalous dimension of the operator $\hat \cO$  and $C_{\cO\bar \cO}^{\hat \cO}$ the OPE coefficient. Inserted into the four point function and expanding in $1/c$ this produces $\log x_{12}^2$ terms which in turn can only arise from $\log u$ terms when the correlator is considered as a function of cross-ratios $u,v$. At $O(c^{-g-1})$, the maximal possible  power of $\log x_{12}^2$ can only arise from terms of the form $ C_{\cO\bar \cO}^{\hat \cO}c^{-g-1} \gamma_1^{g+1} \log^{g+1} x_{12}^2$ in an OPE decomposition,  and in particular it depends entirely on the first non-zero term in the expansion of the anomalous dimension $\gamma_1$ (multiplied by zeroth order OPE coefficients). Thus this maximal $\log^{g+1} u$ power of   $F^{SG;(g)}|_{\lambda^0}$ arises from data which in principle can be extracted from lower order ($g{=}0$ as well as free theory) correlation functions. There is a rather large technical difficulty to overcome however in extracting this data in that there is a large mixing problem to disentangle. Many operators (even entire supermultiplets) have the same free theory quantum numbers. In the supergravity limit we consider here, a key insight from AdS/CFT  is that only operators corresponding to supergravity states survive,  whereas string states become infinitely massive. This simplifies the mixing problem hugely as only operators constructed from single particle half BPS operators are involved in the mixing. 
	Further these can be unmixed by considering four-point functions of all higher charge half BPS correlators (which were obtained in~\cite{Rastelli:2016nze}  and reviewed in section~\ref{sec:hpstrong}). Performing  a super conformal partial wave decomposition~\cite{hep-th/0112251,Arutyunov:2002fh,Doobary:2015gia} of these correlators yields the relevant  data, still in a  mixed form, but then considering all cases yields enough equations to solve the unmixing, giving the $\gamma_1$s and the relevant free OPE coefficients $C_{\cO\bar \cO}^{\hat \cO}$. This unmixing was partially done (enough to obtain the one loop results for  which one only actually needs a partial unmixing) in~\cite{Alday:2017xua,Aprile:2017bgs} and then fully displayed explicitly in~\cite{Aprile:2017xsp}.
	
	Remarkably, a 10d conformal symmetry was discovered to lie hidden in the resulting formulae~\cite{Caron-Huot:2018kta}. It is hidden  in the unmixed  OPE coefficient data whose remarkable structure arises from the decomposition of 10d conformal $SO(2,10)$ representations to $SO(2,4)\times SO(6)$. It is also hidden in the anomalous dimensions $\gamma_1$. We will given more detail of this in the more general situation of section~\ref{sec:hpstrong} (see eg~\eqref{anomdim}).
	
	This 10d symmetry relates (in a somewhat mysterious way, with an eighth order Casimir operator $\DDelta^{(8)}$ playing the role of $G_N$) to the fact that AdS${}_5\times S^5$ is related through a  Weyl transformation to 10d flat space. Again we will give more detail later in a more general setting in section~\ref{sec:hpstrong} (see~\eqref{delta8} for $\DDelta^{(8)}$).
	In practical terms the higher symmetry explains many properties of the unmixed data. 
	The remarkable  structure of the unmixed OPE coefficients found in~\cite{Aprile:2017xsp} then arises  from structure constants in the decomposition of the 10d conformal group $SO(2,10)$ down to the AdS${}_5\times S^5$ symmetry group $SO(2,4)\times SO(6)$. The anomalous dimensions of all  two-particle supergravity operators (labelled by 4 integers) relate to those of  a single family of higher spin currents (labelled by a single integer, the 10d spin). 
	Furthermore this then implies a vast simplification of the formulae for the leading $\log u$ divergence at any loop order which secretly has this simple  10d origin and no longer requires any unmixing, since in 10d there is just a single 10d operator for each even spin $l$. The $O(c^{g+1})$  $\log^{g+1} u$ coefficient then takes the form $(\DDelta^{(8)})^{g}$ acting on a much simpler function which itself can be obtained by summing 10d blocks with known coefficients~\cite{Caron-Huot:2018kta} and has been examined further in~\cite{Bissi:2020wtv,Bissi:2020woe}.

Returning to the derivation of the 1 loop result, having performed the unmixing, this data is  then  inserted back into a superconformal block expansion in the form of the   sum involving the OPE coefficients and  $(\gamma_1)^2$,   for all operators. This  yields the  $\log^2 u$ coefficient of $F^{SG;(1)}|_{\lambda^0}$ in the form of a finite number of terms in  its Taylor expansion. By matching to an appropriate ansatz its analytic form  could then be found. 
Having thus obtained the $\log^2 u$ part of the result, the complete analytic function can then be obtained by matching  to a suitable ansatz, imposing crossing symmetry, and matching with this $\log^2 u$ coefficient~\cite{Aprile:2017bgs}. 
In fact the full solution (not only its leading $\log u$ coefficient) can be simplified using the $\DDelta^{(8)}$ operator and written
\begin{align}
	F^{SG;(1)}|_{\lambda^0} =\frac v{8u^3} \DDelta^{(8)}L^{(2)}_{2222} +\frac14F^{SG;(0)}|_{\lambda^0} \ 
\end{align}
in terms of a preamplitude $L^{(2)}_{2222}$ which is far simpler than $	F^{SG;(1)}|_{\lambda^0}$ itself.
We refer the reader to~\cite{Aprile:2019rep} (eq (227) and following equations) for the explicit expressions. 

\begin{figure}[h!]
	\begin{tikzpicture}[x=4cm,y=1cm]
		\node[label={[label distance=-.3cm]90:$\scriptscriptstyle F^{SG;(0)}|_{\lambda^0}$}] at (0,0) {$\bullet$};
		\node[label={[label distance=-.3cm]90:$\scriptscriptstyle F^{SG;(0)}|_{\lambda^{-3/2}}$}] at (0,3) {$\bullet$};
		\node[label={[label distance=-.5cm]-135:$\scriptscriptstyle F^{SG;(0)}|_{\lambda^{-5/2}}$}] at (0,5) {$\bullet$};
		\node[label={[label distance=-.5cm]-135:$\scriptscriptstyle F^{SG;(1)}|_{\lambda^{1/2}}$}] at (1,-1) {$\bullet$};
		\node[label={[label distance=-.3cm]90:$\scriptscriptstyle F^{SG;(1)}|_{\lambda^{0}}$}] at (1,0) {$\bullet$};
		\node at (0,5) {$\bullet$};
		\node at (0,6) {$\bullet$};
		\node at (0,7) {$\bullet$};
		\node at (0,8) {$\bullet$};
		\node at (1,2) {$\bullet$};
		\node at (1,3) {$\textcolor{red}{\bullet}$};
		\node[label={[label distance=-.3cm]90:$\scriptscriptstyle F^{SG;(2)}|_{\lambda^{0}}$}] at (2,0) {$\textcolor{red}{\bullet}$};
		\node at (2,-3) {$\bullet$};
		\node at (2,-2) {$\bullet$};
		\node at (2,-1) {$\circ$};
		\draw[dashed] (-1/2,5)--node[pos=.42, below, sloped] {$\scriptstyle E(3/2,\tau,\bar \tau)$}(3/2,-3)node {  $R^4$ };
		\draw[dashed] (-1/2,7)--node[pos=.35, below, sloped] {$\scriptstyle E(5/2,\tau,\bar \tau)$}(1+3/2,-3-4+2)node {  $\partial^4 R^4$ };
		\draw[dashed] (-1/2,8)--node[pos=.35, below, sloped] {$\scriptstyle \cE(3,3/2,3/2\tau,\bar \tau)$}(1+3/2,-3-4+2+1)node {  $\partial^6R^4$ };
		\draw[dashed] (-1/2,8+1)--node[pos=.35, below, sloped] {??}(1+3/2,-3-4+2+1+1)node {  $\partial^8 R^4$ };
		\draw[-latex] (0,0)--(2.5,0) node[below]{Powers of $1/c$};
		\draw[-latex] (0,0)--(0,9)node[above]{Powers of $\alpha'=\tfrac1{\sqrt{\lambda}}$};
		\draw[-latex] (0,-2) node[below,text width=2cm] {tree level\\ $F^{SG;{(0)}}$}--(0,-1/2);
		\draw[-latex] (1,-3) node[below,text width=2cm] {one loop\\ $F^{SG;{(1)}}$}--(1,-1-1/2);
		\draw[-latex] (2,-4) node[below,text width=2cm] {two loop\\ $F^{SG;{(2)}}$}--(2,-3-1/4);
		\draw[-latex] (-1/2,0) node[below,text width=2cm] {quantum gravity\\ $\alpha'^0$}--(-1/4,0);
		\draw[-latex] (-1/2,3) node[below,text width=2cm] { $\alpha'^3$\\corrections}--(-1/4,3);
		\node[label=right:{Fully computed}] at (1.2,8) {$\bullet = $};
		\node[label=right:{Partially computed}] at (1.2,7) {$\circ =$};
		\node[label=right:{Not yet computed}] at (1.2,6) {$\textcolor{red}{\bullet} =$};
	\end{tikzpicture}	
	\caption{Plot of known four-point stress-tensor correlators at strong coupling. Quantum gravity loop corrections sit on the $x$ axis with the $y$ axis giving $\alpha'=1/\sqrt{\lambda}$ corrections. Terms below the $x$ axis would then be viewed as counter terms form a quantum gravity perspective. From a string perspective the parameters are $c$ and $\tau=\theta/(2\pi)+4\pi i\sqrt{c}/\lambda$. The diagonal dashed lines are then fixed $c$, varying $\tau$.  }
	\label{fig1}
\end{figure}
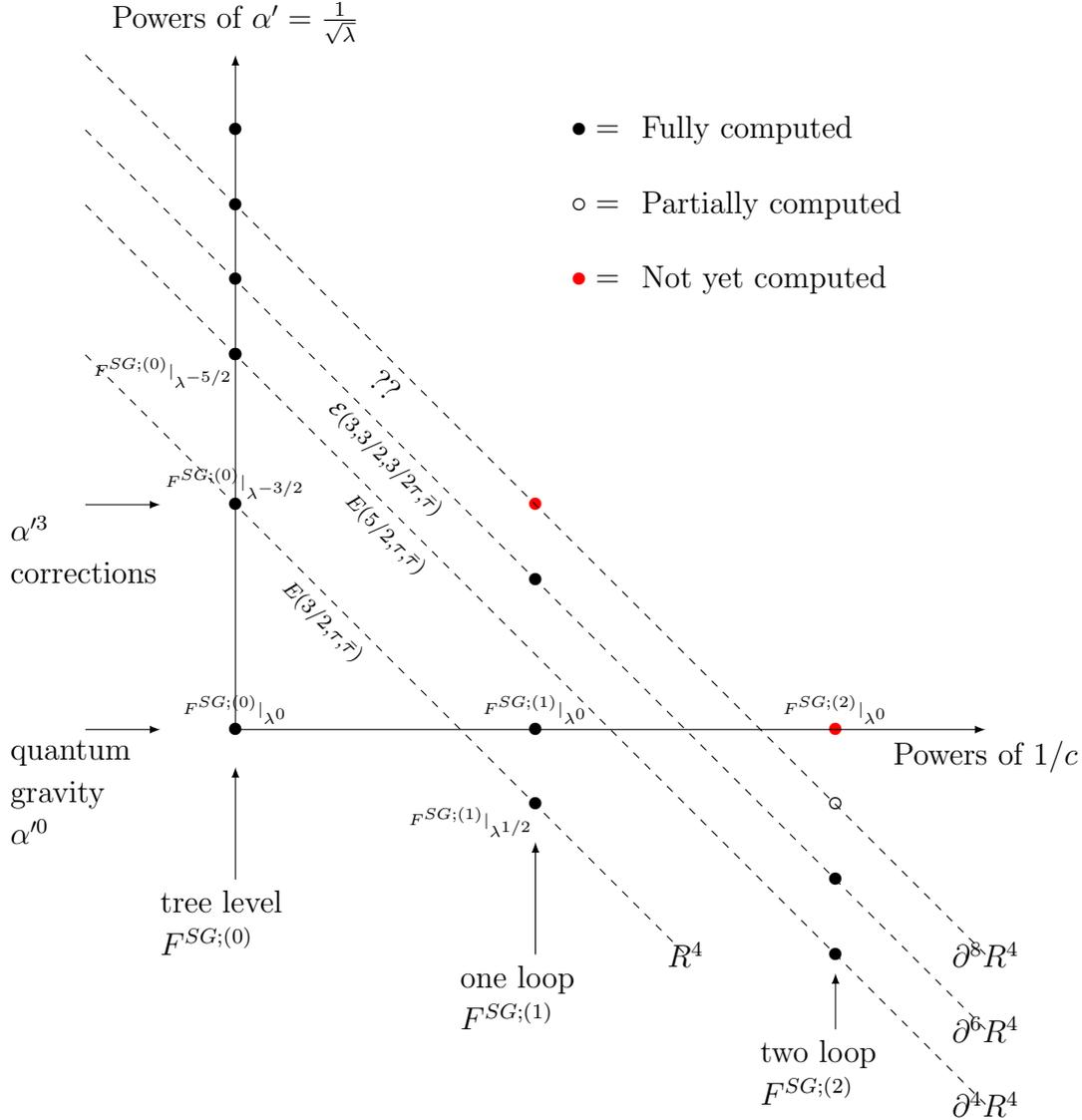

The  superleading term in the one loop supergravity amplitude~\eqref{oneloop}, $F^{SG;(1)}|_{\lambda^{{\scriptscriptstyle{1/2}}}} $, is also known. Its Mellin amplitude is simply a constant  $5\sqrt{\lambda}/8$~\cite{Chester:2019pvm} just like the $\alpha'^3$ tree-level string correction, $M^{(0)}|_{\lambda^{3/2}}$ in~\eqref{m0}. Both terms arise from the same $R^4$ term in the effective action, but now appearing (from the supergravity point of view) as a one loop counter-term. Indeed in string theory the $R^4$ term  comes multiplied by a fully known function of the complex string coupling $\tau=\theta/(2\pi)+4\pi i/g_{YM}^2$, namely the non holomorphic Eisenstein series $E(3/2,\tau,\bar \tau)$~\cite{hep-th/9706175,hep-th/9808061} (see figure~\ref{fig1}). 
This function  contains just two perturbative terms in $\Im(\tau) =\sqrt{c}/\lambda $ corresponding to the tree level $\alpha'^3=1/\lambda^{-3/2}$ correction of~\eqref{m0} and the one loop superleading counter term $\lambda^{1/2}/c$. Indeed all the terms in~\eqref{m0} are now known as full functions of the string coupling in terms of (generalised) Eisenstein series~\cite{hep-th/9706175,hep-th/9808061,hep-th/9910055,hep-th/0510027,1404.2192,1912.13365,2008.02713,2009.01211}.
Furthermore, the same function $\bar D_{4444}$ appears in a {\em third} place,  as the only ambiguity remaining in the computation~\cite{Aprile:2017bgs} of $F^{SG;(1)}|_{\lambda^0}$. This ambiguity was then also fixed by supersymmetric localisation in~\cite{Chester:2019pvm}.

Then the third term in~\eqref{oneloop}, $F^{SG;(1)}|_{\lambda^{{\scriptscriptstyle{-1}}}} $, arises in a similar way. It is proportional to the $1/\lambda^3=\alpha'^6$ tree-level string correction in~\eqref{m0} which in turn arises from a $\partial^6 R^4$ correction to the string theory effective action. Its coefficient is  the generalised non-holomorphic Eisenstein series $\cE(3,3/2,3/2,\tau,\bar \tau)$ which yields a known  perturbative contribution at one loop $O(\lambda^{-1})$~\cite{1404.2192}(see figure~\ref{fig1}).

The fourth term in the expansion~\eqref{oneloop} is also now known at one loop level up to certain ambiguities (in particular an ambiguity arising from a $\partial^8 R^4$ term in the effective action,  see figure~\ref{fig1}). It was found first in Mellin space~\cite{Alday:2018kkw} and then in position space~\cite{Drummond:2019hel}. The  position space correlator has a similar structure to that of supergravity, polynomials in $x, \bar x$ multiplied by polylogarithms (but of weight 3 and below rather than weight four and below) and divided by high order $x{-}\bar x$ poles. A new feature found here is the presence of generalised (but still single valued) polylogarithms. These have $x-\bar x$ in their symbol whereas previously only $x,\bar x, 1{-}x,1{-}\bar x$ occurred. Recall that exactly the same phenomenon
occurred at 3 loops in perturbation theory as discussed below~\eqref{ftoFsimp}.

Very recently the expression $F^{SG;(2)}|_{\lambda^0}$ corresponding to the 2-loop quantum gravity amplitude has been obtained~\cite{Huang:2021xws} using similar methods and found to have the very suggestive form 
\begin{align}\label{2loopQG}
	F^{SG;(2)}|_{\lambda^0}= (\DDelta^{(8)})^2L_{2222}^{(3)} +\frac54 F^{SG;(1)}|_{\lambda^0}-\frac1{16}  F^{SG;(0)}|_{\lambda^0}  +\text{ambiguities}\ 
\end{align}
where there are both tree-level and one-loop ambiguities whose coefficients are unfixed. It would be fascinating to see if this structure of pulling out $\DDelta^{(8)}$ powers persists at higher orders, and to understand why this is the case.

\section {Higher charges and 10d conformal symmetry}
\label{sec:highercharges}

In the previous section we have focussed on  four-point correlators of the lowest charge half BPS operator $\cO:=\Tr(\phi^2)$ where $\phi$ is one of the 3 (complex) scalars in $\cN=4$ SYM. This operator is the first in an infinite class of half BPS operators.  The half BPS operators are members of  separate supermultiplets.   The independent supermultiplets are labelled by an integer $p=2,3,..$. For the four point function of lowest charge operators we could focus on just one operator $\Tr(\phi^2)$ since the  four point function of any other component field  is related to  $\langle \cO \bar \cO \cO\bar \cO \rangle$ by supersymmetric Ward identities. However  this is no longer the case for higher charge operators,  and  we thus need to consider the full scalar sector of $\cN=4$ SYM.%
\footnote{We still do not need to consider other fields beyond scalars (i.e. fermions or the gauge field) at four points though. When moving beyond four points we will need to consider the entire supermultiplet and will use analytic superspace to do this (section~\ref{sec:an}).} 
There are 6 real scalars, $\phi^I,\ I=1,...,6$,  in the theory carrying the fundamental representation of the internal symmetry group $SO(6) \sim SU(4)\subset SU(2,2|4)$. We can deal with these indices by contracting with internal co-ordinates $y_I$, so
\begin{align}\label{phidef}
	\phi^I(x) \rightarrow \phi(x,y)=y_I\phi^I(x)\ .
\end{align} 
The half BPS operators fall into symmetric traceless representations of $SO(6)$ which are then represented as simple products of $\phi(x,y)$ as long as $y_I$ satisfies $y_I y^I = 0$ in order to project out the $SO(6)$ trace.  Recall that the scalars are all in the adjoint of the gauge group $SU(N_c)$ and we take traces over this to obtain gauge invariant operators. We will then focus on {\em single particle} operators
\begin{align}\label{Opdef}
	\cO_p(x,y) = \Tr(\phi^p)+... \qquad \qquad \phi=\phi(x,y),\quad y_I y^I = 0\ .
\end{align}
Single particle operators are equal to single trace operators plus multi-trace $1/N_c$ corrections (hence the dots in the above definition). Single particle operators are uniquely defined to be orthogonal to (have vanishing two-point functions with) all multi-trace operators~\cite{Aprile:2020uxk}. They are equivalent to single trace operators at large $N_c$ but give crucial differences when $1/N_c$ corrections are taken into account~\cite{Aprile:2018efk,Aprile:2019rep,Alday:2019nin}.

Superconformal Ward identities and non-renormalisation theorems now dictate (see discussion leading to and around~\eqref{hiddenprrof2} for the proof) that the  correlator of four arbitrary charge half BPS operators takes the form
\begin{align}\label{corhighercharge}
	\langle \cO_{p_1}\cO_{p_2} \cO_{p_3}  \cO_{p_4}\rangle = \text{free} \ +\  C_{p_1p_2p_3p_4} \frac{I(x_i,y_j)}{\xi^{(4)}} \times F_{p_k}(x_i,y_j;\lambda,c)\, .
\end{align}
Here the normalisation is given by 
\begin{align}
	C_{p_1p_2p_3p_4}=\frac {p_1p_2p_3p_4}{2c} \left(\frac{c}{16\pi^4}\right)^{\frac14\sum {p_i}}	
\end{align}
and
we recall $\xi^{(4)}=x_{13}^4x_{24}^4x_{12}^2 x_{23}^2x_{34}^2x_{14}^2$.  The factor $I(x_i,y_j)$ is the consequence of (the fermionic part of) superconformal symmetry and we'll derive this in section~\ref{sec:an}. It takes the form
\begin{align}\label{intriligator}
	I(x_i,y_j)= &\Big((x_{13}^2 x_{24}^2- x_{14}^2x_{23}^2 )x_{13}^2x_{24}^2y_{12}^2y_{23}^2y_{34}^2 y_{41}^2    + \text{\footnotesize {$(1{\rightarrow} 2 {\rightarrow} 3{\rightarrow} 1)$}} + \text{\footnotesize {$(1{\rightarrow} 3 {\rightarrow} 2{\rightarrow} 1)$}}\Big)  + (x \leftrightarrow y)
\end{align}
where $y_{ij}^2=y_i^Iy_j^I$.
It is completely symmetric under crossing symmetry $S_4$ as well as under the interchange $x \leftrightarrow y$.
It can be rewritten in a simple compact form in terms of cross ratios as:
\begin{align}\label{intriligator2}
	I(x_i,y_j)= &x_{13}^4 x_{24}^4 y_{13}^4 y_{24}^4 (x-y) (x-\bar y) (\bar x-y) (\bar x-\bar y) \ ,
\end{align}
where the cross-ratios $x,\bar x , y,  \bar y$ are defined as
\begin{align}
	x \bar x &= \frac{x_{12}^2 x_{34}^2}{x_{13}^2 x_{24}^2} \qquad &(1{-}x) (1{-}\bar x) = \frac{x_{14}^2 x_{23}^2}{x_{13}^2 x_{24}^2}\notag \\
	y \bar y &= \frac{y_{12}^2 y_{34}^2}{y_{13}^2 y_{24}^2} \qquad &(1{-}y) (1{-}\bar y) = \frac{y_{14}^2 y_{23}^2}{y_{13}^2 y_{24}^2}\ .
\end{align}
Now the correlator~\eqref{corhighercharge} is a homogeneous polynomial of degree $p_j$ in the variable $y_j$ for each $j=1,..,4$,  as one can see directly from~\eqref{phidef},\eqref{Opdef}.
The factor $I(x_i,y_j)$ absorbs 2 $y_j$s for each $j$ and  (crucially) the remainder $f_{p_k}$ is also polynomial in the variables $y_j$  but of reduced homogeneity $p_j-2$. The  function $f_{p_k}$ is also $SU(4)\sim SO(6)$ invariant which implies that the $y$ variables must appear as scalar products only. Thus we can completely parametrise the polynomial (in $y_j$)  $F_{p_k}$,  in terms of functions of the $x_i$ variables only
\begin{align}\label{fcF}
	F_{p_k}(x_i,y_j;\lambda,c)&=\sum_{\{b_{ij}\}} \left(\prod_{i<j} {g_{ij}^{b_{ij}}}\right) F_{\{b_{ij}\}}(x,\bar x;\lambda,c)\notag \\
	\{b_{ij}\}&:=\{b_{ij}=b_{ji}: b_{ii}=0,\quad \sum_i b_{ij} = p_j-2\}\qquad
	g_{ij}:=\frac{y^2_{ij}}{x^2_{ij}}\ .
\end{align}
Note that the correlator is  homogeneous of degree $-p_i$ in each  variable $x_i$ since the operator $O_{p_i}$ has dimension $p_i$ (the fundamental massless scalar in four dimensions has dimension 1). Therefore $F_{p_k}$ has weight ${-}p_i{-}2$ in each  variable $x_i$. This $x$ weight is then absorbed by the explicit factors of $x_{ij}^2$ in~\eqref{fcF} leaving functions  of $x$-cross ratios only $F_{\{b_{ij}\}}(x,\bar x)$.

\subsubsection*{Relation to the simplest correlator}

How does this relate to the simplest correlator considered in section~\ref{sec:simplest}? There we considered the operator $\cO=\Tr(\phi^2)$ where $\phi$ was any complex scalar, for example $\phi=(\phi_1+i \phi_2)/\sqrt2$. This can be obtained from $\phi(x,y)$~\eqref{phidef} by setting $y^I=(1,i,0,0,0,0)/\sqrt2$. Thus the correlator~\eqref{corbasic} of section~\ref{sec:simplest} is simply the charge 2 correlator with a fixed choice of $y$ coordinates  
\begin{align}\label{ychoice}
	\langle \cO \cO \bar \cO \bar \cO \rangle = \langle \cO_2 \cO_2 \cO_2 \cO_2 \rangle|_{\substack{y_1=y_3=(1,+i,0,.,0)/\sqrt2\\y_2=y_4=(1,-i,0,.,0)/\sqrt2}} \ .
\end{align}
One can quickly check that this choice of $y$ coordinates gives $y_{13}^2=y_{24}^2=0$, $y_{12}^2=y_{14}^2=y_{13}^2=y_{34}^2=1$ and so the superconformal factor $I(x_i,y_j)=x_{13}^4x_{24}^4$. Thus the function $F$ in~\eqref{corbasic} is simply the function $F_{2222}$ in~\eqref{corhighercharge}.
Furthermore in this case the decomposition of $F_{2222}$ in~\eqref{fcF} is trivial (it has no $y$ dependence)  and so this equates to the only contributing function $F_{\{b_{ij}\}}$ in~\eqref{fcF} 
\begin{align}
	F_{2222}(x_i,y_j;\lambda,c)=F_{\{000000\}}(x,\bar x;\lambda,c)=	F(x_i;\lambda,c)\ .
\end{align}

\subsection{Loop integrands and 10d symmetry}
\label{pert10d}

A number of higher charge half BPS planar correlators were computed  directly by Feynman graphs to 2-loops in~\cite{Arutyunov:2003ae,Eden:2000mv,Arutyunov:2002fh,Arutyunov:2003ad,D'Alessandro:2005dq,Uruchurtu:2011wh,Chicherin:2014esa}. Then  in~\cite{Chicherin:2015edu,Chicherin:2018avq} the integrands of {\em all} planar half BPS 4-pnt correlators  to five loops were derived using bootstrappy techniques, specifically
superconformal symmetry, the analytic structure, planarity and OPE arguments. The integrals which appear are exactly those of the stress-tensor multiplet correlator~(\ref{Fl})-(\ref{6loops}) but with different coefficients and hence all the integrals to 3-loops are also known analytically from~\cite{Drummond:2013nda}.

The results of~\cite{Chicherin:2015edu,Chicherin:2018avq} display  an interesting structure, that the component correlators are independent of $b_{ij}$ (see~\eqref{fcF}) beyond some minimum value, depending on the loop order.  
Recently this structure has been beautifully understood as due to the presence of a  hidden 10d conformal symmetry. Assuming this continues it then implies that all perturbative correlators can be  understood as derivable directly from the stress-tensor correlator itself~\cite{Caron-Huot:2021usw}. 
Defining the integrands of the higher charge correlators $F_{p_i}$ and their components $F_{\{b_{ij}\}}$ in a similar way to the integrands of the lowest charge correlator~\eqref{Fl} ie
\begin{align}\label{Flhc}
	F_{\{b_{ij}\}} &= \sum_{l=1}^\infty{\left(\frac{\lambda}{4\pi^2}\right)^l}F^{(l)}_{\{b_{ij}\}} \\
	F^{(l)}_{\{b_{ij}\}}(x_1,..,x_{4})	&= \frac{\xi^{(4)}}{l!}\int \frac{d^4x_5}{(-4\pi^2)}..\frac{d^4x_{4+l}}{(-4\pi^2)}f^{(l)}_{\{b_{ij}\}}(x_1,.,x_{4+l})\ \label{Finfhc}
\end{align}
then the higher charge component correlators $ f^{(l)}_{\{b_{ij}\}}$ are obtained directly from the simplest correlator by writing it in terms of 10d conformal invariants and taking the appropriate  coefficient
\begin{align}\label{flb}
	f^{(l)}_{\{b_{ij}\}}(x_{ij}^2) = f^{(l)} ({\bf x}_{ij}^2 )|_{(g_{ij})^{b_{ij}}} 
\end{align}
where on the rhs $f^{(l)}$ are the $f$ graphs of~\eqref{Finf}, the stress-tensor multiplet correlator described in section~\ref{sec:simplest} containing the hidden permutation symmetry. Here ${\bf x}_{ij}^2$ are 10d conformal invariants, constructed from 12 component $SO(2,10)$ variables obtained by appending the $SO(2,4)$ external variable to the $SO(6)$ internal variables ${\bf x}_i=(x_i,y_i)$.  So then  
\begin{align}\label{10dx}
	{\bf x}_{ij}^2:={\bf x}_i.{\bf x}_j=x_i.x_j-y_i.y_j=x_{ij}^2-y_{ij}^2=x_{ij}^2(1-g_{ij})\ .
\end{align}
All the integration variables live in 4d however,  so $y_i=0$ for $i=5,..,4{+}l$. We will discuss later (at the end of section~\ref{sec:proofhidden}) a natural interpretation of~\eqref{flb} with arbitrary $y$ at all points. 

To illustrate we give the simplest example, namely one loop. The function  $f^{(1)}$ is given in~\eqref{1loop} and we have
\begin{align}
	f^{(1)}({\bf x}_{ij}^2)& = {1 \over \prod_{1\leq i<j \leq 5} {\bf x_{ij}^2}}=\frac1{\xi^{(4)}} \frac{x_{13}^2x_{24}^2}{x_{15}^2x_{25}^2x_{35}^2x_{45}^2} \frac1{\prod_{1\leq i<j\leq 4}(1-g_{ij})}\notag \\
	&=\frac1{\xi^{(4)}} \frac{x_{13}^2x_{24}^2}{x_{15}^2x_{25}^2x_{35}^2x_{45}^2} \prod_{1\leq i<j\leq 4}\left(\sum_{b_{ij}=0}^\infty g_{ij}^{b_{ij}} \right)\ .
\end{align}
Then reading off from~\eqref{flb} this gives
\begin{align} 
	\xi^{(4)}	f^{(1)}_{\{b_{ij}\}} = \frac{x_{13}^2x_{24}^2}{x_{15}^2x_{25}^2x_{35}^2x_{45}^2}\ 
\end{align}
so that {\em all} component correlators are equal and given by the one loop box function (with unit coefficient). This is precisely the result derived in~\cite{Arutyunov:2003ae,Chicherin:2015bza}. 

In a similar way the conjectured 10d symmetry~\eqref{flb} correctly reproduces all  the known higher charge BPS correlators to five loops~\cite{Chicherin:2015edu,Chicherin:2018avq} directly from the $f$-graphs~(\ref{2loop}-\ref{f5})~\cite{Caron-Huot:2021usw}.
Furthermore, since we know the stress-tensor multiplet correlators to ten loops, conjecturing this at higher loops immediately gives a  concrete proposal for all  half BPS planar correlators to ten loops. 
Since the derivation of the higher charge correlators to 5 loops in~\cite{Chicherin:2015edu,Chicherin:2018avq} relied solely on the OPE  considerations and planarity, it is presumably also possible to prove the 10d symmetry at the level of the perturbative integrands from these. 

One can rephrase the 10d symmetry in a suggestive way by writing all half BPS operators in terms of a single operator generating all of them
\begin{align}\label{bcO}
	\cbO(x,y) =  \sum_{p=2}^ \infty \frac1p \left(\frac{16\pi^4}c\right)^{p/4}\cO_p(x,y)\ .
\end{align}
Then (in the planar theory) we have that the four point function of {\em all} single particle half BPS operators can be combined into the following beautifully simple {\em master correlator}
\begin{align}\label{master}
	\langle \cbO\cbO\cbO\cbO \rangle=
	\text{free} \ +\ \frac{I(x_i,y_j)}{2c} \times \sum_{l=0}^\infty  \frac{\lambda^l}{(4\pi^2)l!}\int \frac{d^4x_5}{(-4\pi^2)}..\frac{d^4x_{4+l}}{(-4\pi^2)}f^{(l)}({\bf x}_{ij}^2)\, .
\end{align}

It would be interesting to see to what extent this perturbative 10d conformal symmetry survives in the   non-planar regime, especially given the relative simplicity of non planar corrections at strong coupling~\eqref{2loopQG} which is closely related with a 10d conformal symmetry appearing at strong coupling.
So far the  perturbative non-planar corrections (which only start to be non-trivial at 4-loops) are only known for the lowest half BPS correlator~\cite{Eden:2012tu,Fleury:2019ydf}.

\subsection{10d correlator / amplitude duality}

In the context of integrability approaches to correlators (which we will briefly summarise in more generality  in the conclusions) it has been useful to consider  four-point half BPS correlators with  charges taken to certain limits. 
In particular in~\cite{Coronado:2018ypq} a limit of large charge correlators was introduced and called the `simplest' correlator (not to be confused with the stress-tensor correlator which we have referred to as the simplest half BPS correlator). It was shown to be given by the square of a certain octagon form factor $\mathbb{O}$~\cite{Coronado:2018ypq} obtained by  gluing two hexagons~\cite{Basso:2015zoa} together.
Now the large charge limit  corresponds precisely to taking the 10d lightlike limit of the master correlator~\eqref{master} so~\cite{Caron-Huot:2021usw} (taking the master correlator to be a function of ${\bf x}_{ij}^2, x_{ij}^2$ rather than  $x_{ij}^2, y_{ij}^2$)
\begin{align}
	\lim_{{\bf x}_{12}^2,..,{\bf x}_{41}^2\rightarrow 0} 	 \frac{\langle \cbO\cbO\cbO\cbO \rangle|_{1/c}}{{\bf x}_{12}^2{\bf x}_{23}^2{\bf x}_{34}^2{\bf x}_{41}^2 } = \mathbb{O}^2\ .
\end{align}
This is very reminiscent of the correlator/amplitude duality~\eqref{coramp4}. 
Indeed independently of the above discussion of the simplest correlator one could ask
about taking the 10d lightlike limit of the master correlator~\eqref{master} and  it is natural to imagine it as some sort of amplitude. A natural object which then comes to mind is the 4-point amplitude of $\cN=4$ SYM regularised on the coulomb branch, introduced in~\cite{Alday:2009zm}. So the conjecture of~\cite{Caron-Huot:2021usw} is that
\begin{align}\label{bO}
	\lim_{{\bf x}_{12}^2,..,{\bf x}_{41}^2\rightarrow 0} 	 \frac{\langle \cbO\cbO\cbO\cbO \rangle|_{1/c}}{\langle \cbO\cbO\cbO\cbO \rangle|_{1/c,\lambda=0}} = M(x_i,y_i)^2\ 
\end{align}
where $ M(x_i,y_i)^2$ is the Higgs regulated amplitude (with 6d internal variables satisfying $y_i.y_i=0$).
This can then be rephrased as a direct equality between the octagon of~\cite{Coronado:2018ypq} and the four-point amplitude regulated on the Higgs branch of~\cite{Alday:2009zm}
\begin{align}
	\frac{\mathbb{O}(x_i,y_i)}{\mathbb{O}(x_i,y_i)|_{\lambda=0}} =M(x_i,y_i)\ .
\end{align} 
Note that the octagon is known for  arbitrary values of the  coupling~\cite{Belitsky:2020qrm,Belitsky:2020qir}.
Both  sides of this equality yield perfectly finite integrals and so this identification then works at the level of the integrals as well as the integrand and produces new results for integrals from integrability~\cite{Caron-Huot:2021usw}. 

\subsection{Strong coupling and 10d symmetry / effective action}
\label{sec:hpstrong}

\subsubsection*{Tree-level supergravity}

We turn now to higher charge correlators at strong coupling. The tree-level supergravity results of the lowest charge correlator~\eqref{sugra} have been computed for various higher charges by direct supergravity computations in a number of papers~\cite{hep-th/0002170,Arutyunov:2002fh,Arutyunov:2003ae,hep-th/0601148,0709.1365,0811.2320,Uruchurtu:2011wh,1806.09200,1808.06788}. In~\cite{Rastelli:2016nze,Rastelli:2017udc}  analyticity and crossing symmetry arguments were used to  bootstrap the results for {\em all} half BPS four-point functions $\langle \cO_{p_1}\cO_{p_2}\cO_{p_3}\cO_{p_4} \rangle$  in the tree supergravity limit  in Mellin space (recovering and generalising  the previous direct supergravity computations).

These tree-level results for all charges were instrumental in the unmixing needed to bootstrap 1-loop quantum gravity results for the simplest correlator, as mentioned in section~\ref{sec:strong}. We have now introduced the relevant ingredients to  give more details of this now. The supermultiplets which appearing in the large $c$ OPE of two half BPS single particle operators at strong coupling are {\em two-particle operators}: operators in the tensor product of two single particle operators. They have twist (dimension minus spin) $\tau$, spin $l$ and $SU(4)$ representation $[a,b,a]$ and have the form: 
\begin{equation}
	\mathcal{O}_{pq}^{\tau,l;[a,b,a]} = \mathcal{O}_{p} \partial_x^l \Box_x^{\frac12(\tau-p-q)} \partial_y^a \Box_y^{\frac12(p+q-b-2a)} \mathcal{O}_{q} \,, \qquad (p \leq q)\,.
	\label{ops}
\end{equation} 
The  quantum numbers $\tau,l,a,b$ specify the free theory conformal representation of the operator,  then the labels $p,q$  parametrise the free theory degeneracy. The possible values of $p,q$ form the nodes inside  a finite rectangle of height roughly $b/2$ and width $(\tau{-}b{-}2a-4)/2$, rotated by 45 degrees within an integral lattice in $p,q$ space~\cite{Aprile:2018efk}
\begin{align}
	\begin{tikzpicture}[scale=.54]
	\def\prop{.5}
	\def\shifthor{\prop*2}
	\def\ptuno{(\prop*2-\shifthor,\prop*8)}
	\def\ptdue{(\prop*5-\shifthor,\prop*5)}
	\def\pttree{(\prop*9-\shifthor,\prop*15)}
	\def\ptquattro{(\prop*12-\shifthor,\prop*12)}
	%
%	%axis horizontal
%	\draw[-latex, line width=.6pt]		(\prop*2   -\shifthor-1,         \prop*14          -0.5*\shifthor)    --  (\prop*2  -\shifthor  + 0.5,   \prop*14-      0.5*\shifthor) ;
%	\node[scale=.8] (oxxy) at 			(\prop*2   -\shifthor    +0.5,  \prop*16.5     -0.5*\shifthor)  {};
%	\node[scale=.9] [below of=oxxy] {$p$};
%	%
%	%axis vertical
%	\draw[-latex, line width=.6pt] 		(\prop*2   -\shifthor-1,     \prop*14       -0.5*\shifthor)     --  (\prop*2   -\shifthor-1,        \prop*17-      0.5*\shifthor);
%	\node[scale=.8] (oxyy) at 			(\prop*4   -\shifthor*1.7,   \prop*16.8   -0.5*\shifthor) {};
%	\node[scale=.9] [left of= oxyy] {$q$};
	%
	%rectangle
	\draw[] 						\ptuno -- \ptdue;
	\draw[]						\ptuno --\pttree;
	\draw[]						\ptdue --\ptquattro;
	\draw[]						\pttree--\ptquattro;
	%
	%dots
	%
	\foreach \indeyc in {0,1,2,3}
	\foreach \indexc  in {2,...,9}
	\filldraw   					 (\prop*\indexc+\prop*\indeyc-\shifthor, \prop*6+\prop*\indexc-\prop*\indeyc)   	circle (.07);
	%
	%letters
%	%
%	\node[scale=.8] (puntouno) at (\prop*4-\shifthor,\prop*8) {};
%	\node[scale=.8]  [left of=puntouno] {$A$};   
%	%
%	\node[scale=.8] (puntodue) at (\prop*5-\shifthor,\prop*6+.5) {};
%	\node[scale=.8] [below of=puntodue]  {$B$}; 
%	%
%	\node[scale=.8] (puntoquattro) at (\prop*13-\shifthor,\prop*15) {};
%	\node[scale=.8] [below of=puntoquattro] {$C$};
%	%							
%	\node[scale=.8] (puntotre) at (\prop*9-\shifthor,\prop*13) {};
%	\node[scale=.8] [above of=puntotre] {$D$}; 
	%
	%
 \draw[-latex,line width=.6pt] (\prop*2-\shifthor-.2*\shifthor,\prop*8-.2*\shifthor) -- (\prop*5-\shifthor-.2*\shifthor,\prop*5-.2*\shifthor);
 \draw[-latex,line width=.6pt]  (\prop*5-\shifthor-.2*\shifthor,\prop*5-.2*\shifthor) -- (\prop*2-\shifthor-.2*\shifthor,\prop*8-.2*\shifthor) ;
	\node[scale=.8]  at (\prop-.5*\shifthor,\prop*6-.3*\shifthor) {$\mu{-}1$}; 
\draw[-latex,line width=.6pt] (\prop*5-\shifthor+.2*\shifthor,\prop*5-.2*\shifthor) --
(\prop*12-\shifthor+.2*\shifthor,\prop*12-.2*\shifthor);	
	\draw[-latex,line width=.6pt] 
	(\prop*12-\shifthor+.2*\shifthor,\prop*12-.2*\shifthor) -- (\prop*5-\shifthor+.2*\shifthor,\prop*5-.2*\shifthor);	
		\node[scale=.8]  at (\prop*8.5-\shifthor+.7*\shifthor,\prop*8.5-.7*\shifthor) {$t{-}2$}; 
	%									
	%lines
	%
	\foreach \indexc in {3,4}
	\draw (\prop*\indexc-\shifthor, \prop*10-\prop*\indexc ) -- (\prop*\indexc-\shifthor, \prop*6+\prop*\indexc  );
	\foreach \indexc in {5,6,7,8,9}
	\draw (\prop*\indexc-\shifthor, \prop*\indexc ) -- (\prop*\indexc-\shifthor, \prop*6+\prop*\indexc  );
	\foreach \indexc in {10,11,12}
	\draw (\prop*\indexc-\shifthor, \prop*\indexc ) -- (\prop*\indexc-\shifthor, \prop*24-\prop*\indexc  );
\node[] at (15,5)  {$t\equiv (\tau-b)/2-a\,,\quad 
	\mu \equiv   \left\{\begin{array}{ll}
		\bigl\lfloor{\frac{b+2}2}\bigr\rfloor \quad &a+l \text{ even,}\\[.2cm]
		\bigl\lfloor{\frac{b+1}2}\bigr\rfloor \quad &a+l \text{ odd.}
	\end{array}\right.
	$};
\end{tikzpicture}
\label{rectangle}
\end{align}
A conformal partial wave decomposition of the $	\langle \cO_{p_1}\cO_{p_2} \cO_{p_3}  \cO_{p_4}\rangle$ correlator in the free theory and the $\log u$ coefficient of the $1/c$ correction (at strong coupling) then yields the following combinations of OPE coefficients and anomalous dimensions
\begin{align}
	\langle \cO_{p_1}\cO_{p_2} \cO_{p_3}  \cO_{p_4}\rangle_{c^0} &= 	\sum_{\tau,l,a,b} \left(\sum_{p,q} C_{\cO_{p_1}\!\cO_{p_2}\!\cO_{pq}}\, C_{\cO_{p_3}\!\cO_{p_4}\!\cO_{pq}}\right) \times \text{superblock} (\tau,l;[a,b,a])\\
	\langle \cO_{p_1}\cO_{p_2} \cO_{p_3}  \cO_{p_4}\rangle_{\frac1c,\log u,\lambda \rightarrow \infty} &= 	\sum_{\tau,l,a,b}
	\left(\sum_{p,q} C_{\cO_{p_1}\!\cO_{p_2}\!\cO_{pq}}\,\gamma_{\cO_{pq}}\,C_{\cO_{p_3}\!\cO_{p_4}\!\cO_{pq}}\right)
	\text{superblock} (\tau,l;[a,b,a])
	\ .
\end{align}
For fixed quantum numbers, these coefficients give precisely the right number of independent equations to uniquely fix the OPE coefficients $C_{\cO_{p_1}\cO_{p_2}\cO_{pq}}$ and the anomalous dimensions $\gamma_{\cO_{pq}}$.
The resulting OPE coefficients have a remarkable structure~\cite{Aprile:2017xsp,Aprile:2017qoy}
eventually  understood in terms of a 10d conformal symmetry~\cite{Caron-Huot:2018kta} where it arises as the decomposition of 10d conformal $SO(2,10)$ representations down to $SO(2,4)\times SO(6)$ 
as we already mentioned in section~\ref{sec:strong}.
The anomalous dimensions of all 2-particle operators~\eqref{ops} have the simple form~\cite{Aprile:2018efk} 
\begin{equation}\label{anomdim}
	\gamma_{\cO_{pq}}=  {-} \frac{1}{c} \frac{\delta^{(8)}}{ \left(l+2p{-}2{-}a {-} \frac{1+({-})^{a+l}}{2} \right)_6}
\end{equation}
with $\delta^{(8)}{=}M^{(4)}_{t} M^{(4)}_{t{+}l{+}1}$, $M^{(4)}_t{\equiv} (t{-}1) (t{+}a) (t{+}a{+}b{+}1)(t{+}2a{+}b{+}2)$ and $t{\equiv}(\tau{-}b)/2{-}a$.
Note the independence of $q$, indicating degeneracy of the anomalous dimensions at this order, indicated in~\eqref{rectangle} by the vertical lines linking degenerate operators at this order,  and signalling the presence of the 10d symmetry. 
The numerator in~\eqref{anomdim} is the eigenvalue of an eighth order Casimir operator, $\DDelta^{(8)}$, acting on the corresponding superblocks~\cite{Caron-Huot:2018kta} 
\begin{equation}
	\label{delta8}
	\DDelta^{(8)} =  \frac{x {\bar x} y {\bar y}}{(x-{\bar x})(y-{\bar y})}\prod_{i,j=1}^{2} \left(\mathbf{C}_{x_i}^{[+\alpha,+\beta,0]} -\mathbf{C}_{y_j}^{[-\alpha,-\beta,0]}\right) \frac{(x-{\bar x})(y-{\bar y})} {x {\bar x} y {\bar y}}
\end{equation}
where $\alpha=p_{21}/2$, $\beta=p_{34}/2$ and $\mathbf{C}_{x}^{[\alpha,\beta,\gamma]}$ is the elementary $2d$ Casimir
\begin{equation}
	\label{2dcasimir}
	\mathbf{C}_{x}^{[\alpha,\beta,\gamma]}= x^2(1-x)\partial_{x}^2 +x(\gamma-(1+\alpha+\beta)x)\partial_{x} - \alpha\beta x\ .
\end{equation}

These results can then be  succinctly combined together to a single equation giving the master correlator in the tree-level supergravity limit, displaying the same 10d conformal symmetry described for perturbative correlators~\cite{Caron-Huot:2018kta} 
\begin{align}\label{masterstrong}
	\langle \cbO \cbO \cbO\cbO \rangle_{\frac1c,\lambda\rightarrow \infty} = \text{free}- \frac{I(x_i,y_i)}{4c}\frac{D_{2422}(\bx_i)}{{\bx_{13}^2}{\bx_{14}^2}\bx_{34}^2}\, ,
\end{align} 
where as always the bold variables are 10d.
We see that this strong coupling formula exhibits the 10d symmetry in a very similar way to the weak coupling case~\eqref{master}.

Furthermore the free theory, when acted on by the eighth order Casimir, $\DDelta^{(8)}$ (which converts the correlator of primaries to the correlator of certain Lagrangian-type superconformal descendants  $\langle L_p L_q\bar L_r \bar L_s \rangle$ where $L_p \sim Q^4 \cO_p$~\cite{hep-th/0610280,Caron-Huot:2018kta}) also possesses this 10d conformal symmetry! Defining a master descendant operator $\bf L$ similarly to $\cbO$~\eqref{bO}, this carries the same conformal  representation as a 10d massless (dimension 4) scalar field! Its four-point correlator  decomposes into 10d conformal blocks of 10d higher spin currents,  $\cbO_{l_{10d}}\equiv {\bf L}{\bm \partial}^{l_{10d}} {\bf L}$, one for each spin
\begin{align}\label{10d1}
	\langle {\bf L}{\bf L}\bar {\bf L} \bar {\bf L} \rangle_{c^0} = \frac1 {{\bf x}_{14}^8{\bf x}_{23}^8}+\frac1 {{\bf x}_{13}^8{\bf x}_{24}^8}= \sum_{l_{10d}} C_{{\bf LL}\cbO_{l_{10d}}}C_{ {\bf \bar L\bar L}\cbO_{l_{10d}}} \times \text{10dblock}_{l_{10d}}({\bf x}_i)\ .
\end{align}
The 10d part of the  tree level master correlator~\eqref{masterstrong} on the other hand has a similar 10d  block decomposition 
\begin{align}\label{10d2}
	\frac{D_{2422}(\bx_i)}{{\bx_{13}^2}{\bx_{14}^2}\bx_{34}^2}= \sum_{l_{10d}} C_{{\bf{LLO}}_{l_{10d}}}C_{{\bf{\bar L\bar LO}}_{l_{10d}}} \frac1{(l_{10d})_6} \times \text{10dblock}_{l_{10d}}({\bf x}_i)\ .
\end{align}
We therefore see that the ratio of the 10d block coefficients in~\eqref{10d1} and~\eqref{10d2} is $1/{(l_{10d})_6}$.
This then explains the form of the anomalous dimensions~\eqref{anomdim}. Since there is only a single 10d representation for each 10d spin, there is no 10d unmixing problem.  The $\delta^{(8)}$   arises from the  action of $\DDelta^{(8)}$ to obtain the  descendant correlator, whereas the denominator arises from $1/(l_{10d})_6$ with the 10d higher spin currents $l_{10d}=l+2p{-}2{-}a {-} \frac{1+({-})^{a+l}}{2}$. 
This then  also then gives a very more direct way of obtaining the  leading, log${}^g$, divergence of higher $g$-loop correlators, which arise from taking $(1/(l_{10d})_6)^g$, expanding in 10d bocks and acting with $g$ powers of $\DDelta^{(8)}$~\cite{Caron-Huot:2018kta}. 

Note that in the strong coupling case the 10d symmetry has a stronger motivation than at weak coupling, since here it seems to be intimately related to the fact that $AdS_5 \times S^5$ is conformally equivalent to 10d flat space,  and a supergravity argument for the symmetry could perhaps be envisaged starting from this point.

\subsubsection*{Tree level string corrections and 10d symmetry}

The first $(\alpha'^3=1/\lambda^{3/2})$ string corrections to the strong coupling half BPS four-point correlators have been computed for arbitrary charges in~\cite{Alday:2017xua,Binder:2019jwn,Drummond:2019odu}. 
In~\cite{Abl:2021mxo} it was shown that these corrections for all half BPS correlators can be rewritten in a way which explicitly  exhibits the 10d conformal symmetry, after applying a certain 6th order Casimir of the conformal and internal ($SO(2,4)\times SO(6)$) symmetry groups acting at points 1 and 2, $\mathcal{C}^{(6) }_{1,2}$. All half BPS operators at this order are then given by the succinct formula 
\begin{align}\label{preamp2}
	\langle \cbO \cbO \cbO \cbO \rangle|_{c^{-1}\lambda^{-3/2}}&\propto  \cI(x_i,y_i) \times \mathcal{C}^{(6) }_{1,2} \left[\frac{  D_{4411}( {\bf x}_i)}{\bx_{34}^6} \right]_{10d}\ .
\end{align}
Here the $\log u$ part of the 10d function $D_{4411}( {\bf x}_i)$, which is the part controlling the anomalous dimension, is equal to a single spin 0, 10d conformal block.
The obvious generalisation of this structure for the next order $\alpha'^5$ correction would then be to write the master correlator in terms of  tenth order Casimirs acting on 10d spin 2 and spin 0 blocks. The order of the Casimir is directly related to the mass dimension  of the correction (as was already observed in the supergravity case in~\cite{Caron-Huot:2018kta}) whereas the maximum spin dependence arises from the relation to the effective action following arguments of~\cite{Heemskerk:2009pn} applied to 10d. It seems though that the $\alpha'^5$ corrections,  found for all charges in~\cite{Drummond:2020dwr},  can   {\em not} be rewritten in this above 10d form~\cite{Abl:2021mxo} and this is consistent with the observations of the breaking of 10d symmetry found in the spectrum of operators at this order in~\cite{Drummond:2019odu}.

On the other hand a clear 10d (but non-conformal) structure {\em has} recently been discovered in the higher charge correlators corresponding to string corrections~\cite{Abl:2020dbx}. The master correlator generating all half BPS correlators can be written in terms of  AdS$\times$ S integrals mimicking scalar Witten diagrams, but as integrals over a full 10d  AdS${}_5\times S^5$ space-time rather than just the AdS part.
This suggests that there is a simple scalar 10d effective action describing  the quartic terms of all string corrections to IIB supergravity on AdS${}_5\times S^5$. 
The first string correction~\eqref{preamp2}  arises from a simple $\phi^4$ term
\begin{align}
	S_{\alpha'^3} =\frac1{8.4!} \left(\frac{\alpha'}2\right)^3\times 2\zeta_3 \times \int_{\text{AdS}\times\text{S}} d^{10} {\bf z} \,\phi({\bf z})^4\ .
\end{align}
To obtain the corresponding CFT correlators, we mimic the standard AdS/CFT procedure  but in a fully 10d AdS${}_5\times S^5$ covariant way,
using generalised AdS${}\times S$ bulk-to-boundary propagators. In this way we obtain the  
AdS$\times$S Witten diagram for this contact interaction, yielding the following proposal for the $\alpha'^3$ corrections to all higher charge correlators via the master correlator:
\begin{align}\label{apcubed}
	\langle\cbO\cbO\cbO\cbO \rangle|_{c^{-1}\lambda^{-3/2}} &\sim  \frac{ 2\zeta_3}{8.4!} \left(\frac{\alpha'}2\right)^3  \frac{(\cC_4)^4}{(-2)^{16}}\,\cI(x_i,y_i)\, \int_{\text{AdS}\times\text{S}}  
	\frac{{d^{10} {\bf z}}}{({\bf z}.\bx_1)^4({\bf z}.\bx_2)^4({\bf z}.\bx_3)^4({\bf z}.\bx_4)^4}\ .
\end{align}
Here we use 12 component variables $\bf z$ to define the bulk AdS$\times$S space, the first 6 components are embedding space variables for AdS${}_5$ and the next 6  for $S^5$. These then contract with the 12 component variables $\bx=(x,y)$  (with $x$   6-component embedding space variables for Minkowski space and $y$ the 6 $SO(6)$  internal variables) using  an $SO(2,10)$ metric as ${\bf z}.\bx$. The AdS$\times$ S generalisation of the bulk to boundary propagator for a dimension $\DDelta$ scalar in 10d is then $\frac{\cC_\DDelta}{(-2)^\DDelta}({\bf z}.\bx)^{-\DDelta}$ with the normalisation $\mathcal{C}_{\DDelta} =\frac{\GGamma(\DDelta)}{2{\pi}^{d/2} \GGamma(\DDelta-d/2+1)}$. Extracting the individual higher charge  component correlators from~\eqref{apcubed} is then a straightforward Taylor expansion, and there is also a simple formula for obtaining the corresponding Mellin transform (see~\cite{Abl:2020dbx} for more details).

Higher order corrections can similarly be read off from $\phi^4$ terms with derivatives in the effective action. The derivatives  are AdS$\times$S covariant and so don't commute. 
The relation to the flat space Virasoro amplitude -- which can also be viewed  via a quartic  effective action with derivatives -- is then very direct:  simply replace the covariant derivatives with flat space ones. However, the uplift from flat space to curved space is not unique, due to the non commuting derivatives. Remarkably the ambiguities in this process correspond precisely to ambiguities one obtains  from bootstrap approaches to obtaining higher charge correlators~\cite{Aprile:2020mus}. Some of these ambiguities have been fixed by localisation~\cite{Binder:2019jwn}.
So for example at $O(\alpha'^5)$ one considers a scalar effective action with 4-derivative terms and below. The  scalar 10d effective action can be written
\begin{align}
	S_{\alpha'^5}=\frac{ \zeta_5}{8.4!}\left( \frac{\alpha'}2\right)^5 \int_{\text{AdS}\times\text{S}} d^{10} {\bf z} \Big(3 	(\nabla\phi.\nabla\phi)(\nabla\phi. \nabla\phi)-9  \nabla^{2}\nabla_{\mu}\phi\nabla^{\mu}\phi\phi^{2}-30  \phi^4 \Big)\ ,
	\label{alphap5corr}
\end{align}
where $\nabla_\mu$ are AdS${}\times S$ covariant derivatives. Replacing the scalars with  bulk to boundary propagators as in~\eqref{apcubed}   converts this directly into a formula for the master correlator  giving all higher charge half BPS correlators  at $O(\alpha'^5)$. In~\eqref{alphap5corr}, the coefficient of the first term  is fixed by comparing with the flat space effective action (arising from the Virasoro amplitude) whereas the coefficients of the remaining two terms need other mechanisms to fix them (but we emphasise that there are just two coefficients unfixed by the effective action and flat space limit needed to fix {\em all} half BPS operators). This result agrees precisely with the Mellin space formulae derived by $\cN=4$ bootstrap techniques  in~\cite{Drummond:2020dwr}.
This agreement continues at higher orders and has been checked to $O(\alpha'^7)$.  
Some of the remaining  coefficients arising from ambiguities can be fixed by supersymmetric localisation in $\cN=4$ SYM~\cite{Binder:2019jwn}.

There are a number of interesting open problems arising from this 10d effective action. The first  problem is to derive the scalar effective action from first principles in supergravity. It is known that
IIB supergravity linearised on the AdS$\times$S superspace background
is described by a single chiral scalar superfield with a certain fourth order constraint~\cite{Heslop:2000np} just as on flat space~\cite{Howe:1983sra}. 
It presumably then makes sense to integrate a superpotential consisting of a holomorphic function of this scalar in chiral AdS$\times$S superspace. 
But this does then lead to the question of  the existence of an effective  chiral superpotential describing the full nonlinear theory. 
Such an object has been discussed before, notably  in~\cite{deHaro:2002vk,bh,Rajaraman:2005up}. An obstruction to its existence  was found (after some initial confusion) in~\cite{bh,deHaro:2002vk}. However, despite this it was shown in~\cite{Rajaraman:2005up} that all terms  in the full non-linear effective action  consisting of the curvature and  five-form field strength
{\em are} correctly reproduced by such a superpotential.
The second problem is to understand the relation between this 10d structure and the afore-mentioned 10d conformal symmetry. The  above results would suggest  that the covariant derivatives in the effective action at $O(\alpha'^5)$ break the conformal symmetry. Intriguingly though  there is a different combination of  the three terms in the effective action at this order~\eqref{alphap5corr} for which the corresponding Witten diagram integrand has 10d conformal symmetry~\cite{Abl:2021mxo}, reminiscent of the symmetry of the perturbative integrand of~\cite{Caron-Huot:2021usw} discussed in section~\ref{pert10d}.

\subsubsection*{Loop corrections}

Finally quantum gravity loop corrections $O(1/c)$ have also been computed for various higher charge correlators in position space~\cite{Aprile:2019rep} and Mellin space~\cite{Alday:2019nin}.
The first string $(\alpha'^3)$ correction is also known at one loop  order for some higher charge correlators in Mellin space~\cite{Drummond:2020uni}. 
For both string and supergravity one loop correlators,  the higher dimensional 10d conformal symmetry has a clear imprint on the results -- they can be written in a vastly simpler way by implementing the differential operators arising from the symmetry --  but the precise way this symmetry impacts on these corrections is not yet clear.

\section{Higher points}
\label{sec:higherpoints}

In the final section of this review we turn to correlators of more  than four half BPS operators. They are classified by their Grassmann odd degree. The four point loop integrands arise from higher point correlators  with maximal Grassmann odd degree (maximally nilpotent). Going below the maximally nilpotent case (which can be viewed as 5 and higher  point   loop level integrands) there is  far less   known, nevertheless we will review in detail the structure  superconformal symmetry imposes and the explicitly known results.

\subsection{Analytic superspace}
\label{sec:an}

In order to fully understand half BPS correlators beyond four points (and indeed to derive the results we have already used without proof at four-points) the most direct method is to use analytic superspace. 
Analytic superspace was first introduced for $\cN=2$ supersymmetric theories in~\cite{Galperin:1984av}. Later this was generalised to arbitrary $\cN$ and placed within the mathematical formalism of super flag manifolds, coset spaces of the superconformal group $SU(2,2|\cN)$ in~\cite{Howe:1995md,Hartwell:1994rp}. 
A particularly useful $\cN=4$ analytic superspace is provided by the  maximal super Grassmannian, $\Gr(2|2,4|4)$, the space of $(2|2)$ planes in $(4|4)$ dimensions~\cite{Howe:2001je,Heslop:2001dr,Heslop:2001gp,Heslop:2001zm,Heslop:2002hp,Heslop:2003xu,Doobary:2015gia}. This space is particularly natural for describing half BPS correlators and in particular manifesting the superconformal symmetry. 
This space  is a very natural generalisation of both 4d Minkowski space viewed as a Grassmannian,  as well as the internal space ($y$s)
used in the previous section for dealing with higher charge half BPS operators (see~\eqref{phidef}). Thus we will begin by reviewing in more detail these bosonic spaces from various points of view, before considering the full superconformal symmetry in analytic superspace.

\subsubsection*{Equivalent descriptions of $4d$ Minkowski space}
\label{sec:Mink}

There are several equivalent ways to view 4d complexified%
\footnote{We consider complexified (super)spaces as coset spaces of the complexified (super)conformal group $SL(4|\cN;\com)$ as it makes things much simpler to describe and it's easy to go back to the real forms at the end if needed.}  Minkowski space, $\mathbb{R}^{3,1} \xrightarrow{\com} \mathbb{C}^4$: the Grassmannian of 2 planes in a 4d vector space $\Gr(2,4)$; a flag manifold coset space of the superconformal group; the embedding space formalism (Klein quadric $\subset \mathbb{P}^5$); and  lines in projective twistor space. They are straightforwardly related to each other.
We will use	 Minkowski space indices $\mu=0,..,3$, Weyl spinor indices $\alpha=1,2$, $\dot \alpha =1,2$, twistor indices $\Alpha=1,..,4$ with $v^\Alpha=(v^\alpha,v_{\dot \alpha})$ which carry the fundamental of the conformal group $SU(2,2)\xrightarrow{\com}  SL(4;\com)$ (the 4d conformal group is  $SU(2,2) \sim SO(2,4)$) and finally we also have embedding space indices $\mathrm{I}=-1,0,..,4$ which carry the fundamental of $SO(2,4)\xrightarrow{\com}  SO(6;\com)$.  These various formalisms are  related to each other as follows:
\begin{align}\label{diag}
	\begin{tikzpicture}
		\node (coset) at (-2,2) {Coset space : $SL(4;\com)\ni x_\Alpha{}^\Beta \sim \left(\begin{array}{cc}
				\delta_\alpha{}^\beta&x_{\alpha\dot \beta}\\
				0& \delta^{\dot \alpha}{}_{\dot \beta}
			\end{array}\right)$};
		\draw [draw=black] (-0,2.6) rectangle (1.8,2);
		\node (Grassmannian) at (-2,0) {Grassmannian $\Gr(2,4)$: $x_\alpha{}^\Beta \sim (\delta_\alpha{}^\beta,x_{\alpha\dot \beta})$};
		\node (coords) at (2.5,-2) {Minkowski space: $x_{\alpha \dot \alpha}= x_\mu \sigma^\mu_{\alpha \dot \alpha}$};
		\node (Embedding) at (-5,-2) {Embedding space: $x^\mathrm{I} \sim x^{\mathrm{AB}} =  x_\alpha{}^\mathrm{A} x_\beta{}^\mathrm{B}\epsilon^{\alpha \beta}$};
		\draw[->] (.9,2) -- (.9,0.3);
		\draw[->] (-1,-.3) -- (-2,-1.7);
		\draw[->] (1.2,-.3) -- (4.5,-1.8);
	\end{tikzpicture}
\end{align}

{\em  Minkowski space} coordinates $x_\mu \in \com^4$ have spinorial form as a $2\times 2$ matrix $x_{\alpha \dot \alpha}= x_\mu \sigma^\mu_{\alpha \dot \alpha}$. 
{ The Grassmannian} $\Gr(2,4)$ can be viewed as the equivalence class of  $2\times 4$ matrices modulo the left action of $GL(2)$:  
$
\Gr(2,4)=\{ x_\alpha{}^\Alpha \sim m_\alpha{}^\beta x_\beta{}^\Alpha: m_\alpha{}^\beta \in GL(2)\}.
$
A natural way to fix a representative of the $GL(2)$ equivalence class is to use up the $GL(2)$ by fixing  the left $2\times 2$ block of the $2\times 4$ matrix $x_\alpha{}^\Alpha$ to the identity. This leaves a remaining $2\times 2$ matrix which we can then identify with the spinorial form of  coordinates of Minkowski space $x_\alpha{}^\Beta \sim (\delta_\alpha{}^\beta,x_{\alpha \dot \beta})$.

{\em Twistor space } is the vector space $\com^4$ and the Grassmannian is geometrically the space of 2 planes through the origin in the 4 dimensional twistor space.  To see this note that a  2-plane through the origin can be defined by giving any two independent twistors in the 2-plane. These are the rows of the $2\times 4$ matrix, $(x_1)^\mathrm{A}$ and $(x_2)^\mathrm{A}$. Then any such choice of two twistors in the plane will be  related by a $GL(2)$ transformation. It is common to view the  2-plane in twistor space rather  as a line in projective twistor space $\mathbb{P}^3$.

{\em  The coset space} of $SL(4;\com)$: 
\begin{align}\label{cosetmink}
	\text{coset space:}\ \  \{x_\Alpha{}^\Beta \sim h_\Alpha{}^\Gamma x_\Gamma{}^\Beta: x_\Alpha{}^\Beta \in SL(4;\com), \  h = \left(\begin{array}{cc}
		m&0\\n&p
	\end{array}\right)\in SL(4;\com)\}
\end{align}
(where $m,n,p$ are $2\times 2$ matrices) is equivalent to the Grassmannian $\Gr(2,4)$.
To see this note that we can use up the matrices $n,p$ of $h$ to set the bottom two rows of the matrix $x_\Alpha{}^\Beta$ to $(0_2,1_2)$ so there is a coset representative yielding Minkowski coordinates given as the upper triangular matrix at the top of~\eqref{diag}.  
%\begin{align}\label{coord}
%x_\Alpha{}^\Beta \sim 	\left(\begin{array}{cc}
	%		\delta_\alpha{}^\beta&x_{\alpha\dot \beta}\\
	%		0& \delta^{\dot \alpha}{}_{\dot \beta}
	%	\end{array}\right)\,.
%\end{align}
Then the top two rows of the matrix, together with the remaining equivalence transformation $m$ comprise precisely the Grassmannian $\Gr(2,4)$.

{\em  Embedding space} can be derived by noting that an alternative approach for dealing with the $GL(2)$ equivalence, rather than simply fixing it (which in particular breaks the linear action of the conformal group) is to instead take two copies of $x_\alpha{}^\Alpha$ and contract the $\alpha$ indices with an $\epsilon$ tensor to form $x^{\mathrm{AB}}:=x_\alpha{}^\Alpha x_\beta{}^\Beta \epsilon^{\alpha \beta}$. These coordinates $x^{\Alpha\Beta}$ are known as Pl\"ucker coordinates for the Grassmannian.  Since $x^{\mathrm{AB}}$ is antisymmetric in its 4-indices it forms the 6 rep of the conformal group $SU(2,2)$ which is the fundamental of $SO(2,4)$. This can be manifested using appropriate 6d $SO(2,4)$ sigma matrices $x^I=x^{\mathrm{AB}}\sigma^\mathrm{I}_{\mathrm{AB}}$. The coordinates $x^I$ then transform linearly under the conformal group $SO(2,4)$ and  are invariant under $SL(2) \subset GL(2)$.  However there is a $\com^*\subset GL(2)$ transformation remaining and thus the $x^\mathrm{I}$ are projective coordinates in a six-dimensional space, $x^I \in \mathbb{P}^5$ known in this context as the embedding space. The coordinates $x^\mathrm{I}$ are not independent. We can see that $x^{[\mathrm{AB}}x^{\Gamma\Delta]}=x_\alpha{}^{[\Alpha} x_\beta{}^\Beta x_\gamma{}^\Gamma x_\delta{}^{\Delta]} \epsilon^{\alpha \beta}\epsilon^{\gamma \delta}=0$ which becomes $x_\mathrm{I}x^\mathrm{I}=0$ where the $\mathrm{I}$ index is raised and lowered using an $SO(2,4)$ metric in the real case. Thus the coordinates $x^\mathrm{I}$ live in a hyperbolic surface (or in fact simply a complex sphere as we are considering the complex case) inside $\mathbb{P}^5$.

Finally note that there is an orthogonal plane to any Grassmannian element $x^\perp_{\Alpha \dot \beta}$ which satisfies $x x^\perp = 0$ and which can be thought of as the right hand half of the inverse of the coset matrix $(x^{-1})_\Alpha{}^\Beta$ so  $x^\perp_{\Alpha \dot \beta}=(x^{-1})_{\Alpha \dot \beta}$.  Its equivalence class is obtained by $p^{-1}$ acting from the right,  $x^\perp \sim x^\perp p^{-1}$. 
This orthogonal plane then allows us to form $SL(4)$ invariant objects $(x_{ij})_{\alpha \dot \alpha}:=x_{i\alpha}{}^\Alpha x^\perp_{jA \dot \alpha}$ which transform locally on the left and right with different  $GL(2)$s,  $(x_{ij}) \sim M x_{ij} P^{-1}$. If  the coset representative at the top of~\eqref{diag} is chosen then this becomes the difference of Minkowski space coordinates: $x_{ij \alpha \dot \alpha}\sim x_{i \alpha \dot \alpha}-x_{j \alpha \dot \alpha}$.
We can form the $SL(2)$ invariant combination
$x_{ij}^2 = \det (x_{ij})$. This transforms as $x_{ij}^2 \sim x_{ij}^2 \det M_i \det^{-1}P_j$.

\subsubsection*{Internal $SU(4)$ space}

In section~\ref{sec:highercharges} we introduced coordinates $y^I$ with $ I =1..6$ satisfying $y^Iy_I=0$~\eqref{Opdef} in order  to deal with the  six fundamental scalars $\phi^I$ in $\cN=4$ SYM. Remarkably, these coordinates can themselves be viewed in exactly the same way as the embedding space coordinates $x^\mathrm{I}$  for Minkowski space. In the real case the $x^\mathrm{I}$ transform under the fundamental of the conformal group $SO(2,4)$ whereas the $y^I$ transform under the fundamental of the internal group $SO(6)$ , but after complexifying both groups are $SO(6;\com)$.
In fact {\em every aspect} of the equivalent descriptions in section~\ref{sec:an} of Minkowski space above can be repeated for the internal space. For the internal space we will use equivalent Latin versions of the same Greek indices used for Minkowski space:  they are transforming under different copies of the same complexified groups. So in particular we have an internal  Grassmannian $Gr(2,4)$, described by a $2\times4$ matrix  with elements $y_a{}^A$.\footnote{These are usually labelled $u$ in the literature on $\cN=4$ harmonic/ analytic superspace, eg $u_a{}^i$ in~\cite{Howe:1995md,Hartwell:1994rp,Howe:2001je,Eden:2001ec} and $u^{+a}_A$ in~\cite{Chicherin:2015bza}. They are viewed as elements of the coset space $SU(4)/(S(U(2)\times U(2)))$ which  on complexification becomes the Grassmannian.} These are then related to the coordinates $y^I$ in exact analogy to the $x^\mathrm{I}$~\eqref{diag}, i.e. $y_I\sigma^I_{ab}=y_a{}^Ay_b{}^B\epsilon^{ab}$. 
We have the orthogonal Grassmannian $y^\perp_{A\dot b}$ with $y y^\perp=0$ just as for Minkowski space.
We also have explicit coordinates for the coset space,  $y_{a\dot a}$, analogous to the spinor rep of Minkowski space coordinates. In short everything we say about Minkowski space has a precise analogy for the internal space and indeed after complexification this analogy is actually a precise equivalence locally.

\subsubsection*{Analytic superspace}
\label{sec:analytic}
Finally we come to analytic superspace itself which combines the above two $\Gr(2,4)$ Grassmannian descriptions of Minkowski space and the internal space into a super Grassmannian $\Gr(2|2,4|4)$,  the space of $(2|2)$ dimensional super planes inside a $(4|4)$ dimensional vector space.  
So we introduce new superindices, calligraphic Latin versions  of the corresponding Minkowski (Greek) or internal (Latin) indices. So $\ca=(\alpha|a),\dot \ca=(\dot \alpha| \dot a)$ are $(2|2)$ superindices and $\cA$ are $(2|4|2)$ (a convenient permutation of the $(4|4)$ vector space) superindices with $V^\cA=(V^\ca,V_{\dot \ca})=(V^\alpha|V^a,V_{\dot a}|V_{\dot \alpha})$. Then the super Grassmannian is 
\begin{align}
	\Gr(2|2,4|4) = \{ X_{\ca}{}^\cA \sim M_{\ca}{}^{\cb} X_{\cb}{}^{\cA}: M_{\ca}{}^{\cb} \in GL(2|2)\}\ 
\end{align}
and using the $GL(2|2)$ we can  fix the first block to the identity to arrive at the coordinates $X_{\ca \dot \ca}$ of analytic superspace
$	X_{\ca}{}^\cB \sim (\delta_\ca{}^\cb,X_{\ca \dot \cb})\ .
$
The coordinates of analytic superspace include the coordinates of Minkowski space $x_{\alpha \dot \alpha}$ and of the internal  space $y_{a \dot a}$ as well as Grassmann odd superspace variables $\rho_{\alpha \dot a}$ and $\bar \rho_{\dot \alpha a }$ in the supermatrix
\begin{align}\label{coords}
	X_{\ca \dot \ca} = 	\left(\begin{array}{c|c}
		x_{\alpha \dot \alpha}&\rho_{\alpha \dot a}
		\\ \hline
		\bar \rho_{a \dot \alpha} & y_{a \dot a}
	\end{array}\right)\ .
\end{align}
Just as for Minkowski space,  this super Grassmannian can also be viewed as the top half of the supercoset space (compare with~\eqref{cosetmink})
\begin{align}\label{supercoset}
	\text{supercoset space:}\ \  \left\{X_\cA{}^\cB \sim H_\cA{}^\cB X_\cC{}^\cB: X_\cA{}^\cB \in SL(4|4;\com), \  H_\cA{}^\cB = \left(\begin{array}{cc}
		M_\ca{}^\cb&0\\ N^{\dot \ca \cb} &P^{\dot \ca}{}_{\dot \cb}
	\end{array}\right)\right\}
\end{align}
with $M,N,P \in GL(2|2)$ so that we obtain the same coordinates via the coset representative
\begin{align}\label{cosetrep}
	X_\cA{}^\cB \sim 	\left(\begin{array}{cc}
		\delta_\ca{}^\cb&X_{\ca\dot \cb}\\
		0& \delta^{\dot \ca}{}_{\dot \cb}
	\end{array}\right)\,.
\end{align}

We also mention here that we also have  available to us the inverse of the supercoset matrix $(X^{-1})_{\cA}{}^\cB$. The non-trivial part of this is the right hand half of the matrix $X^\perp_{\cA\dot \cb}:=(X^{-1})_{\cA\dot \cb}$ which in the Grassmannian language represents the plane perpendicular to $X_\ca{}^\cA$, \begin{align}
	(X)_\ca{}^\cA  X^\perp_{\cA \dot \cb} =0\ .
\end{align}
Note that the orthogonal complement transforms under the $GL(2|2)$ $P^{\dot \ca}{}_{\dot \cb}$ rather than $M_\ca{}^\cb$ in $H$~\eqref{supercoset}, so we have the equivalence transformation $  X^\perp_{\cA \dot \cb} \sim  X^\perp_{\cA \dot \ca}P^{\dot \ca}{}_{\dot \cb}$.

\subsection{Half BPS operators in analytic superspace}

Analytic superspace can be used to solve the Ward identities for correlators of any operator in the theory~\cite{Heslop:2001zm,Heslop:2003xu} but is especially suited to describing half BPS operators which are simply scalar fields on this space. The simplest half BPS representation contains the fundamental fields of the theory, the six scalars $\phi(x,y)$ (see~\eqref{phidef}), 4 complex fermions $\lambda^{\alpha  A}(x)$ and their conjugates and the Yang-Mills field strength tensor $F_{\mu \nu}(x)$ in spinor form 
\begin{align}\label{W}
	W(x,y,\rho,\bar \rho)= \phi  + \lambda^{\alpha A}  y^\perp_{A\dot a} \rho_{\alpha  \dot b}\epsilon^{\dot a\dot b}
	+ \bar \lambda^{\dot \alpha}_A y_a{}^A \bar \rho_{b \dot \alpha }\epsilon^{ab}
	+ F^{\alpha \beta} \rho^2_{\alpha \beta}+ \bar F^{\dot \alpha \dot \beta} \bar \rho^2_{ \dot \alpha\dot \beta } \ .
\end{align}
A key point is that the multiplet is an unconstrained superfield on analytic superspace.  It has a  finite expansion in the Grassmann odd variables $\rho, \bar \rho$ since they are nilpotent, but also in the internal variables $y$. This is because the internal space is the complexification of a compact space (to be compared with the $x$ which are non-compact and thus produce an infinite expansion).   
The multiplet transforms in the adjoint representation of the gauge group and is not itself a a gauge-invariant operator. However all half BPS operators are built out out of $W$. For example the simplest stress tensor multiplet is
\begin{align}\label{O2}
	\cO_2 := \Tr(W^2) = \Tr (\phi^2) + ... + \rho^4 L + ... + \bar \rho^4 \bar L+ .. + (\rho^2)_{\alpha \beta} (\bar \rho^2)_{\dot \alpha \dot \beta} T^{\alpha \beta \dot \alpha \dot \beta}+... \ .
\end{align}
where we just display a few of the component fields in the multiplet. The lowest weight state $\Tr(\phi^2)$, the chiral Lagrangian $L$ and its conjugate, and the stress-tensor multiplet $T^{\mu \nu}$ in the spinor representation, from which the multiplet gets its name.

\subsection{Ward identities (part 1)}

\label{sec:WI1}

Having set up the analytic superspace formalism, we can now turn to the question of solving the superconformal Ward identities for correlation functions of half BPS operators in $\cN=4$ SYM. The presentation here is similar to that in~\cite{Heslop:2003xu} but we here do this explicitly in the language of the super Grassmannian whereas there it was done in coordinate language.
In the Grassmannian formulation, the superconformal transformations are linear transformations on the $\cA$  indices. An $n$-point correlator  is a function of $n$ elements of the Grassmannian $\Gr(2|2,4|4)$, $(X_1)_\ca{}^\cA,..,(X_n)_\ca{}^\cA$ or their orthogonal complements. It must be {\em invariant} under $SL(4|4)$, and have transformation properties under the local $GL(2|2)$ transformations $M_\ca{}^\cb, P^{\dot \ca}{}_{\dot \cb}$ in $H$~\eqref{supercoset} which are specified by the operator inserted at each point. 
To obtain $SL(4|4)$ invariant quantities we simply need to contract the $\cA$ indices, the only possibility is to pair planes with their  orthogonal complements at different points: 
\begin{align}
	(X_{ij})_{\ca \dot \cb} =	(X_i)_{\ca}{}^\cA ( X^\perp_j)_{\cA \dot \cb}\ .
\end{align}
These then transform as $(X_{ij})\sim M X_{ij} P^{-1}$.
Note that if we fix to the coordinates $X_\ca{}^\cB \sim (\delta_\ca{}^\cb,X_{\ca \dot \cb})$ then the orthogonal Grassmannian is $( X^\perp)_{\cA\dot \cb} \sim (-X_{\ca \dot \cb},\delta^{\dot \ca}{}_{\dot\cb},)$ and $(X_{ij})_{\ca \dot \cb}$ is simply the difference of coordinates $(X_{ij})_{\ca \dot \cb} \sim (X_{i})_{\ca \dot \cb} -(X_{j})_{\ca \dot \cb}$. 

Now in the Grassmannian language the $(X_{ij})_{\ca \dot \cb}$ have already  solved the superconformal $SL(4|4)$ Ward identities! However, the catch is that we now also need to ensure 
the result transforms correctly under the two  $GL(2|2)$ transformations at each point $M_i,P_i$ (so the result actually lives on the Grassmannian). 
Up to this point the presentation is valid for correlators of any operators in the theory - different operators have different transformation properties under the two $GL(2|2)$s.  Specialising to half BPS correlators $\cO_p(X_i)$ these transform under $M,P$ only via a scaling of $\sdet M= \sdet^{-1}P$. (This latter equality arises from the fact that we are considering $SL(4|4)$ rather than $GL(4|4)$ and so $1=\sdet H= \sdet M \sdet P$ (see~\eqref{supercoset}).)
This scaling under $\sdet M= \sdet^{-1}P$ can be accounted for by taking the $\sdet$ of $X_{ij}$,  $g_{ij}:=\sdet(X_{ij}^{-1})$. These are superpropagators in analytic superspace, proportional to the 2-point function of two fundamental scalars and if we consider the coordinates~\eqref{coords} and switch off the Grassmann odd coordinates we find $g_{ij}= y_{ij}^2/x_{ij}^2$ 
matching up with the notation we introduced previously (see~\eqref{fcF}). Thus by taking appropriate powers and products of the superpropagators we can straightforwardly obtain a function which transforms correctly as a half BPS correlator  under the local $GL(2|2)$s $M,P$. 

It remains therefore to multiply this prefactor 
by   quantities which are {\em invariant} under the two $GL(2|2)$s $M,P$ (with $\sdet M= \sdet^{-1}P$). To obtain these we need to  contract away all the  $\ca,\dot \ca$ indices, recalling that they must contract `locally' with indices associated at the same point.  
To contract away the indices we make  use of the inverse $(X_{ij}^{-1})^{\dot \ca \ca}:=((X_{ji})_{ \ca \dot \ca })^{-1}$ to get upstairs $\ca,\cb$ indices with which to contract. Then simply link them together pointwise to arrive at traces such as  $\str(X_{ij}X_{jk}^{-1}X_{kl}X_{li}^{-1})$ or  $\sdet$s of similar objects e.g. $\sdet(X_{ij}X_{jk}^{-1}X_{kl}X_{li}^{-1})$. A systematic analysis~\cite{Heslop:2002hp,Heslop:2003xu}
shows that the Ward identities can be reduced to finding functions of the $(2|2)\times (2|2)$ matrices $(Z_i)_{\ca}{}^\cb=(X_{12}X_{2i}^{-1}X_{i3}X_{31}^{-1})_{\ca}{}^\cb$ for $i=4,5,..,n$,  invariant under $SL(2|2)$ transformations%
\footnote{\label{footnote:svg}They only need to be invariant under $SL(2|2)$ rather than $GL(2|2)$. A non-trivial transformation under $\sdet(M_i)$ can be cancelled by appropriate products of $g_{ij}$s (recalling that $\sdet M= \sdet^{-1}P$).}
$
$
$G$, so $f(Z_i)=f(G^{-1}Z_i G)$.%
\footnote{\label{an123}The quickest way to see that the Ward identities reduce to this is to adapt the standard CFT argument of using conformal transformations to fix $X_1,X_2,X_3$ to $1,\infty,0$ which in the Grassmannian language becomes
	$
	(X_1)_{\ca}{}^\cB \rightarrow (\delta_\ca{}^\cb,\delta_{\ca \dot \cb}),\   (X_2)_{\ca}{}^\cB \rightarrow  (0,\delta_{\ca \dot \cb})	,\  (X_3)_{\ca}{}^\cB \rightarrow  (\delta_\ca{}^\cb,0) \, .	
	$
	Then the remaining transformations on the remaining  coordinates $X_4,X_5,..$ leaving $X_1,X_2,X_3$ invariant are $G^{-1}X_iG$. One can check that $X_i=Z_{i}$ when $X_1,X_2,X_3$ are fixed thus.
}

\subsection{Deriving the structure of four-point correlators}

Before continuing the discussion of Ward identities, let us consider this first at  four points.
The above description on solving the Ward identities implies  that the correlator of four half BPS operators takes the form (with $Z:=Z_4$)
\begin{align}
	\langle \cO_{p_1}\cO_{p_2}\cO_{p_3}\cO_{p_4} \rangle = \text{prefactor} \times f(Z) \qquad f(Z)=f(G^{-1} Z G)\ ,
\end{align}
where the prefactor is an appropriate monomial in $g_{ij}$ absorbing the transformations of the operators.
For a bosonic matrix, a function $f(Z)=f(G^{-1} Z G)$ is simply a symmetric function (meaning symmetric under permutation) of its eigenvalues. For a super matrix it is   a {\em supersymmetric function} of its eigenvalues. A supersymmetric function has two sets of eigenvalues, and is doubly supersymmetric in these two sets and satisfies the additional constraint that on setting an eigenvalue from one set equal to the eigenvalue of the other set, the dependence on the eigenvalue disappears~\cite{stembridge1985characterization}. %
%%
%\footnote{See \href{https://www.symmetricfunctions.com/superSymmetricSchur.htm}{https://www.symmetricfunctions.com/superSymmetricSchur.htm}.} 
%	
In the current context the eigenvalues of the $(2|2)\times (2|2)$ matrix $Z$ we denote  $(x,\bar x|y,\bar y)$ and this becomes
\begin{align}\label{sufun}
	f(x,\bar x|y,\bar y)=f(\bar x,x|y,\bar y)=f(x,\bar x|\bar y,y) \qquad \partial_tf(x,t,|y,t)=0\ ,
\end{align}
which are the four-point Ward identities first derived in~\cite{Dolan:2001tt,Arutyunov:2002fh}.
This has general solution (which can be found for example by considering an appropriate basis of such functions eg super Schur polynomials, see~\cite{Heslop:2002hp,Doobary:2015gia})
\begin{align} \label{decomp}
	f(x,\bar x|y,\bar y)=&a+\left[\left(\frac{(x-y)(x-\bar y)(\bar x-y)}{(x-\bar x)(y-\bar y)}b(x,y) + x\leftrightarrow \bar x \right)+y\leftrightarrow \bar y \right]\notag\\
	&+ (x-y)(x-\bar y)(\bar x-y)(\bar x-\bar y)c(x,\bar x|y,\bar y),
\end{align}
where the functions $b,c$ are (doubly) symmetric and non singular at $x=y$.
The constant $a$ is just the value of $f$ when $y=x$ and $\bar y=\bar x$ and  the function $b(x,y)$ is related to $f$ at  $\bar y=\bar x$ only 
\begin{align}\label{ab}
	a=f(x,\bar x| x,\bar x) \qquad  b(x,y)=\frac{a-f(x,\bar x|y,\bar x)}{x-y}\,.
\end{align}
One can easily check that any supersymmetric function (i.e. satisfying~\eqref{sufun}) decomposes uniquely as~\eqref{decomp}.%
\footnote{Note that subtracting the first line on the rhs of~\eqref{decomp} from $f$ and setting  $y=x$ gives zero when using the definitions in~\eqref{ab}  for {\em any} doubly symmetric function $f$. Thus the factor $x-y$ can be pulled out and by symmetry the other three factors multiplying the function $c$.}

On top of this we must also impose that the full correlator is polynomial in the $Y_i$ variables. By choosing the prefactor judiciously this means the function $f$ must be polynomial in $y,\bar y$ of degree determined by the weights of the external operators.
This, then yields the most general solution of the $\cN=4$ superconformal Ward identities. 
As we will see, it turns out that only the function  $c$ can depend on the coupling~\cite{hep-th/0009106} and so the $a$ and $b$ pieces may as well be replaced by the free theory. This then reproduces the structure quoted in~\eqref{corhighercharge}.

\subsection{Ward identities (part 2): nilpotent invariants}
\label{sec:WI2}

Returning to the general $n$-point case,
we  have the most general solution to the superconformal Ward identities in terms of functions  $f(Z_i)=f(G^{-1}Z_i G)$ with $G \in SL(2|2)$. At first sight one would imagine such functions can only be built from $\sdet$ or $\str$ of products of the $Z_i$ or their inverses as was the case at four points. 
Indeed this is the solution derived through a number of early papers on the subject~\cite{hep-th/9509140,hep-th/9607060,hep-th/9611074,hep-th/9808162,hep-th/9611075}. But there is a puzzle which was first pointed out in~\cite{Intriligator:1998ig}. The puzzle is that the above procedure produces functions  $f(Z_i)=f(G^{-1}Z_i G)$ for $G \in GL(2|2)$ not only $SL(2|2)$. This then in turn means the correlation functions are covariant under the larger group $GL(4|4)$, rather than just  the superconformal group $SL(4|4)$, since the result will be invariant even when $\sdet(M)\neq\sdet(P)^{-1}$ (see~\eqref{supercoset} and footnote~\ref{footnote:svg}).%
\footnote{ Note that somewhat confusingly any matrix proportional to the identity has unit $\sdet$ and thus lies inside $SL(4|4)$. Such a transformation acts trivially on the coordinates and on all operators, thus only $PSL(4|4)$  can act non-trivially. However this is not really relevant for the above discussion: one can consider the superconformal group to be $SL(4|4)$ rather than $PSL(4|4)$. The same discussion arises for $GL(4|4)$ versus $PGL(4|4)$. }
The additional `bonus' symmetry (transformations in $GL(4|4)$ but not in $SL(4|4)$) leaves the bosonic variables $x_{\alpha \dot \alpha},y_{a \dot a}$ invariant and scales the odd variables $\rho \rightarrow \lambda \rho$ and $\bar \rho \rightarrow \lambda^{-1} \bar \rho$, so that $\rho \bar \rho$ is invariant. The problem  is that loop corrections to correlators are correlators with an  insertion of the Lagrangian operator. But the Lagrangian  sits in the half BPS multiplet $\cO_2$ at $O(\rho^4)$~\eqref{O2}. Thus the $l$-loop  correction  of a bosonic correlator must be $\propto \rho^{4l}$ and  will not be $GL(4|4)$ invariant but only $SL(4|4)$ invariant. Put another way, if all correlators are $GL(4|4)$ invariant then they can not have any loop corrections and there is no interacting theory!

The resolution of this puzzle is simply that there exist functions $f(Z_i)=f(G^{-1}Z_i G)$ invariant under $G \in SL(2|2)$ but not under $G \in GL(2|2)$ --  so called nilpotent   $O(\rho^{4l})$ invariants~\cite{Eden:1999gh,Howe:1999hz,Heslop:2003xu,Eden:2011we,Eden:2012tu,Chicherin:2015bza}.
This  possibility arises from the existence of a non-trivial constant tensor (introduced in~\cite{Heslop:2003xu} and denoted $\cE$). This tensor  is invariant under $SL(2|2)$ transformations but not under $GL(2|2)$. It is the analogue for the group $SL(2|2)$ of the more familiar $SL(n)$ completely  antisymmetric tensor $\epsilon_{A_1..A_n}$ which is invariant under $SL(n)$ transformations, but scales under $GL(n)$. Unlike the completely antisymmetric tensor however,  $\cE$  has both upstairs and downstairs indices,   an equal number of both, allowing for an alternative way to produce invariants by contracting the indices of $Z_i$s with $\cE$. The upstairs and downstairs indices of $\cE$ are anti-symmetrised differently and so one needs at least two different $Z_i$ to obtain a non-zero answer, thus they only exist for $n\geq5$. Note that the resulting $SL(2|2)$ invariant will scale under $\sdet(M_1)$. This can be cancelled by taking appropriate factors of $g_{ij}$ when $\sdet(M)= \sdet(P^{-1})$ to produce an $SL(4|4)$ (but not $GL(4|4)$) invariant object.

In~\cite{Eden:2011we,Eden:2012tu,Chicherin:2015bza} an alternative method was used for obtaining these nilpotent invariants which reveals a crucial hidden permutation symmetry of the invariants.
In this approach one obtains invariants by simply writing down {\em any} function on analytic superspace, and then integrating over the action of the superconformal group on this function. Note that this approach will also work for a (compact) bosonic group (for example consider $U(1)$ invariant functions of $z,\bar z$ by integrating arbitrary functions of $ze^{i\theta}$ and   $\bar ze^{-i\theta}$ over $\theta$) but becomes extremely useful for supersymmetric groups.
It is very useful at this point to make a simplifying assumption, namely we set the coordinates $\bar \rho = 0$ and thus break half of the superconformal symmetry. This assumption was motivated by the duality between supercorrelators and superamplitudes~\cite{Alday:2010zy,Eden:2010zz,Eden:2010ce,Eden:2011yp,Adamo:2011dq,Eden:2011ku} (for which this half supersymmetry is broken) and is present from the start in the twistor approach to correlators~\cite{Adamo:2011dq,Chicherin:2014uca} (although this can be restored in that context using the closely related Lorentz Harmonic Chiral formalism~\cite{Chicherin:2016fac,Chicherin:2016fbj,Chicherin:2016soh}). The assumption makes the resulting expressions much simpler, but means they are only correct up to   $\bar \rho$ corrections.
With this assumption then, the super Grassmannian has coordinates
\begin{align}
	X_{\ca}{}^\cA = \left(\begin{array}{c|c}
		x_{\alpha}{}^{\Alpha}&\theta_{\alpha}{}^{A}
		\\ \hline
		0 & y_{a}{}^A
	\end{array}\right)\ .
\end{align}
This includes the Minkowski space Grassmannian $x_{\alpha}{}^\Alpha$ and the internal  space Grassmannian $y_{a}^A$ as well as Grassmann odd variables $\theta_{\alpha}{}^A$ in the $(2|2)\times (4|4)$ matrix. 
Now    the local $SL(2|2)$ transformation is broken to the upper block triangular part by our assumption $\bar \rho=0$. The remaining Grassmann odd part, $\chi_\alpha{}^a$, leaves $x,y$ invariant and transforms $\theta_\alpha{}^A \rightarrow \theta_\alpha{}^A\ + \chi_\alpha{}^a y_a{}^A$.  Thus the variable $\rho_{\alpha \dot a} = \theta_\alpha{}^A y^\perp_{A\dot a}$ is invariant under  $\chi_\alpha{}^a$ and only transforms under the bosonic $SL(2)$ subgroups of $SL(2|2)$ as dictated by its indices. Note that this variable is equal to the $\rho_{\alpha \dot a}$ of~\eqref{coords} when we fix the coset representative~\eqref{cosetrep}.

Now consider the remaining Grassmann odd superconformal transformation which leaves $\bar \rho$ invariant. It is   generated by $\Xi_{\Alpha}{}^{B}$ (this combines supersymmetry $Q^\alpha_B$ generated by  $\Xi_{\alpha}{}^{B}$ and the special superconformal transformation $\bar S_{\dot \alpha B}$ generated by $\Xi^{\dot \alpha B}$) acting as 
\begin{align}
	X_{\ca}{}^\cB = \left(\begin{array}{c|c}
		x_{\alpha}{}^{\Beta}&\theta_{\alpha}{}^{B}
		\\ \hline
		0 & y_{a}{}^B
	\end{array}\right) \rightarrow  \left(\begin{array}{c|c}
		x_{\alpha}{}^{\Alpha}&\theta_{\alpha}{}^{A}
		\\ \hline
		0 & y_{a}{}^A
	\end{array}\right)\left(\begin{array}{c|c}
		\delta_\Alpha{}^\Beta&\Xi_{\Alpha}{}^{B}
		\\ \hline
		0 & \delta_{A}{}^B
	\end{array}\right)\ .
\end{align}
So $x,y$ are invariant but $\theta_\alpha{}^B \rightarrow \theta_\alpha{}^B + 	x_{\alpha}{}^{\Alpha}\Xi_{\Alpha}{}^{B}$
and so $\rho \rightarrow \hat \rho(\Xi)$ where
\begin{align}\label{rhohat}
	\hat \rho_{\alpha \dot b}
	(\Xi):=\rho_{\alpha \dot b} + 	x_{\alpha}{}^{\Alpha}\Xi_{\Alpha}{}^{B}y^\perp_{B \dot b}\ .
\end{align}

Now we can build nilpotent objects invariant  under $Q,\bar S$  by taking a function of the transformed $\hat \rho_i$ and integrating over the superconformal transformations $\Xi$
\begin{align}\label{intchi}
	\int d^{16}\Xi \, \,F(\hat \rho_i(\Xi))\ .
\end{align}
We have now dealt with all the (local and global) Grassmann odd transformations. 
However we still need to deal with the local $SL(2)$ symmetries indicated by  any remaining $\alpha, \dot a$ indices from the $(\rho_i)_{\alpha \dot a}$. We can contract them with $(x_i)_{\alpha}{}^\Alpha$ and $(y_i^\perp)_{A \dot a}$ locally, so at the same point $i$.
Then we are left with  remaining global transformations, the bosonic conformal transformation  and internal $SL(4)$ symmetries carried by   $\Alpha,A$. These can be absorbed using the objects $x_i^{\Alpha \Beta}:=x_{i\alpha}{}^\Alpha x_{i\beta}{}^\Beta \epsilon^{\alpha \beta}$ and $\epsilon_{\Alpha_1 \Alpha_2 \Alpha_3 \Alpha_4}$ for the conformal group and the analogous objects $y_i^{AB}:=y_{ia}{}^A y_{ib}{}^B \epsilon^{ab}$ and $\epsilon_{A_1 A_2 A_3 A_4}$. Finally we need to make sure they have the right charge under the determinants of the local $GL(2)$ groups (we only considered $SL$ so far not $GL$). This can be achieved using appropriate powers of $x_{ij}^2$ and $y_{ij}^2$.

We will look at the simplest cases of doing this explicitly in the next sections.

\subsection{Proof of hidden symmetry and new prediction}
\label{sec:proofhidden}

We  now have all the ingredients needed to prove the hidden symmetry of section~\ref{sec:hidden}. Firstly we relate the four point loop corrections to  higher point correlators via insertion of the Lagrangian
\begin{align}\label{loopexp}
	\langle \cO_2 \cO_2 \cO_2 \cO_2 \rangle &= \sum_{l=0}^\infty \frac{\lambda^l}{(4\pi^2)^ll!}	\int d^4x_{5}..d^4x_{4+l} \Big\langle \cO_2 \cO_2 \cO_2 \cO_2 \overbrace{L .. L}^{l} \Big\rangle|_{\lambda=0}\notag\\ &=\sum_{l=0}^\infty \frac{\lambda^l}{(4\pi^2)^ll!}	\int d^4x_{5}..d^4x_{4+l}d^4\rho_{5}..d^4\rho_{4+l}	\Big\langle \overbrace{\cO_2 .. \cO_2}^{l+4} \Big\rangle|_{\lambda=0}
\end{align}
So we see that the integrand of $l$-loop four point function is derived from  the $4+l$ point correlator at order $\rho^{4l}$. These are variously known as maximally nilpotent or maximally $U(1)_Y$ violating correlators and they are known at tree-level up to 14 points, since they are equivalent to the 4 point correlator integrands described in section~\ref{sec:hidden} and known to ten loops. Recently the first few  cases have  also been found at large $N_c$ to all orders in the string coupling~\cite{Green:2020eyj}.

According to~\eqref{intchi} this $4+l$ point correlator at $O(\rho^{4l})$ (and with $\bar \rho_i=0$ which will be implicit from now on) can be obtained in terms of a function of $\hat \rho_i(\Xi)$ integrated over the 16 $\Xi$s (for $l>0$). The integration over $\Xi$ will reduce the Grassmann degree by 16 and so to obtain a correlator of $O(\rho^{4l})$ we need to integrate a function of $O(\rho^{4(l+4)})$. But since we only have $4(l+4)$ odd variables $\rho_i$,  the only function available is simply the product of all the variables. We thus arrive at the very simple solution of the Ward identities:
\begin{align}\label{hiddenprrof}
	\Big\langle \overbrace{\cO_2 .. \cO_2}^{l+4} \Big \rangle\Big|_{\rho^{4l},\lambda=0} = \frac{8c}{(-4\pi^2)^{4+l}}\left(\int d^{16} \Xi \,  \hat \rho_1^4 \,..\, \hat \rho_{4+l}^4\right) \times f^{(l)}(x_1,..,x_{4+l})\ . 
\end{align}
Note that since all the indices of the $\rho_{\alpha \dot a}$ indices are contracted, there is no remaining bosonic $SL(2)$ symmetry to deal with. The only remaining symmetries to consider are related to the $y$ and $x$ weight at each point (the scaling under the  $\det$ of the local bosonic $GL(2)$s).
But nicely the $y_i$ weight of $\rho_1^4 \,..\, \rho_{4+l}^4$  is four at each point, which is exactly the correct weight for the correlator $\langle \cO_2...\cO_2 \rangle$. This means any function multiplying the integral must be independent of the $y$s (cross ratios in $y$ would inevitably introduce non polynomial terms).
We are thus left by multiplication of an $x$ dependent function, which is $SL(2)$ invariant (and so must be a function of $x_{ij}^2$).

Now the correlator in~\eqref{hiddenprrof} is crossing symmetric. On the lhs each operator is a function  $\cO_2(x_i,y_i,\rho_i)$ and so this crossing symmetry means the rhs is invariant under the simultaneous permutation $x_i\leftrightarrow x_j$, $y_i\leftrightarrow y_j$, $\rho_i\leftrightarrow \rho_j$. However since the first factor in~\eqref{hiddenprrof} is also invariant under this exchange, it means that the function $f$ is invariant under interchange of $x_i\leftrightarrow x_j$, for any $i,j$.  This is precisely the hidden permutation symmetry of section~\ref{sec:hidden}. 

Let us now explicitly see how this all relates back to previous expressions for the four-point  correlator in section~\ref{sec:hidden}. First plug~\eqref{hiddenprrof} into~\eqref{loopexp}. Recalling~\eqref{rhohat}, we see that  the integrals over $\int d^4\rho_{5}..d^4\rho_{4+l}$ are trivial and we are left with 
\begin{align}
	\int d^{16}\Xi \hat \rho_1^4  \hat\rho_2^4  \hat\rho_3^4  \hat\rho_4^4 = \int d^{16}\Xi \prod_{i=1}^4 (x_{i\alpha}{}^{\Alpha}\Xi_{\Alpha}{}^{B}y^\perp_{iB \dot b} )^4 = \det x_{i \alpha}^\Alpha y^\perp_{iB \dot b}
	=I(x_i,y_j)
	\ .
\end{align} 
The penultimate equality gives  the Grassmann integral as the determinant of $x_{i \alpha}^A y^\perp_{iB \dot b}$ viewed as a $16\times 16 $ matrix where  one 16 dimensional vector space is labelled by the multi-index $ i \alpha \dot b$ and the other by the multi-index ${}^\Alpha{}_B$.
Then the last equality is the statement that evaluating this determinant yields the expression~\eqref{intriligator}. In practice the best way to check the last equality is to fix the coordinates using conformal symmetry similarly to footnote~\ref{an123} in analytic space. Specifically we can use conformal and internal symmetry to fix  
$
(x_1)_{\alpha}{}^\Beta \rightarrow (\delta_\alpha{}^\beta,\delta_{\alpha \dot \beta}),\   (x_2)_{\alpha}{}^\Beta \rightarrow  (0,\delta_{\alpha \dot \beta})	,\  (x_3)_{\alpha}{}^\Beta \rightarrow  (\delta_\alpha{}^\beta,0) \, ,	
$ and then fix $(x_4)_{\alpha}{}^\Beta \rightarrow (\delta_\alpha{}^\beta,(x_4)_{\alpha \dot \beta} )$ with $(x_4)_{\alpha \dot \beta} =$diag$(x,\bar x)$. Then take the same for the $y$s, which gives 
$
(y^\perp_1)_{A\dot \beta}\rightarrow (-\delta_{\alpha \dot \beta},\delta^{\dot \alpha}{}_{ \dot \beta}),\   (y^\perp_2)_{A\dot \beta} \rightarrow  (\delta_{\alpha \dot \beta},0)	,\  (y^\perp_3)_{A\dot \beta} \rightarrow  (0,\delta^{\dot \alpha}{}_{ \dot \beta}), \ (y^\perp_4)_{A\dot \beta} \rightarrow  (-(y_4)_{\alpha \dot \beta},\delta^{\dot \alpha}{}_{ \dot \beta}) $ 	
with  $(y_4)_{a \dot b} =$diag$(y,\bar y)$. Then plugging these values into the  $16\times 16 $ matrix  $x_{i \alpha}^A y^\perp_{iB \dot b}$ one can compute the determinant to be $(x{-}y)(x{-}\bar y)(\bar x{-}y)(\bar x{-}\bar y)$. This equals the expressions for $I(x_i,y_j)$ in~\eqref{intriligator2} and in turn~\eqref{intriligator} when the above choices for $x_i,y_j$ are made.

So the conclusion of this is that the four point function of four $\cO_2$ operators has the perturbative form
\begin{align}
	\langle \cO_2 \cO_2 \cO_2 \cO_2 \rangle &= \langle \cO_2 \cO_2 \cO_2 \cO_2 \rangle|_{\lambda=0}+\frac{8c}{(4\pi^2)^{4}}\sum_{l=1}^\infty \left(\!\!\frac{\lambda}{4\pi^2}\!\!\right)^l \int \frac{d^4x_{5}}{(-4\pi^2)}..\frac{d^4x_{4+l}}{(-4\pi^2)} I(x_i,y_j) f^{(l)}(x_i)\ 
\end{align}
with the hidden permutation symmetry $f^{(l)}(..,x_i,..,x_j,..)=f^{(l)}(..,x_j,..,x_i,..)$. 
Inputting  the choice of $y_i$s in~\eqref{ychoice} to yield the component correlator $\langle \cO \bar \cO \cO \bar \cO \rangle$, this reduces to the form stated in equations~(\ref{corbasic})-(\ref{Finf}).

Note that following the argument of this subsection through, but starting with higher charge correlators $	\langle \cO_{p_1}\cO_{p_2}\cO_{p_3}\cO_{p_4} \rangle$ we find that their $l$-loop corrections are given in terms of mixed charge $(4{+}l)$-point half BPS  correlators  $	\langle \cO_{p_1}\cO_{p_2}\cO_{p_3}\cO_{p_4} \cO_2..\cO_2\rangle$ at $O(\rho^{4l})$.
\begin{align}\label{loopexp2}
	\langle \cO_{p_1}\cO_{p_2}\cO_{p_3}\cO_{p_4}\rangle=\sum_{l=0}^\infty \frac{\lambda^l}{(4\pi^2)^ll!}	\int d^4x_{5}..d^4x_{4+l}d^4\rho_{5}..d^4\rho_{4+l}\Big\langle \cO_{p_1}\cO_{p_2}\cO_{p_3}\cO_{p_4} \overbrace{\cO_2 .. \cO_2}^{l} \Big\rangle\Big|_{\lambda=0}\ .
\end{align}
These will then  in turn be given by a formula like~\eqref{hiddenprrof}
\begin{align}\label{hiddenprrof2}
	\Big\langle \cO_{p_1}\cO_{p_2}\cO_{p_3}\cO_{p_4} \overbrace{\cO_2 .. \cO_2}^{l} \Big\rangle\Big|_{\rho^{4l},\lambda=0} =  \frac{C_{p_1p_2p_3p_4}}{(-4\pi^2)^{l}} \left(\int d^{16} \Xi \,  \hat \rho_1^4 \,..\, \hat \rho_{4+l}^4\right) \times f^{(l)}_{p_k}(x_i,y_j)\  
\end{align}
where $f_{p_k}^{(l)}$ are loop integrands in the expansion of $F_{p_k}$ in~\eqref{corhighercharge} (the precise relation is just as in~\eqref{Fl},\eqref{Finf}). This then explains the structure for the four-point correlators quoted without proof in~\eqref{corhighercharge} as well as being consistent with the solution of the four-point Ward identities~\eqref{decomp} whilst implying that  there are  no loop corrections to  $a,b(x,y)$. This thus proves the partial non-renormalisation result of the four point correlator discussed below~\eqref{ab} and first proven in~\cite{hep-th/0009106}. 

But the 10d symmetry observed for four-point higher charge integrands~\cite{Caron-Huot:2021usw} and reviewed in section~\ref{pert10d} suggests that we can go further. We propose  the following simple generalisation giving all $n$-point, arbitrary charge, maximally nilpotent, half BPS, planar, Born-level correlators in a  compact formula
\begin{align}\label{hiddenprrof3}
	\Big\langle  \overbrace{\cbO .. \cbO}^{4+l} \Big\rangle\Big|_{\rho^{4l},\lambda=0} =  \frac{(4c)^{-l/2}}{2c}  \left(\int d^{16} \Xi \,  \hat \rho_1^4 \,..\, \hat \rho_{4+l}^4\right) \times f^{(l)}({\bf x}_{ij}^2)\ .
\end{align}
In the case that $y_5=..=y_{4+l}=0$ this formula reproduces  the loop corrections to arbitrary charge four point correlators in the 10d symmetric form~\eqref{master} (using \eqref{hiddenprrof2} and \eqref{loopexp2}). 
But it also gives a proposal for  correlators with higher charge versions of the Lagrangian operator that have not previously been considered to our knowledge. This could presumably be proven  using OPE considerations in a similar manner to the proof we hinted at above~\eqref{bcO} of the four-point perturbative 10d symmetry.

\subsection{Bootstrapping higher point correlators}

We saw the simplest class of nilpotent invariant in the previous section,  obtained by integrating all $4n$ $\hat\rho$ variables of an  $n$-point correlator  over $\Xi$ and producing an invariant with the maximal  number $4(n{-}4)$  $\rho$ variables. We called this the maximally nilpotent invariant.

We now consider the next simplest class, the next to maximally nilpotent invariant,  following~\cite{Chicherin:2015bza}. The internal $SU(4)$ group contains a centre given by diag$(i,i,i,i)$ which acts as $\rho \rightarrow i \rho,\bar \rho \rightarrow -i \bar \rho$ and leaving $x,y$ invariant. Therefore  to obtain an $O(\rho^K)$ invariant we must have  $K=4k$. So the next simplest example is $O(\rho^{4(n-5)})$, obtained by integrating $4(n-1)$ $\hat \rho$ variables. It is simpler to view this by removing 4 $\hat \rho$ variables from the maximal case. We thus arrive at the following set of $Q,\bar S$ next-to-maximally-nilpotent invariants:
\begin{align}\label{Rr}
	\int d^{16}\Xi \,  
	{\partial\over\partial\hat\rho_{i\a \dot a}}
	{\partial\over\partial\hat\rho_{j\b \dot b}}
	{\partial\over\partial\hat\rho_{k\gamma \dot c}}
	{\partial\over\partial\hat\rho_{l\delta \dot d}}\left(
	\hat\rho_1^4 \dots \hat\rho_n^4\right).
\end{align}
These transform locally under the various $SL(2)$s as indicated by the $\alpha, \dot a$ indices. As discussed at the end of section~\ref{sec:WI2} we can remove these by contracting with $x_{i\alpha}{}^A$ and $ (y^\perp_{i}){}_{A\dot a}$ at the appropriate points leaving left over global conformal $\Alpha,\Beta$ and internal $A,B$ indices which themselves need to be contracted with the only available objects  $x_i^{\Alpha \Beta}$,  $\epsilon_{\Alpha_1 \Alpha_2 \Alpha_3 \Alpha_4}$, $y_i^{AB}$ and $\epsilon_{A_1 A_2 A_3 A_4}$. 
However the $y$ weight is  exactly right for the correlator of $\cO_2$ at each point and therefore we can not use  $y_i^{AB}$, since doing so would require using a non-polynomial $y_i$ to reduce the charge to the correct value.  So we have to contract the $A,B$ indices with  $\epsilon_{A_1 A_2 A_3 A_4}$. We are thus left with the basis 
%\begin{align}\label{Rr}
%	\cI_{ijkl}^{\Alpha_1 \Alpha_2 \Alpha_3 \Alpha_4} = &  x_{i\alpha}^{\Alpha_1}x_{j\beta}^{\Alpha_2}x_{k\gamma}^{\Alpha_3}x_{l\delta}^{\Alpha_4}\epsilon^{ABCD} (y^\perp_{i}){}_{A\dot a} (y^\perp_{j}){}_{B\dot b} (y^\perp_{k}){}_{C\dot c} (y^\perp_{l}){}_{D\dot d} \int d^{16}\Xi \,  
%	{\partial\over\partial\hat\rho_{i\a \dot a}}
%	{\partial\over\partial\hat\rho_{j\b \dot b}}
%	{\partial\over\partial\hat\rho_{k\gamma \dot c}}
%	{\partial\over\partial\hat\rho_{l\delta \dot d}}\left(
%	\hat\rho_1^4 \dots \hat\rho_n^4\right)\ 
%\end{align}
\begin{align}\label{Rr2}
	\cI_{i_1i_2i_3i_4}^{\Alpha_1 \Alpha_2 \Alpha_3 \Alpha_4} = &  \epsilon^{A_1A_2A_3A_4} \int d^{16}\Xi \,  
	{(y^\perp\partial_{\hat\rho}x)_{i_1A_1}^{\Alpha_1}}{(y^\perp\partial_{\hat\rho}x)_{i_2A_2}^{\Alpha_2}}{(y^\perp\partial_{\hat\rho}x)_{i_3A_3}^{\Alpha_3}}{(y^\perp\partial_{\hat\rho}x)_{i_4A_4}^{\Alpha_4}}\left(
	\hat\rho_1^4 \dots \hat\rho_n^4\right)\ 
\end{align}
where
\begin{align}
	(y^\perp\partial_{\hat\rho}x)_{iA}^{\Alpha}:=(y_i^\perp)_{A \dot a} x_{i\alpha}^\Alpha \frac{\partial}{\partial\hat\rho}_{i \alpha \dot a} \ .
\end{align}
Thus the correlator can be written in terms of functions $f^{ijkl}_{\Alpha_1 \Alpha_2 \Alpha_3 \Alpha_4}$ of $x$ only
\begin{align}
	\Big\langle \overbrace{\cO_2 .. \cO_2}^{n} \Big \rangle|_{\rho^{4(n-5)}}=\sum \cI_{i_1i_2i_3i_4}^{\Alpha_1 \Alpha_2 \Alpha_3 \Alpha_4} \times f^{i_1i_2i_3i_4}_{\Alpha_1 \Alpha_2 \Alpha_3 \Alpha_4}(x_{i}^{\Alpha\Beta}) \ .
\end{align}

The invariants satisfy various symmetries and constraints. Firstly, from the definition they are symmetric under simultaneous interchange of position and $SL(4)$ index so
\begin{align}
	\cI_{i_1i_2i_3i_4}^{\Alpha_1 \Alpha_2 \Alpha_3 \Alpha_4} = \cI_{i_{\sigma_1}i_{\sigma_2}i_{\sigma_3}i_{\sigma_4}}^{\Alpha_{\sigma_1} \Alpha_{\sigma_2} \Alpha_{\sigma_3} \Alpha_{\sigma_4}} \qquad \sigma \in S_4\ .
\end{align}
This then implies further symmetries when there are repeated positions, eg if $i_1=i_2$ then it is symmetric in $A_1 ,A_2$ etc. If three positions coincide  then the invariant vanishes 
$\cI_{iiij}^{\Alpha_1 \Alpha_2 \Alpha_3 \Alpha_4}=0$ since it contains the factor $(y_i^\perp)_{A \dot a}(y_i^\perp)_{B \dot b}(y_i^\perp)_{C \dot c} \epsilon^{ABCD}  =0$.
The crossing symmetry of the correlator means it is invariant under the simultaneous interchange $(x_i,y_i, \rho_i)\rightarrow(x_j,y_j, \rho_j)$  and so the coefficient functions satisfy the crossing symmetry 
\begin{align}
	f_{\Alpha\Beta\Gamma\Delta}^{ijkl}(x_{\sigma_1},..,x_{\sigma_n})=	f_{\Alpha\Beta\Gamma\Delta}^{\sigma_i\sigma_j\sigma_k\sigma_l}(x_{1},..,x_n) \qquad \sigma \in S_n\ .
\end{align}

We can thus write the general next-to-maximally nilpotent correlator  in terms of 8 scalar functions $f^{(a)}(x)$ as
\begin{align}\label{nmn}
	&\Big\langle \overbrace{\cO_2 .. \cO_2}^{n} \Big \rangle|_{\rho^{4(n-5)}}\notag \\&=
	\cI_{1122}^{\Alpha\Beta\Gamma\Delta} \,x_{3\Alpha \Gamma}x_{4\Beta \Delta}f^{(1)}_{34}(x)
	{+} \cI_{1123}^{\Alpha\Beta\Gamma\Delta}\left(x_{3\Alpha \Gamma}x_{2\Beta \Delta}f^{(2)}_{32}(x){+}\,x_{3\Alpha \Gamma}x_{4\Beta \Delta}f^{(3)}_{34}(x){+}\,x_{4\Alpha \Gamma}x_{5\Beta \Delta}f^{(4)}_{45}(x)\right)
	\notag \\&{+}\cI_{1234}^{\Alpha\Beta\Gamma\Delta}\left(x_{2\Alpha \Gamma}x_{1\Beta \Delta}f^{(5)}_{21}(x){+}\,x_{2\Alpha \Gamma}x_{5\Beta \Delta}f^{(6)}_{25}(x){+}\,x_{5\Alpha \Gamma}x_{6\Beta \Delta}f^{(7)}_{56}(x){+}\,\epsilon_{\Alpha \Beta \Gamma \Delta}f^{(8)}(x)\right)  {+} \text{$S_n$ perm.} 
\end{align}
The functions $f^{(a)}_{ij}$ are scalar functions of $x$  of conformal weight 5 at points $i,j$ and 4 at all other points. 

Finally, the invariants are not all independent but rather satisfy the constraint
\begin{align}\label{sczero}
	\sum_{i=1}^n \cI_{ijkl}^{\Alpha\Beta\Gamma\Delta}  = 0
	\qquad (\text{for all } j,k,l,\Alpha,\Beta,\Gamma,\Delta)\ .
\end{align}
This can   be understood  by noticing from~\eqref{rhohat} that 
\begin{align}
	{\partial \over \partial \Xi_\Alpha{}^B} = \sum_{i=1}^n x_{i\alpha}^{\Alpha } (y^\perp)_{iB\dot b}{\partial
		\over \partial \hat \rho_{i\alpha \dot b} }\ 
\end{align}
and so the sum in~\eqref{sczero} using~\eqref{Rr2} gives an expression of the form $\int d^{16}\Xi \,\partial_\Xi f(\Xi)$ which vanishes. 

The above expression~\eqref{nmn} is a non-perturbative statement valid for all $\lambda,c$. At zeroth order in $\lambda$ however, we know that all the coefficient functions are rational and can only have simple poles in $x_{ij}^2$, putting further constraints on the result. 
Remarkably, for $n=6$ this system is constrained enough to fix the result up to just 4 unfixed constants~\cite{Chicherin:2015bza}. These constants can then be fixed up to a single overall coefficient by considering OPE limits. We refer the reader to~\cite{Chicherin:2015bza} for the explicit result.
Apart from this case there is very little known for half BPS correlators beyond the 4-point/ maximally nilpotent sector. One loop, ie $O(\lambda)$, $n$-point correlators at $O(\rho^0)$ were studied in~\cite{0812.3341} and recently  the 5-point $O(\rho^0)$ correlator was also obtained at strong coupling~\cite{Goncalves:2019znr}.
We note here that the form~\eqref{nmn} is also  valid for {\em the interacting part of the} $n=5$ non-nilpotent $O(\rho^0)$ correlator to all orders in $\lambda$
and it would be interesting to explore the non perturbative structure at 5 points in more detail.%
\footnote{Thanks to Congkao Wen for discussions on this point.}
It is also tempting to propose lifting~\eqref{nmn} to a formula for all half BPS correlators of all charges, by simply promoting the correlator on the lhs to a master correlator, and the variables of the functions $f_{ij}^{(k)}$ to 10d variables just as we did in~\eqref{hiddenprrof3}.

\subsection{Supercorrelator/ superamplitude duality}
\label{sec:superampcor}

The correlator / amplitude duality, already discussed in the context of four-point amplitudes in section~\ref{sec:ampcor}, extends to stress-tensor  correlators at any number of points~\cite{Eden:2011yp,Eden:2011ku}. Indeed this extends in principle to correlators of higher charge half BPS operators as well as non BPS operators~\cite{Adamo:2011dq,Chicherin:2016qsf}.    Focussing on the stress-tensor correlators, the duality states that in the planar, $c\rightarrow \infty$  limit 
\begin{align}
	\lim_{\substack{(X_{i})_{2}^\cA\rightarrow Z_{i}^\cA\\
		(X_{{i+1}})_{1}^\cA \rightarrow Z_{i}^\cA}}	\frac{\langle \cO_2 ..\cO_2  \rangle_{\phantom{\lambda^0,\theta^0}\!\!\!\!\!\!\!\!\!\!\!\!}}{\langle \cO_2 ..\cO_2  \rangle|_{\lambda^0,\theta^0}\!\!\!\!\!\!\!\!\!\!\!\!}\quad =\quad \cA^2\ .
\end{align}
Both sides are functions of the 't Hooft coupling $\lambda$. On the lhs the correlators are functions of $n$ copies of analytic superspace (see section~\ref{sec:an}) but we turn off the conjugate Grassmann odd variables, so $\bar \theta=0$ (in fact $\bar \theta$ can be turned back on via a simple shift of variables~\cite{Chicherin:2016fac,Chicherin:2016fbj}). On the rhs, the superamplitude (divided by tree level MHV amplitude) $\cA$ is taken in momentum super twistor space~\cite{Hodges:2009hk,Mason:2009qx} whose $(4|4)$ component variables are $Z_i^\cA = (z_i^\Alpha,\chi_i^A)$. These relate to the analytic superspace variables in the super Grassmannian formulation~\eqref{W} with coordinates $X_{i\alpha}^\cA$ (the  internal $y$ coordinates arising from the other half of the Grassmannian, $X_{ia}^\cA$,  drop out in the ratio on the lhs) as $(X_{i})_{2}^\cA=(X_{{i+1}})_{\,1}^\cA= Z_{i}^\cA$. Geometrically the $X_{i\alpha}^\cA $ define a line in supertwistor space, and the limit is one in which the lines intersect consecutively with intersection points defining the $Z_{i}^\cA$. 

Expanding both sides in $\lambda$ (the loop expansion) as well as in the Grassmann odd variables, gives versions of the duality  refined by Grassmann degree and/ or loop order:
\begin{align}\label{refdu}
	\lim_{X_{i2}^\cA\rightarrow X_{{i+1}\,1}^\cA }	\frac{\langle \cO_2 ..\cO_2  \rangle|_{\lambda^l\theta^{4k}\!\!\!\!\!\!\!\!\!\!\!\!}}{\langle \cO_2 ..\cO_2  \rangle|_{\lambda^0,\theta^0}\!\!\!\!\!\!\!\!\!\!\!\!}\quad\quad  =\ (\cA^2)^{(l)}_{k}\ ,
\end{align}
where we define the square of the superamplitude at $l$ loops and Grassmann degree $4k$, $(\cA^2)^{(l)}_{k}$, by expanding the superamplitude to that  order, explicitly 
\begin{align}
	 (\cA^2)^{(l)}_{k}\ := \  \sum_{l'=0}^l\sum_{k'=0}^k \cA^{(l')}_{k'} \cA^{(l-l')}_{k-k'}
\end{align}
where $\cA^{(l)}_{k}$ is the $l$ loop, N${}^k$MHV superamplitude. 

Now this duality holds at the level of the integrand. However as alluded to previously, loop level stress-tensor multiplet correlator integrands are themselves tree-level higher point correlators
\begin{align}\label{intril}
	\langle\, \overbrace{\cO_2 ..\cO_2}^n  \,\rangle|_{\lambda^l\theta^{4k}} = \int d^4x_{n+1}d^4 \rho_{n+1}..d^4x_{n+l}d^4 \rho_{n+l} \,\langle \,\overbrace{\cO_2 ..\cO_2}^{n+l} \, \rangle|_{\lambda^{0}\theta^{4(k+l)}}\ .
\end{align} 
Putting this together with the duality~\eqref{refdu} we see that a single {\em tree level} correlator gives  many different squared amplitude {\em loop level} integrands within it by taking different polygonal lightlike limits
\begin{align}\label{genll}
	\begin{tikzpicture}
\node at (0,0) {$\langle\,\overbrace{\cO_2 ..\cO_2}^{n} \, \rangle|_{\lambda^{0}\theta^{4k}}$};
\draw[-latex] (1,0) --node[midway,sloped,above]{\qquad \qquad \qquad $n$-point lightlike limit} (9,2);
\node at (10,2){$ (\cA_n^2)^{(0)}_{k}$};		
\draw[-latex] (1,0) --node[midway,sloped,above]{\qquad \qquad \qquad  $(n{-}1)$-point lightlike limit} (9,.3);
\node at (10,.3){$ (\cA_{n-1}^2)^{(1)}_{k-1}$};		
\draw[-latex] (1,0) --node[midway,sloped,above]{\qquad \qquad \qquad  $(n{-}k)$-point lightlike limit} (9,-2);
\node at (10,-2){$ (\cA_{n-k}^2)^{(k)}_{0}$};	
\node at (10,-1){$\vdots$};	
\node at (7,-.3){$\vdots$};	
\end{tikzpicture}
\end{align}

  A special case of this occurs when we consider the maximally nilpotent case $k=n-4$ which is related via~\eqref{intril} to the 4-point $(n{-}4)$-loop correlator.
This thus explains  the duality stated in~\eqref{npll} between $n$ point amplitudes and the {\em four}-point $(n{+}l{-}4)$-loop amplitude arises.  As mentioned there, this duality between the simplest and by far best understood correlator, the four-point correlator, may by itself  contain enough information to fix {\em all} amplitudes~\cite{Heslop:2018zut}. 
But from~\eqref{genll} we see it is just one example of the duality and higher point versions will give amplitudes more directly.

\subsection{The twistor approach}
\label{sec:twistor}

Twistor space provides a practical method for computing stress-tensor multiplet correlators explicitly (at least when $\bar \theta$ is turned off). This was first considered in~\cite{Adamo:2011dq} for arbitrary operators and then developed further to give  concrete practical Feynman rules for stress-tensor correlators in~\cite{Chicherin:2014uca}. The case of correlators of more general operators was then taken up again and made more precise in~\cite{Chicherin:2016qsf}.
In this formulation an $n$-point stress-tensor multiplet of Grassmann odd degrees $\theta^{4k}$ is obtained by summing over all $n$-point graphs with $n+k$ edges. The vertices can have arbitrary degree $>1$. Colour indices are associated to the graph in the standard way for adjoint representations (eg gluons). Each edge between vertex $i,j$ corresponds to the superpropagator $g_{ij}=y_{ij}^2/x_{ij}^2$. Each vertex $i$ with incoming edges from $j_1,..,j_p$ corresponds to  an $R$ vertex $R(i;j_1,..,j_p)$. 
\begin{align}
	\begin{tikzpicture}[roundnode/.style={circle, fill=black, inner sep=0pt, minimum size=1.5mm}]
		\node[roundnode,label=$i$] at (-7,0) {};
		\node[roundnode,label=$j$] at (-6,0) {};
		\draw[thick] (-6,0) -- (-7,0);
		\foreach \i [count=\ni] in {120, 60, ..., -120}{
			\node[label=\i:{$j_\ni$}] at  (\i:1.0cm) (u\ni) {};
			\draw (0,0) -- (\i:1.2cm);
		}
		\node[label=-180:{$j_p$}] at  (-180:1.0cm) (up) {};
		\draw (0,0) -- (-180:1.2cm);
		\node  at  (-4.3,0)   {$=\frac{y_{ij}^2}{x_{ij}^2}\tr(T^{a_i}T^{a_j})$};
		\node[rotate=120]  at  (-1,-.6)   {$\dots$};
		\node[roundnode] at (0,0) {}; 	
		\node  at  (4.5,0)   {$=R(i;j_1j_2..j_p)\tr(T^{a_{j_1}}..T^{a_{j_p}})$};
		\node[roundnode] at (0,0) {}; 	
	\end{tikzpicture}
\end{align}
where
\begin{align}
	R(i;jk)&=1\notag\\
	R(i;j_1j_2j_3)&= -{\delta^2\Big(\langle\sigma_{ij_1}\sigma_{ij_2}\rangle
		A_{ij_3}+\langle\sigma_{ij_2}\sigma_{ij_3}\rangle
		A_{ij_1}+\langle\sigma_{ij_3}\sigma_{ij_1}\rangle A_{ij_2}\Big) \over  \langle{\sigma_{i j_1} \sigma_{i j_2}}\rangle  \,\langle{\sigma_{i j_2} \sigma_{i j_3}}\rangle\,\langle{\sigma_{i j_3} \sigma_{i j_1}}\rangle }\notag\\
	R(i;j_1..j_p)&= R(i;j_1j_2j_3)R(i;j_1j_3j_4)..R(i;j_1j_{p-1}j_p)
\end{align}
\begin{align}\label{A-y}
	A_{ij}^{a}  = \left[{\sigma_{ji}^{\alpha}\rho_{j\alpha b'}} +{\sigma_{ij}^{\alpha}\rho_{i\alpha b'}} +\theta_*^A y^\perp_{jAb'}\right] (y_{ij}^{-1})^{b'a}\qquad 
	\sigma_{ij}^\alpha = \epsilon^{\alpha\beta}{\vev{x_{i\beta} z_*x_{j1} x_{j2}} \over \vev{x_{i1} x_{i2} x_{j1} x_{j2}}} \ .
\end{align}
Here $Z_*^\cA =(z_*^\Alpha,\theta_*^A)$ is a reference supertwistor. Although each individual diagram depends on $Z_*$, this drops out after summing over all diagrams.  

This twistor Feynman diagram approach works at the planar (for which one restricts to planar graphs) or non-planar level. In fact this approach has been used to fully fix the four-loop non-planar four-point function~\cite{Fleury:2019ydf} (as discussed in section~\ref{sec:np}).

In the planar case this approach gives a very direct, diagrammatic  way of seeing the afore-mentioned amplitude/correlator duality. In the consecutive light-like limit $x_{i\,i+1}^2 \rightarrow 0$ and so this will project onto  Feynman diagrams which contain edges $(i,i{+}1)=y_{i\,i+1}^2/x_{i\,i+1}^2$ thus containing an $n$ cycle. The remaining part of the graph is planar and thus has a well-defined part `inside' and `outside' the $n$-cycle. The graphs which can appear `inside' the $n$-cycle are precisely the graphs which appear in a planar $n$-gon lightlike Wilson loop in twistor space~\cite{Mason:2010yk,Adamo:2011dq}.
Similarly for the `outside'. Thus summing over all valid planar correlator graphs containing an $n$-cycle is the same as summing over all planar graphs contributing to the product of Wilson loops. Furthermore i can be shown that the expressions for the graphs reduce to the expressions for the corresponding Wilson loop diagrams in the limit~\cite{Adamo:2011dq,Chicherin:2014uca}.

\subsection{Correlahedron}

The above twistor space approach to correlation functions together with its close relation to amplitudes suggest a geometric description of correlators, mimicking the geometric description of amplitudes discovered in~\cite{Arkani-Hamed:2013jha,Arkani-Hamed:2013kca} known as the amplituhedron (see SAGEX review chapter 7~\cite{chapter7}). 
From~\eqref{intril} we only need to consider tree-level correlators since the loop level ones are obtained from these.

This suggests the following simple (to write down anyway) proposal for a $(n,k)$ {\em correlahedron}~\cite{Eden:2017fow}
\begin{align}\label{eq:41}
	\Big\{ Y \in \Gr(n{+}k,n{+}k{+}4), X_i \in \Gr(2,n{+}k{+}4): \
	\langle Y X_i X_j \rangle > 0  \Big \}\ .
\end{align}
Here the external data $X_i,
i=1,..,n$ are themselves 2-planes, $X_i \in \Gr(2,n{+}k{+}4)$, and are
equivalent to points in chiral superspace, bosonised in the standard amplituhedron way. The  geometrical reduction of this geometry corresponding to taking consecutive lightlike limits gives a geometry known as the {\em squared amplituhedron} which contains the amplituhedron geometry with other almost disconnected pieces.  The amplituhedron geometry in question can be tree or loop depending on whether all the $X$ variables take place in the lightlike limit or if there are some left free to become the loop variables. This resulting {\em squared amplituhedron} geometry is that of the amplituhedron but without an additional topological winding condition~\cite{Arkani-Hamed:2017vfh}.
It has been thoroughly checked and indeed carefully proved in many cases~\cite{Dian:2021idl} that for the maximally nilpotent, $n{=}k{+}4$, case the squared amplituhedron indeed gives the square of the amplitude via the oriented canonical form. The different almost-disconnected pieces, arising from different winding sectors and dubbed `amplituhedron-like' in~\cite{Dian:2021idl}, correspond to the different amplitude products appearing in the sum of~\eqref{npll}.
Direct checks of the correlahedron proposal itself (rather than its limits) are much more difficult, even in this simplest $n{=}k{+}4$ case. Indeed it is clear that one will need some generalisation of  the canonical form to obtain the correlator from the geometry.
For $k{<}n{-}4$ things are less clear even at the squared amplituhedron level. The proposal~\eqref{eq:41}  needs more detail in this case, namely constraints on the positions of the $X$s which would for example yield convexity in the lightlike limit.

\section{Conclusions}

We have attempted a review of all that is known about half BPS correlators in $\cN=4$ SYM. This has been  a huge area of research over the last two decades and new features are being discovered even now. Inevitably many areas have lacked detail and some key areas have been neglected entirely. To conclude therefore we will mention briefly some of these neglected areas as well as a few closely related areas.

\subsubsection*{Conformal bootstrap}

There is a large current research program aiming at bootstrapping CFTs using crossing symmetry at four points with unitarity and the OPE. This programme has also been applied to half BPS correlators in $\cN=4$ SYM giving fully non-perturbative  constraints in  the space of allowable correlation functions in $\cN=4$ SYM~\cite{Beem:2013qxa,Beem:2016wfs,Bissi:2020jve,Chester:2021aun}.

\subsubsection*{Integrability}

A key development in planar $\cN=4$ SYM has been the presence of integrability in various quantities, giving ways of computing them non-perturbatively at finite coupling. This initially focussed on two point correlators of non BPS operators, however 
in the last few years integrability has been applied to higher point functions beginning with three-points and the introduction of the hexagon approach~\cite{Basso:2015zoa}.
Then~\cite{Eden:2016xvg,Fleury:2016ykk,Eden:2018vug} showed that four and higher point functions can be obtained by appropriately gluing together the hexagons. This then allows an integrability approach to  the study of the half BPS correlators studied here.
This has been pursued in a number of works~\cite{Basso:2017khq,Eden:2017ozn,Fleury:2017eph,Coronado:2018ypq,Bargheer:2019kxb,deLeeuw:2019tdd,Belitsky:2019fan,Bargheer:2019exp,DeLeeuw:2019dak,Belitsky:2020qrm,Belitsky:2020qir,Aprile:2020luw}. See SAGEX review chapter 9~\cite{chapter9} for a review of the related integrability approach to amplitudes.

\subsubsection*{Finite $N_c$ and other gauge groups}

A sector of the half BPS correlator family that has been neglected  in this review is that of going deep into the finite $N_c$ sector, and beyond single particle  operators. Indeed we started by saying that the first non-trivial correlator of half BPS operators was at four points, since lower point correlators are independent of the coupling. However they do   depend in a highly non-trivial and interesting way on the number of colours in the gauge group, $N_c$. The two-point functions of half BPS operators were diagonalised using a Schur polynomial basis of half BPS operators~\cite{Corley:2001zk} and their two-point and extremal  higher-point correlators computed. 
More recently, the space of non-extremal higher point correlators in the free theory at arbitrary $N_c$ have also been explored~\cite{Aprile:2020uxk}. Here the single particle basis - corresponding to single-particle supergravity states via AdS/CFT -- was employed.

The exploration of half BPS correlators in $\cN=4$ SYM with other (non $SU(N_c)$) gauge groups has  also been recently initiated both in the free theory~\cite{Lewis-Brown:2018dje} and at strong coupling~\cite{Alday:2021vfb,Dorigoni:2022zcr}.

\subsubsection*{Integrated correlators}

Certain infrared finite observables of the type used to describe events in colliders can be computed in conformal field theories~\cite{Hofman:2008ar}. In $\cN=4$ SYM they  can be computed from the stress tensor four-point functions of section~\ref{sec:simplest} and this has been investigated in  a series of papers~\cite{Belitsky:2013xxa,Belitsky:2013bja,Belitsky:2013ofa,Korchemsky:2021okt}.

As well as obtaining the above physical objects by integrating known four-point correlators, recently it has been shown how to obtain certain other types of integrated correlators directly to all orders in the coupling from localisation in $\cN=4$ SYM~\cite{Chester:2019jas,Chester:2020vyz,Dorigoni:2021rdo,Dorigoni:2021guq,Dorigoni:2021bvj,Dorigoni:2022zcr} (see also SAGEX review chapter 10~\cite{chapter10}).

\subsubsection*{Generalised correlators/ form factors involving half BPS operators}

Various generalisations of half BPS correlators have been considered in great detail, eg correlator of a null polygonal
Wilson loop and half BPS operators~\cite{1107.5702,1110.0758,1207.4316,1209.0227,1301.0149,2112.06956,2202.05596}
and form factors involving  half BPS operators~\cite{1402.1300,1406.1443} (see also SAGEX review chapter 1~\cite{chapter1}.

\subsubsection*{Other theories}

We have focussed exclusively on $\cN=4$ SYM which is by far the most studied and best understood  theory with half BPS correlators, but there are many other interesting theories one could similarly consider. The two most obvious cases to consider are the 6d theory with (2,0) supersymmetry and its cousin the 3d theory with $\cN=8$ supersymmetry. These are the theories on the worldsheet  of the M5 and M2 branes respectively in M theory. They have no Lagrangian description and indeed are inherently non-perturbative, but in many respects look quite similar to $\cN=4$ at least in the AdS/CFT context. However it also seems that many of the additional symmetries that have helped in $\cN=4$ do not seem to have an obvious analogue for these theories.
Furthermore they are technically much more complicated eg the solution of the superconformal Ward identities is much more involved. Nevertheless much impressive work has been done computing half BPS correlators  in the 6d theory~\cite{Park:1998nra,Corrado:1999pi,Howe:2000nq,Eden:2001wg,Arutyunov:2002ff,Heslop:2004du,Beem:2015aoa,Rastelli:2017ymc,Heslop:2017sco,Chester:2018dga,Abl:2019jhh,Alday:2020lbp,Alday:2020tgi,Alday:2020dtb,Lemos:2021azv,Lambert:2021fsl} 
as well as the 3d theory~\cite{Zhou:2017zaw,Chester:2018aca,Binder:2018yvd,Alday:2021ymb}
and the one loop quantum gravity corrections to the four-point supergravity amplitude on AdS${}_7\times S^4$ and AdS${}_4\times S^7$ (or their ${\mathbb{Z}_2}$ orbifolds) have even been computed~\cite{Alday:2020tgi,Alday:2021ymb} overcoming a number of these technical difficulties.

A number of theories with non maximal   supersymmetry whose half BPS correlators exhibit higher dimensional conformal symmetry (which does not seem to be present for the 3d $\cN=8$  or the 6d (2,0) theories) have been investigated recently in 1d (dual  to theories on AdS${}_2\times S^2$)~\cite{Abl:2021mxo} and 2d (dual  to theories on  AdS${}_3\times S^3$)~\cite{Rastelli:2019gtj,Giusto:2020neo,Wen:2021lio,Aprile:2021mvq} as well as theories dual  to   AdS${}_5\times S^3$~\cite{Alday:2021odx,Drummond:2022dxd}.
Also of recent interest has been the theories which live inside defects in QFT, and in this context many results for  half BPS operators in the 1d theory living on  a Wilson line within $\cN=4$ SYM have been found~\cite{1706.00756,1802.05201,1806.01862,1811.02369,1812.04593,2001.11039,2103.10440,2107.08510, 2108.13432,2110.13126,2112.10780,2202.07627,Cavaglia:2022qpg}.

Finally, the basic tools of the superconformal bootstrap  are the  superconformal blocks, and those  of half BPS correlators in many of these and other theories have recently been given in a universal formalism~\cite{Aprile:2021pwd} (see also the earlier work~\cite{Dolan:2004mu}) and found to be equal to  certain objects  in the theory of symmetric polynomials / CMS wave functions on BC super root systems, generalising connections made for scalar blocks in~\cite{Dolan:2003hv,Isachenkov:2017qgn}.

\section*{Acknowledgments}

I would like to thank Francesco Aprile,  Shai Chester, Burkhard Eden, Grisha Korchemsky and  Hynek Paul for useful comments and improvements on an earlier draft and also
all my amazing collaborators over many years on these and related topics: Theresa Abl, Fernando Alday, Raquel Ambrosio, Francesco Aprile, Jake Bourjaily, Andi Brandhuber, Dmitry Chicherin, Gabriele Dian, Reza Doobary, James Drummond, Claude Duhr, Burkhard Eden, Timothy Goddard, Paul Howe, Gregory Korchemsky, Arthur Lipstein, Lionel Mason, Hynek Paul, Jeffrey Pennington, Francesco Sanfilippo, Michele Santagata, Jakub Sikorowski, Vladimir Smirnov, Emery Sokatchev, Alastair Stewart, Charles Taylor, Viet Tran and  Gabriele Travaglini.

This work was supported by the European Union's Horizon 2020 research
and innovation programme under the Marie Sk\l{}odowska-Curie grant
agreement No.~764850 {\it ``\href{https://sagex.org}{SAGEX}''}. I
also acknowledge support from the Science and Technology Facilities
Council (STFC) Consolidated Grant ST/P000371/1.\\

\bibliography{heslop.bib}

\providecommand{\newblock}{}
\begin{thebibliography}{100}
\expandafter\ifx\csname url\endcsname\relax
  \def\url#1{{\tt #1}}\fi
\expandafter\ifx\csname urlprefix\endcsname\relax\def\urlprefix{URL }\fi
\providecommand{\eprint}[2][]{\url{#2}}

\bibitem{hep-th/9807098}
D'Hoker E, Freedman D~Z and Skiba W 1999 {\em Phys. Rev. D\/} {\bf 59} 045008
  (\textit{Preprint} \eprint{hep-th/9807098})

\bibitem{Intriligator:1998ig}
Intriligator K~A 1999 {\em Nucl. Phys. B\/} {\bf 551} 575--600
  (\textit{Preprint} \eprint{hep-th/9811047})

\bibitem{hep-th/9808162}
Howe P~S, Sokatchev E and West P~C 1998 {\em Phys. Lett. B\/} {\bf 444}
  341--351 (\textit{Preprint} \eprint{hep-th/9808162})

\bibitem{hep-th/9903094}
Gonzalez-Rey F, Kulik B and Park I~Y 1999 {\em Phys. Lett. B\/} {\bf 455}
  164--170 (\textit{Preprint} \eprint{hep-th/9903094})

\bibitem{hep-th/9905020}
Intriligator K~A and Skiba W 1999 {\em Nucl. Phys. B\/} {\bf 559} 165--183
  (\textit{Preprint} \eprint{hep-th/9905020})

\bibitem{Eden:1999gh}
Eden B, Howe P~S and West P~C 1999 {\em Phys. Lett. B\/} {\bf 463} 19--26
  (\textit{Preprint} \eprint{hep-th/9905085})

\bibitem{hep-th/9907088}
Skiba W 1999 {\em Phys. Rev. D\/} {\bf 60} 105038 (\textit{Preprint}
  \eprint{hep-th/9907088})

\bibitem{hep-th/9910197}
Penati S, Santambrogio A and Zanon D 1999 {\em JHEP\/} {\bf 12} 006
  (\textit{Preprint} \eprint{hep-th/9910197})

\bibitem{hep-th/0009106}
Eden B, Petkou A~C, Schubert C and Sokatchev E 2001 {\em Nucl. Phys. B\/} {\bf
  607} 191--212 (\textit{Preprint} \eprint{hep-th/0009106})

\bibitem{Gonzalez-Rey:1998wyj}
Gonzalez-Rey F, Park I~Y and Schalm K 1999 {\em Phys. Lett. B\/} {\bf 448}
  37--40 (\textit{Preprint} \eprint{hep-th/9811155})

\bibitem{Eden:1998hh}
Eden B, Howe P~S, Schubert C, Sokatchev E and West P~C 1999 {\em Nucl. Phys.
  B\/} {\bf 557} 355--379 (\textit{Preprint} \eprint{hep-th/9811172})

\bibitem{Eden:1999kh}
Eden B, Howe P~S, Schubert C, Sokatchev E and West P~C 1999 {\em Phys. Lett.
  B\/} {\bf 466} 20--26 (\textit{Preprint} \eprint{hep-th/9906051})

\bibitem{Eden:2000mv}
Eden B, Schubert C and Sokatchev E 2000 {\em Phys. Lett. B\/} {\bf 482}
  309--314 (\textit{Preprint} \eprint{hep-th/0003096})

\bibitem{Bianchi:2000hn}
Bianchi M, Kovacs S, Rossi G and Stanev Y~S 2000 {\em Nucl. Phys. B\/} {\bf
  584} 216--232 (\textit{Preprint} \eprint{hep-th/0003203})

\bibitem{Eden:2011we}
Eden B, Heslop P, Korchemsky G~P and Sokatchev E 2012 {\em Nucl. Phys. B\/}
  {\bf 862} 193--231 (\textit{Preprint} \eprint{1108.3557})

\bibitem{Eden:2012tu}
Eden B, Heslop P, Korchemsky G~P and Sokatchev E 2012 {\em Nucl. Phys. B\/}
  {\bf 862} 450--503 (\textit{Preprint} \eprint{1201.5329})

\bibitem{Bourjaily:2015bpz}
Bourjaily J~L, Heslop P and Tran V~V 2016 {\em Phys. Rev. Lett.\/} {\bf 116}
  191602 (\textit{Preprint} \eprint{1512.07912})

\bibitem{Bourjaily:2016evz}
Bourjaily J~L, Heslop P and Tran V~V 2016 {\em JHEP\/} {\bf 11} 125
  (\textit{Preprint} \eprint{1609.00007})

\bibitem{Alday:2010zy}
Alday L~F, Eden B, Korchemsky G~P, Maldacena J and Sokatchev E 2011 {\em
  JHEP\/} {\bf 09} 123 (\textit{Preprint} \eprint{1007.3243})

\bibitem{Alday:2007hr}
Alday L~F and Maldacena J~M 2007 {\em JHEP\/} {\bf 06} 064 (\textit{Preprint}
  \eprint{0705.0303})

\bibitem{Drummond:2007aua}
Drummond J~M, Korchemsky G~P and Sokatchev E 2008 {\em Nucl. Phys. B\/} {\bf
  795} 385--408 (\textit{Preprint} \eprint{0707.0243})

\bibitem{Brandhuber:2007yx}
Brandhuber A, Heslop P and Travaglini G 2008 {\em Nucl. Phys. B\/} {\bf 794}
  231--243 (\textit{Preprint} \eprint{0707.1153})

\bibitem{Bern:2008ap}
Bern Z, Dixon L~J, Kosower D~A, Roiban R, Spradlin M, Vergu C and Volovich A
  2008 {\em Phys. Rev. D\/} {\bf 78} 045007 (\textit{Preprint}
  \eprint{0803.1465})

\bibitem{Drummond:2008aq}
Drummond J~M, Henn J, Korchemsky G~P and Sokatchev E 2009 {\em Nucl. Phys. B\/}
  {\bf 815} 142--173 (\textit{Preprint} \eprint{0803.1466})

\bibitem{Mason:2010yk}
Mason L~J and Skinner D 2010 {\em JHEP\/} {\bf 12} 018 (\textit{Preprint}
  \eprint{1009.2225})

\bibitem{Caron-Huot:2010ryg}
Caron-Huot S 2011 {\em JHEP\/} {\bf 07} 058 (\textit{Preprint}
  \eprint{1010.1167})

\bibitem{Eden:2010zz}
Eden B, Korchemsky G~P and Sokatchev E 2011 {\em JHEP\/} {\bf 12} 002
  (\textit{Preprint} \eprint{1007.3246})

\bibitem{Eden:2010ce}
Eden B, Korchemsky G~P and Sokatchev E 2012 {\em Phys. Lett. B\/} {\bf 709}
  247--253 (\textit{Preprint} \eprint{1009.2488})

\bibitem{Eden:2011yp}
Eden B, Heslop P, Korchemsky G~P and Sokatchev E 2013 {\em Nucl. Phys. B\/}
  {\bf 869} 329--377 (\textit{Preprint} \eprint{1103.3714})

\bibitem{Adamo:2011dq}
Adamo T, Bullimore M, Mason L and Skinner D 2011 {\em JHEP\/} {\bf 08} 076
  (\textit{Preprint} \eprint{1103.4119})

\bibitem{Eden:2011ku}
Eden B, Heslop P, Korchemsky G~P and Sokatchev E 2013 {\em Nucl. Phys. B\/}
  {\bf 869} 378--416 (\textit{Preprint} \eprint{1103.4353})

\bibitem{Ambrosio:2013pba}
Ambrosio R~G, Eden B, Goddard T, Heslop P and Taylor C 2015 {\em JHEP\/} {\bf
  01} 116 (\textit{Preprint} \eprint{1312.1163})

\bibitem{Heslop:2018zut}
Heslop P and Tran V~V 2018 {\em JHEP\/} {\bf 07} 068 (\textit{Preprint}
  \eprint{1803.11491})

\bibitem{Arkani-Hamed:2018rsk}
Arkani-Hamed N, Langer C, Yelleshpur~Srikant A and Trnka J 2019 {\em Phys. Rev.
  Lett.\/} {\bf 122} 051601 (\textit{Preprint} \eprint{1810.08208})

\bibitem{Langer:2019iuo}
Langer C and Yelleshpur~Srikant A 2019 {\em JHEP\/} {\bf 04} 105
  (\textit{Preprint} \eprint{1902.05951})

\bibitem{dhs}
Dian G, Heslop P and Stewart A {\em In preparation\/}

\bibitem{Fleury:2019ydf}
Fleury T and Pereira R 2020 {\em JHEP\/} {\bf 03} 003 (\textit{Preprint}
  \eprint{1910.09428})

\bibitem{Chicherin:2014uca}
Chicherin D, Doobary R, Eden B, Heslop P, Korchemsky G~P, Mason L and Sokatchev
  E 2015 {\em JHEP\/} {\bf 06} 198 (\textit{Preprint} \eprint{1412.8718})

\bibitem{Usyukina:1993ch}
Usyukina N~I and Davydychev A~I 1993 {\em Phys. Lett. B\/} {\bf 305} 136--143

\bibitem{Drummond:2013nda}
Drummond J, Duhr C, Eden B, Heslop P, Pennington J and Smirnov V~A 2013 {\em
  JHEP\/} {\bf 08} 133 (\textit{Preprint} \eprint{1303.6909})

\bibitem{Goncharov:2010jf}
Goncharov A~B, Spradlin M, Vergu C and Volovich A 2010 {\em Phys. Rev. Lett.\/}
  {\bf 105} 151605 (\textit{Preprint} \eprint{1006.5703})

\bibitem{Eden:2012rr}
Eden B 2012  (\textit{Preprint} \eprint{1207.3112})

\bibitem{brown2004single}
Brown F~C 2004 {\em Comptes Rendus Mathematique\/} {\bf 338} 527--532

\bibitem{Schnetz:2021ebf}
Schnetz O 2021  (\textit{Preprint} \eprint{2111.11246})

\bibitem{Eden:2016dir}
Eden B and Smirnov V~A 2016 {\em JHEP\/} {\bf 10} 115 (\textit{Preprint}
  \eprint{1607.06427})

\bibitem{Borinsky:2021gkd}
Borinsky M and Schnetz O 2021  (\textit{Preprint} \eprint{2105.05015})

\bibitem{Eden:2012fe}
Eden B, Heslop P, Korchemsky G~P, Smirnov V~A and Sokatchev E 2012 {\em Nucl.
  Phys. B\/} {\bf 862} 123--166 (\textit{Preprint} \eprint{1202.5733})

\bibitem{1607.02195}
Gon\c{c}alves V 2017 {\em JHEP\/} {\bf 03} 079 (\textit{Preprint}
  \eprint{1607.02195})

\bibitem{1608.04222}
Eden B and Paul F 2016  (\textit{Preprint} \eprint{1608.04222})

\bibitem{1710.06419}
Georgoudis A, Goncalves V and Pereira R 2018 {\em JHEP\/} {\bf 11} 184
  (\textit{Preprint} \eprint{1710.06419})

\bibitem{chapter5}
Papathanasiou G 2022 {\em J. Phys. A\/} {\bf 55} 443006 (\textit{Preprint} \eprint{2203.13016})

\bibitem{Bissi:2022mrs}
Bissi A, Sinha A and Zhou X 2022  (\textit{Preprint} \eprint{2202.08475})

\bibitem{Maldacena:1997re}
Maldacena J~M 1998 {\em Adv. Theor. Math. Phys.\/} {\bf 2} 231--252
  (\textit{Preprint} \eprint{hep-th/9711200})

\bibitem{Gubser:1998bc}
Gubser S~S, Klebanov I~R and Polyakov A~M 1998 {\em Phys. Lett. B\/} {\bf 428}
  105--114 (\textit{Preprint} \eprint{hep-th/9802109})

\bibitem{Witten:1998qj}
Witten E 1998 {\em Adv. Theor. Math. Phys.\/} {\bf 2} 253--291
  (\textit{Preprint} \eprint{hep-th/9802150})

\bibitem{Liu:1998ty}
Liu H and Tseytlin A~A 1999 {\em Phys. Rev. D\/} {\bf 59} 086002
  (\textit{Preprint} \eprint{hep-th/9807097})

\bibitem{hep-th/9903196}
D'Hoker E, Freedman D~Z, Mathur S~D, Matusis A and Rastelli L 1999 {\em Nucl.
  Phys. B\/} {\bf 562} 353--394 (\textit{Preprint} \eprint{hep-th/9903196})

\bibitem{hep-th/9911222}
D'Hoker E, Mathur S~D, Matusis A and Rastelli L 2000 {\em Nucl. Phys. B\/} {\bf
  589} 38--74 (\textit{Preprint} \eprint{hep-th/9911222})

\bibitem{hep-th/0002170}
Arutyunov G and Frolov S 2000 {\em Phys. Rev. D\/} {\bf 62} 064016
  (\textit{Preprint} \eprint{hep-th/0002170})

\bibitem{Arutyunov:2002fh}
Arutyunov G, Dolan F~A, Osborn H and Sokatchev E 2003 {\em Nucl. Phys. B\/}
  {\bf 665} 273--324 (\textit{Preprint} \eprint{hep-th/0212116})

\bibitem{hep-th/0112251}
Dolan F~A and Osborn H 2002 {\em Nucl. Phys. B\/} {\bf 629} 3--73
  (\textit{Preprint} \eprint{hep-th/0112251})

\bibitem{0907.0151}
Heemskerk I, Penedones J, Polchinski J and Sully J 2009 {\em JHEP\/} {\bf 10}
  079 (\textit{Preprint} \eprint{0907.0151})

\bibitem{1011.1485}
Penedones J 2011 {\em JHEP\/} {\bf 03} 025 (\textit{Preprint}
  \eprint{1011.1485})

\bibitem{1410.4717}
Alday L~F, Bissi A and Lukowski T 2015 {\em JHEP\/} {\bf 06} 074
  (\textit{Preprint} \eprint{1410.4717})

\bibitem{Goncalves:2014ffa}
Gon\c{c}alves V 2015 {\em JHEP\/} {\bf 04} 150 (\textit{Preprint}
  \eprint{1411.1675})

\bibitem{Binder:2019jwn}
Binder D~J, Chester S~M, Pufu S~S and Wang Y 2019 {\em JHEP\/} {\bf 12} 119
  (\textit{Preprint} \eprint{1902.06263})

\bibitem{Chester:2020dja}
Chester S~M and Pufu S~S 2021 {\em JHEP\/} {\bf 01} 103 (\textit{Preprint}
  \eprint{2003.08412})

\bibitem{Chester:2019jas}
Chester S~M, Green M~B, Pufu S~S, Wang Y and Wen C 2020 {\em JHEP\/} {\bf 11}
  016 (\textit{Preprint} \eprint{1912.13365})

\bibitem{Chester:2020vyz}
Chester S~M, Green M~B, Pufu S~S, Wang Y and Wen C 2021 {\em JHEP\/} {\bf 04}
  212 (\textit{Preprint} \eprint{2008.02713})

\bibitem{Abl:2020dbx}
Abl T, Heslop P and Lipstein A~E 2021 {\em JHEP\/} {\bf 04} 237
  (\textit{Preprint} \eprint{2012.12091})

\bibitem{Aprile:2017bgs}
Aprile F, Drummond J~M, Heslop P and Paul H 2018 {\em JHEP\/} {\bf 01} 035
  (\textit{Preprint} \eprint{1706.02822})

\bibitem{Rastelli:2016nze}
Rastelli L and Zhou X 2017 {\em Phys. Rev. Lett.\/} {\bf 118} 091602
  (\textit{Preprint} \eprint{1608.06624})

\bibitem{Doobary:2015gia}
Doobary R and Heslop P 2015 {\em JHEP\/} {\bf 12} 159 (\textit{Preprint}
  \eprint{1508.03611})

\bibitem{Alday:2017xua}
Alday L~F and Bissi A 2017 {\em Phys. Rev. Lett.\/} {\bf 119} 171601
  (\textit{Preprint} \eprint{1706.02388})

\bibitem{Aprile:2017xsp}
Aprile F, Drummond J~M, Heslop P and Paul H 2018 {\em JHEP\/} {\bf 02} 133
  (\textit{Preprint} \eprint{1706.08456})

\bibitem{Caron-Huot:2018kta}
Caron-Huot S and Trinh A~K 2019 {\em JHEP\/} {\bf 01} 196 (\textit{Preprint}
  \eprint{1809.09173})

\bibitem{Bissi:2020wtv}
Bissi A, Fardelli G and Georgoudis A 2021 {\em Phys. Rev. D\/} {\bf 104}
  L041901 (\textit{Preprint} \eprint{2002.04604})

\bibitem{Bissi:2020woe}
Bissi A, Fardelli G and Georgoudis A 2021 {\em J. Phys. A\/} {\bf 54} 324002
  (\textit{Preprint} \eprint{2010.12557})

\bibitem{Aprile:2019rep}
Aprile F, Drummond J, Heslop P and Paul H 2020 {\em JHEP\/} {\bf 03} 190
  (\textit{Preprint} \eprint{1912.01047})

\bibitem{Chester:2019pvm}
Chester S~M 2020 {\em JHEP\/} {\bf 04} 193 (\textit{Preprint}
  \eprint{1908.05247})

\bibitem{hep-th/9706175}
Green M~B, Gutperle M and Vanhove P 1997 {\em Phys. Lett. B\/} {\bf 409}
  177--184 (\textit{Preprint} \eprint{hep-th/9706175})

\bibitem{hep-th/9808061}
Green M~B and Sethi S 1999 {\em Phys. Rev. D\/} {\bf 59} 046006
  (\textit{Preprint} \eprint{hep-th/9808061})

\bibitem{hep-th/9910055}
Green M~B, Kwon H~h and Vanhove P 2000 {\em Phys. Rev. D\/} {\bf 61} 104010
  (\textit{Preprint} \eprint{hep-th/9910055})

\bibitem{hep-th/0510027}
Green M~B and Vanhove P 2006 {\em JHEP\/} {\bf 01} 093 (\textit{Preprint}
  \eprint{hep-th/0510027})

\bibitem{1404.2192}
Green M~B, Miller S~D and Vanhove P 2015 {\em Commun. Num. Theor. Phys.\/} {\bf
  09} 307--344 (\textit{Preprint} \eprint{1404.2192})

\bibitem{1912.13365}
Chester S~M, Green M~B, Pufu S~S, Wang Y and Wen C 2020 {\em JHEP\/} {\bf 11}
  016 (\textit{Preprint} \eprint{1912.13365})

\bibitem{2008.02713}
Chester S~M, Green M~B, Pufu S~S, Wang Y and Wen C 2021 {\em JHEP\/} {\bf 04}
  212 (\textit{Preprint} \eprint{2008.02713})

\bibitem{2009.01211}
Green M~B and Wen C 2021 {\em JHEP\/} {\bf 02} 042 (\textit{Preprint}
  \eprint{2009.01211})

\bibitem{Alday:2018kkw}
Alday L~F 2021 {\em JHEP\/} {\bf 04} 005 (\textit{Preprint}
  \eprint{1812.11783})

\bibitem{Drummond:2019hel}
Drummond J~M and Paul H 2021 {\em JHEP\/} {\bf 03} 038 (\textit{Preprint}
  \eprint{1912.07632})

\bibitem{Huang:2021xws}
Huang Z and Yuan E~Y 2021  (\textit{Preprint} \eprint{2112.15174})

\bibitem{Aprile:2020uxk}
Aprile F, Drummond J~M, Heslop P, Paul H, Sanfilippo F, Santagata M and Stewart
  A 2020 {\em JHEP\/} {\bf 11} 072 (\textit{Preprint} \eprint{2007.09395})

\bibitem{Aprile:2018efk}
Aprile F, Drummond J, Heslop P and Paul H 2018 {\em Phys. Rev. D\/} {\bf 98}
  126008 (\textit{Preprint} \eprint{1802.06889})

\bibitem{Alday:2019nin}
Alday L~F and Zhou X 2020 {\em JHEP\/} {\bf 09} 008 (\textit{Preprint}
  \eprint{1912.02663})

\bibitem{Arutyunov:2003ae}
Arutyunov G and Sokatchev E 2003 {\em Nucl. Phys. B\/} {\bf 663} 163--196
  (\textit{Preprint} \eprint{hep-th/0301058})

\bibitem{Arutyunov:2003ad}
Arutyunov G, Penati S, Santambrogio A and Sokatchev E 2003 {\em Nucl. Phys.
  B\/} {\bf 670} 103--147 (\textit{Preprint} \eprint{hep-th/0305060})

\bibitem{D'Alessandro:2005dq}
D'Alessandro M and Genovese L 2006 {\em Nucl. Phys. B\/} {\bf 732} 64--88
  (\textit{Preprint} \eprint{hep-th/0504061})

\bibitem{Uruchurtu:2011wh}
Uruchurtu L~I 2011 {\em JHEP\/} {\bf 08} 133 (\textit{Preprint}
  \eprint{1106.0630})

\bibitem{Chicherin:2014esa}
Chicherin D and Sokatchev E 2014 {\em JHEP\/} {\bf 11} 139 (\textit{Preprint}
  \eprint{1408.3527})

\bibitem{Chicherin:2015edu}
Chicherin D, Drummond J, Heslop P and Sokatchev E 2016 {\em JHEP\/} {\bf 08}
  053 (\textit{Preprint} \eprint{1512.02926})

\bibitem{Chicherin:2018avq}
Chicherin D, Georgoudis A, Gon\c{c}alves V and Pereira R 2018 {\em JHEP\/} {\bf
  11} 069 (\textit{Preprint} \eprint{1809.00551})

\bibitem{Caron-Huot:2021usw}
Caron-Huot S and Coronado F 2022 {\em JHEP\/} {\bf 03} 151 (\textit{Preprint}
  \eprint{2106.03892})

\bibitem{Chicherin:2015bza}
Chicherin D, Doobary R, Eden B, Heslop P, Korchemsky G~P and Sokatchev E 2016
  {\em JHEP\/} {\bf 03} 031 (\textit{Preprint} \eprint{1506.04983})

\bibitem{Coronado:2018ypq}
Coronado F 2019 {\em JHEP\/} {\bf 01} 056 (\textit{Preprint}
  \eprint{1811.00467})

\bibitem{Basso:2015zoa}
Basso B, Komatsu S and Vieira P 2015  (\textit{Preprint} \eprint{1505.06745})

\bibitem{Alday:2009zm}
Alday L~F, Henn J~M, Plefka J and Schuster T 2010 {\em JHEP\/} {\bf 01} 077
  (\textit{Preprint} \eprint{0908.0684})

\bibitem{Belitsky:2020qrm}
Belitsky A~V and Korchemsky G~P 2020 {\em JHEP\/} {\bf 07} 219
  (\textit{Preprint} \eprint{2003.01121})

\bibitem{Belitsky:2020qir}
Belitsky A~V and Korchemsky G~P 2021 {\em JHEP\/} {\bf 04} 257
  (\textit{Preprint} \eprint{2006.01831})

\bibitem{hep-th/0601148}
Dolan F~A, Nirschl M and Osborn H 2006 {\em Nucl. Phys. B\/} {\bf 749} 109--152
  (\textit{Preprint} \eprint{hep-th/0601148})

\bibitem{0709.1365}
Berdichevsky L and Naaijkens P 2008 {\em JHEP\/} {\bf 01} 071
  (\textit{Preprint} \eprint{0709.1365})

\bibitem{0811.2320}
Uruchurtu L~I 2009 {\em JHEP\/} {\bf 03} 133 (\textit{Preprint}
  \eprint{0811.2320})

\bibitem{1806.09200}
Arutyunov G, Klabbers R and Savin S 2018 {\em JHEP\/} {\bf 09} 023
  (\textit{Preprint} \eprint{1806.09200})

\bibitem{1808.06788}
Arutyunov G, Klabbers R and Savin S 2018 {\em JHEP\/} {\bf 09} 118
  (\textit{Preprint} \eprint{1808.06788})

\bibitem{Rastelli:2017udc}
Rastelli L and Zhou X 2018 {\em JHEP\/} {\bf 04} 014 (\textit{Preprint}
  \eprint{1710.05923})

\bibitem{Aprile:2017qoy}
Aprile F, Drummond J~M, Heslop P and Paul H 2018 {\em JHEP\/} {\bf 05} 056
  (\textit{Preprint} \eprint{1711.03903})

\bibitem{hep-th/0610280}
Drummond J~M, Gallot L and Sokatchev E 2007 {\em Phys. Lett. B\/} {\bf 645}
  95--100 (\textit{Preprint} \eprint{hep-th/0610280})

\bibitem{Drummond:2019odu}
Drummond J~M, Nandan D, Paul H and Rigatos K~S 2019 {\em JHEP\/} {\bf 12} 173
  (\textit{Preprint} \eprint{1907.00992})

\bibitem{Abl:2021mxo}
Abl T, Heslop P and Lipstein A~E 2022 {\em JHEP\/} {\bf 03} 076
  (\textit{Preprint} \eprint{2112.09597})

\bibitem{Heemskerk:2009pn}
Heemskerk I, Penedones J, Polchinski J and Sully J 2009 {\em JHEP\/} {\bf 10}
  079 (\textit{Preprint} \eprint{0907.0151})

\bibitem{Drummond:2020dwr}
Drummond J~M, Paul H and Santagata M 2020  (\textit{Preprint}
  \eprint{2004.07282})

\bibitem{Aprile:2020mus}
Aprile F, Drummond J~M, Paul H and Santagata M 2021 {\em JHEP\/} {\bf 11} 109
  (\textit{Preprint} \eprint{2012.12092})

\bibitem{Heslop:2000np}
Heslop P and Howe P~S 2001 {\em Phys. Lett. B\/} {\bf 502} 259--264
  (\textit{Preprint} \eprint{hep-th/0008047})

\bibitem{Howe:1983sra}
Howe P~S and West P~C 1984 {\em Nucl. Phys. B\/} {\bf 238} 181--220

\bibitem{deHaro:2002vk}
de~Haro S, Sinkovics A and Skenderis K 2003 {\em Phys. Rev. D\/} {\bf 67}
  084010 (\textit{Preprint} \eprint{hep-th/0210080})

\bibitem{bh}
Berkovits N and Howe P

\bibitem{Rajaraman:2005up}
Rajaraman A 2005 {\em Phys. Rev. D\/} {\bf 72} 125008 (\textit{Preprint}
  \eprint{hep-th/0505155})

\bibitem{Drummond:2020uni}
Drummond J~M, Glew R and Paul H 2021 {\em JHEP\/} {\bf 12} 072
  (\textit{Preprint} \eprint{2008.01109})

\bibitem{Galperin:1984av}
Galperin A, Ivanov E, Kalitsyn S, Ogievetsky V and Sokatchev E 1984 {\em Class.
  Quant. Grav.\/} {\bf 1} 469--498 [Erratum: Class.Quant.Grav. 2, 127 (1985)]

\bibitem{Howe:1995md}
Howe P~S and Hartwell G~G 1995 {\em Class. Quant. Grav.\/} {\bf 12} 1823--1880

\bibitem{Hartwell:1994rp}
Hartwell G~G and Howe P~S 1995 {\em Int. J. Mod. Phys. A\/} {\bf 10} 3901--3920
  (\textit{Preprint} \eprint{hep-th/9412147})

\bibitem{Howe:2001je}
Howe P~S and West P~C 2001 {\em Class. Quant. Grav.\/} {\bf 18} 3143--3158
  (\textit{Preprint} \eprint{hep-th/0105218})

\bibitem{Heslop:2001dr}
Heslop P~J and Howe P~S 2001 {\em Phys. Lett. B\/} {\bf 516} 367--375
  (\textit{Preprint} \eprint{hep-th/0106238})

\bibitem{Heslop:2001gp}
Heslop P~J and Howe P~S 2002 {\em Nucl. Phys. B\/} {\bf 626} 265--286
  (\textit{Preprint} \eprint{hep-th/0107212})

\bibitem{Heslop:2001zm}
Heslop P~J 2002 {\em Class. Quant. Grav.\/} {\bf 19} 303--346
  (\textit{Preprint} \eprint{hep-th/0108235})

\bibitem{Heslop:2002hp}
Heslop P~J and Howe P~S 2003 {\em JHEP\/} {\bf 01} 043 (\textit{Preprint}
  \eprint{hep-th/0211252})

\bibitem{Heslop:2003xu}
Heslop P~J and Howe P~S 2004 {\em JHEP\/} {\bf 01} 058 (\textit{Preprint}
  \eprint{hep-th/0307210})

\bibitem{Eden:2001ec}
Eden B and Sokatchev E 2001 {\em Nucl. Phys. B\/} {\bf 618} 259--276
  (\textit{Preprint} \eprint{hep-th/0106249})

\bibitem{stembridge1985characterization}
Stembridge J~R 1985 {\em Journal of algebra\/} {\bf 95} 439--444

\bibitem{Dolan:2001tt}
Dolan F~A and Osborn H 2002 {\em Nucl. Phys. B\/} {\bf 629} 3--73
  (\textit{Preprint} \eprint{hep-th/0112251})

\bibitem{hep-th/9509140}
Howe P~S and West P~C 1999 {\em Int. J. Mod. Phys. A\/} {\bf 14} 2659--2674
  (\textit{Preprint} \eprint{hep-th/9509140})

\bibitem{hep-th/9607060}
Howe P~S and West P~C 1996 {\em Phys. Lett. B\/} {\bf 389} 273--279
  (\textit{Preprint} \eprint{hep-th/9607060})

\bibitem{hep-th/9611074}
Howe P~S and West P~C 1996 {Is N=4 Yang-Mills theory soluble?} {\em {2nd
  International Sakharov Conference on Physics}\/} pp 295--301
  (\textit{Preprint} \eprint{hep-th/9611074})

\bibitem{hep-th/9611075}
Howe P~S and West P~C 1997 {\em Phys. Lett. B\/} {\bf 400} 307--313
  (\textit{Preprint} \eprint{hep-th/9611075})

\bibitem{Howe:1999hz}
Howe P~S, Schubert C, Sokatchev E and West P~C 2000 {\em Nucl. Phys. B\/} {\bf
  571} 71--90 (\textit{Preprint} \eprint{hep-th/9910011})

\bibitem{Chicherin:2016fac}
Chicherin D and Sokatchev E 2017 {\em JHEP\/} {\bf 02} 062 (\textit{Preprint}
  \eprint{1601.06803})

\bibitem{Chicherin:2016fbj}
Chicherin D and Sokatchev E 2017 {\em JHEP\/} {\bf 03} 048 (\textit{Preprint}
  \eprint{1601.06804})

\bibitem{Chicherin:2016soh}
Chicherin D and Sokatchev E 2017 {\em J. Phys. A\/} {\bf 50} 205402
  (\textit{Preprint} \eprint{1603.08478})

\bibitem{Green:2020eyj}
Green M~B and Wen C 2021 {\em JHEP\/} {\bf 02} 042 (\textit{Preprint}
  \eprint{2009.01211})

\bibitem{0812.3341}
Drukker N and Plefka J 2009 {\em JHEP\/} {\bf 04} 001 (\textit{Preprint}
  \eprint{0812.3341})

\bibitem{Goncalves:2019znr}
Gon\c{c}alves V, Pereira R and Zhou X 2019 {\em JHEP\/} {\bf 10} 247
  (\textit{Preprint} \eprint{1906.05305})

\bibitem{Chicherin:2016qsf}
Chicherin D and Sokatchev E 2017 {\em J. Phys. A\/} {\bf 50} 275402
  (\textit{Preprint} \eprint{1605.01386})

\bibitem{Hodges:2009hk}
Hodges A 2013 {\em JHEP\/} {\bf 05} 135 (\textit{Preprint} \eprint{0905.1473})

\bibitem{Mason:2009qx}
Mason L~J and Skinner D 2009 {\em JHEP\/} {\bf 11} 045 (\textit{Preprint}
  \eprint{0909.0250})

\bibitem{Arkani-Hamed:2013jha}
Arkani-Hamed N and Trnka J 2014 {\em JHEP\/} {\bf 10} 030 (\textit{Preprint}
  \eprint{1312.2007})

\bibitem{Arkani-Hamed:2013kca}
Arkani-Hamed N and Trnka J 2014 {\em JHEP\/} {\bf 12} 182 (\textit{Preprint}
  \eprint{1312.7878})

\bibitem{chapter7}
Herrmann E and Trnka J 2022 {\em J. Phys. A\/} {\bf 55} 443008 (\textit{Preprint} \eprint{2203.13018})

\bibitem{Eden:2017fow}
Eden B, Heslop P and Mason L 2017 {\em JHEP\/} {\bf 09} 156 (\textit{Preprint}
  \eprint{1701.00453})

\bibitem{Arkani-Hamed:2017vfh}
Arkani-Hamed N, Thomas H and Trnka J 2018 {\em JHEP\/} {\bf 01} 016
  (\textit{Preprint} \eprint{1704.05069})

\bibitem{Dian:2021idl}
Dian G and Heslop P 2021 {\em JHEP\/} {\bf 11} 074 (\textit{Preprint}
  \eprint{2106.09372})

\bibitem{Beem:2013qxa}
Beem C, Rastelli L and van Rees B~C 2013 {\em Phys. Rev. Lett.\/} {\bf 111}
  071601 (\textit{Preprint} \eprint{1304.1803})

\bibitem{Beem:2016wfs}
Beem C, Rastelli L and van Rees B~C 2017 {\em Phys. Rev. D\/} {\bf 96} 046014
  (\textit{Preprint} \eprint{1612.02363})

\bibitem{Bissi:2020jve}
Bissi A, Manenti A and Vichi A 2021 {\em JHEP\/} {\bf 05} 111
  (\textit{Preprint} \eprint{2010.15126})

\bibitem{Chester:2021aun}
Chester S~M, Dempsey R and Pufu S~S 2021  (\textit{Preprint}
  \eprint{2111.07989})

\bibitem{Eden:2016xvg}
Eden B and Sfondrini A 2017 {\em JHEP\/} {\bf 10} 098 (\textit{Preprint}
  \eprint{1611.05436})

\bibitem{Fleury:2016ykk}
Fleury T and Komatsu S 2017 {\em JHEP\/} {\bf 01} 130 (\textit{Preprint}
  \eprint{1611.05577})

\bibitem{Eden:2018vug}
Eden B, Jiang Y, de~Leeuw M, Meier T, le~Plat D and Sfondrini A 2018 {\em
  JHEP\/} {\bf 11} 097 (\textit{Preprint} \eprint{1806.06051})

\bibitem{Basso:2017khq}
Basso B, Coronado F, Komatsu S, Lam H~T, Vieira P and Zhong D~l 2019 {\em
  JHEP\/} {\bf 07} 082 (\textit{Preprint} \eprint{1701.04462})

\bibitem{Eden:2017ozn}
Eden B, Jiang Y, le~Plat D and Sfondrini A 2018 {\em JHEP\/} {\bf 02} 170
  (\textit{Preprint} \eprint{1710.10212})

\bibitem{Fleury:2017eph}
Fleury T and Komatsu S 2018 {\em JHEP\/} {\bf 02} 177 (\textit{Preprint}
  \eprint{1711.05327})

\bibitem{Bargheer:2019kxb}
Bargheer T, Coronado F and Vieira P 2019 {\em JHEP\/} {\bf 08} 162
  (\textit{Preprint} \eprint{1904.00965})

\bibitem{deLeeuw:2019tdd}
de~Leeuw M, Eden B, Le~Plat D and Meier T 2020 {\em Phys. Part. Nucl. Lett.\/}
  {\bf 17} 678--686 (\textit{Preprint} \eprint{1907.07014})

\bibitem{Belitsky:2019fan}
Belitsky A~V and Korchemsky G~P 2020 {\em JHEP\/} {\bf 05} 070
  (\textit{Preprint} \eprint{1907.13131})

\bibitem{Bargheer:2019exp}
Bargheer T, Coronado F and Vieira P 2019  (\textit{Preprint}
  \eprint{1909.04077})

\bibitem{DeLeeuw:2019dak}
De~Leeuw M, Eden B, Le~Plat D, Meier T and Sfondrini A 2020 {\em JHEP\/} {\bf
  09} 039 (\textit{Preprint} \eprint{1912.12231})

\bibitem{Aprile:2020luw}
Aprile F and Vieira P 2020 {\em JHEP\/} {\bf 12} 206 (\textit{Preprint}
  \eprint{2007.09176})

\bibitem{chapter9}
Chicherin D and Korchemsky G~P 2022 443010 (\textit{Preprint} \eprint{2203.13020})

\bibitem{Corley:2001zk}
Corley S, Jevicki A and Ramgoolam S 2002 {\em Adv. Theor. Math. Phys.\/} {\bf
  5} 809--839 (\textit{Preprint} \eprint{hep-th/0111222})

\bibitem{Lewis-Brown:2018dje}
Lewis-Brown C and Ramgoolam S 2018 {\em JHEP\/} {\bf 11} 035 (\textit{Preprint}
  \eprint{1804.11090})

\bibitem{Alday:2021vfb}
Alday L~F, Chester S~M and Hansen T 2021 {\em JHEP\/} {\bf 12} 159
  (\textit{Preprint} \eprint{2110.13106})

\bibitem{Dorigoni:2022zcr}
Dorigoni D, Green M~B and Wen C 2022  (\textit{Preprint} \eprint{2202.05784})

\bibitem{Hofman:2008ar}
Hofman D~M and Maldacena J 2008 {\em JHEP\/} {\bf 05} 012 (\textit{Preprint}
  \eprint{0803.1467})

\bibitem{Belitsky:2013xxa}
Belitsky A~V, Hohenegger S, Korchemsky G~P, Sokatchev E and Zhiboedov A 2014
  {\em Nucl. Phys. B\/} {\bf 884} 305--343 (\textit{Preprint}
  \eprint{1309.0769})

\bibitem{Belitsky:2013bja}
Belitsky A~V, Hohenegger S, Korchemsky G~P, Sokatchev E and Zhiboedov A 2014
  {\em Nucl. Phys. B\/} {\bf 884} 206--256 (\textit{Preprint}
  \eprint{1309.1424})

\bibitem{Belitsky:2013ofa}
Belitsky A~V, Hohenegger S, Korchemsky G~P, Sokatchev E and Zhiboedov A 2014
  {\em Phys. Rev. Lett.\/} {\bf 112} 071601 (\textit{Preprint}
  \eprint{1311.6800})

\bibitem{Korchemsky:2021okt}
Korchemsky G, Sokatchev E and Zhiboedov A 2021  (\textit{Preprint}
  \eprint{2106.14899})

\bibitem{Dorigoni:2021rdo}
Dorigoni D, Green M~B and Wen C 2021 {\em JHEP\/} {\bf 11} 132
  (\textit{Preprint} \eprint{2109.08086})

\bibitem{Dorigoni:2021guq}
Dorigoni D, Green M~B and Wen C 2021 {\em JHEP\/} {\bf 05} 089
  (\textit{Preprint} \eprint{2102.09537})

\bibitem{Dorigoni:2021bvj}
Dorigoni D, Green M~B and Wen C 2021 {\em Phys. Rev. Lett.\/} {\bf 126} 161601
  (\textit{Preprint} \eprint{2102.08305})

\bibitem{chapter10}
Dorigoni D, Green M~B and Wen C 2022 443011 (\textit{Preprint} \eprint{2203.13021})

\bibitem{1107.5702}
Alday L~F, Buchbinder E~I and Tseytlin A~A 2011 {\em JHEP\/} {\bf 09} 034
  (\textit{Preprint} \eprint{1107.5702})

\bibitem{1110.0758}
Engelund O~T and Roiban R 2012 {\em JHEP\/} {\bf 05} 158 (\textit{Preprint}
  \eprint{1110.0758})

\bibitem{1207.4316}
Alday L~F, Heslop P and Sikorowski J 2013 {\em JHEP\/} {\bf 03} 074
  (\textit{Preprint} \eprint{1207.4316})

\bibitem{1209.0227}
Engelund O~T and Roiban R 2013 {\em JHEP\/} {\bf 03} 172 (\textit{Preprint}
  \eprint{1209.0227})

\bibitem{1301.0149}
Alday L~F, Henn J~M and Sikorowski J 2013 {\em JHEP\/} {\bf 03} 058
  (\textit{Preprint} \eprint{1301.0149})

\bibitem{2112.06956}
Arkani-Hamed N, Henn J and Trnka J 2022 {\em JHEP\/} {\bf 03} 108
  (\textit{Preprint} \eprint{2112.06956})

\bibitem{2202.05596}
Chicherin D and Henn J~M 2022  (\textit{Preprint} \eprint{2202.05596})

\bibitem{1402.1300}
Penante B, Spence B, Travaglini G and Wen C 2014 {\em JHEP\/} {\bf 04} 083
  (\textit{Preprint} \eprint{1402.1300})

\bibitem{1406.1443}
Brandhuber A, Penante B, Travaglini G and Wen C 2014 {\em JHEP\/} {\bf 08} 100
  (\textit{Preprint} \eprint{1406.1443})

\bibitem{chapter1}
Brandhuber A, Plefka J and Travaglini G 2022 443002 (\textit{Preprint}
  \eprint{2203.13012})

\bibitem{Park:1998nra}
Park J~H 1999 {\em Nucl. Phys. B\/} {\bf 539} 599--642 (\textit{Preprint}
  \eprint{hep-th/9807186})

\bibitem{Corrado:1999pi}
Corrado R, Florea B and McNees R 1999 {\em Phys. Rev. D\/} {\bf 60} 085011
  (\textit{Preprint} \eprint{hep-th/9902153})

\bibitem{Howe:2000nq}
Howe P~S 2001 {\em Phys. Lett. B\/} {\bf 503} 197--204 (\textit{Preprint}
  \eprint{hep-th/0008048})

\bibitem{Eden:2001wg}
Eden B, Ferrara S and Sokatchev E 2001 {\em JHEP\/} {\bf 11} 020
  (\textit{Preprint} \eprint{hep-th/0107084})

\bibitem{Arutyunov:2002ff}
Arutyunov G and Sokatchev E 2002 {\em Nucl. Phys. B\/} {\bf 635} 3--32
  (\textit{Preprint} \eprint{hep-th/0201145})

\bibitem{Heslop:2004du}
Heslop P~J 2004 {\em JHEP\/} {\bf 07} 056 (\textit{Preprint}
  \eprint{hep-th/0405245})

\bibitem{Beem:2015aoa}
Beem C, Lemos M, Rastelli L and van Rees B~C 2016 {\em Phys. Rev. D\/} {\bf 93}
  025016 (\textit{Preprint} \eprint{1507.05637})

\bibitem{Rastelli:2017ymc}
Rastelli L and Zhou X 2018 {\em JHEP\/} {\bf 06} 087 (\textit{Preprint}
  \eprint{1712.02788})

\bibitem{Heslop:2017sco}
Heslop P and Lipstein A~E 2018 {\em JHEP\/} {\bf 02} 004 (\textit{Preprint}
  \eprint{1712.08570})

\bibitem{Chester:2018dga}
Chester S~M and Perlmutter E 2018 {\em JHEP\/} {\bf 08} 116 (\textit{Preprint}
  \eprint{1805.00892})

\bibitem{Abl:2019jhh}
Abl T, Heslop P and Lipstein A~E 2019 {\em JHEP\/} {\bf 04} 038
  (\textit{Preprint} \eprint{1902.00463})

\bibitem{Alday:2020lbp}
Alday L~F and Zhou X 2020 {\em Phys. Rev. Lett.\/} {\bf 125} 131604
  (\textit{Preprint} \eprint{2006.06653})

\bibitem{Alday:2020tgi}
Alday L~F, Chester S~M and Raj H 2021 {\em JHEP\/} {\bf 01} 133
  (\textit{Preprint} \eprint{2005.07175})

\bibitem{Alday:2020dtb}
Alday L~F and Zhou X 2021 {\em Phys. Rev. X\/} {\bf 11} 011056
  (\textit{Preprint} \eprint{2006.12505})

\bibitem{Lemos:2021azv}
Lemos M, van Rees B~C and Zhao X 2022 {\em JHEP\/} {\bf 01} 022
  (\textit{Preprint} \eprint{2105.13361})

\bibitem{Lambert:2021fsl}
Lambert N, Lipstein A, Mouland R and Richmond P 2022 {\em JHEP\/} {\bf 02} 151
  (\textit{Preprint} \eprint{2109.04829})

\bibitem{Zhou:2017zaw}
Zhou X 2018 {\em JHEP\/} {\bf 08} 187 (\textit{Preprint} \eprint{1712.02800})

\bibitem{Chester:2018aca}
Chester S~M, Pufu S~S and Yin X 2018 {\em JHEP\/} {\bf 08} 115
  (\textit{Preprint} \eprint{1804.00949})

\bibitem{Binder:2018yvd}
Binder D~J, Chester S~M and Pufu S~S 2020 {\em JHEP\/} {\bf 04} 052
  (\textit{Preprint} \eprint{1808.10554})

\bibitem{Alday:2021ymb}
Alday L~F, Chester S~M and Raj H 2022 {\em JHEP\/} {\bf 02} 005
  (\textit{Preprint} \eprint{2107.10274})

\bibitem{Rastelli:2019gtj}
Rastelli L, Roumpedakis K and Zhou X 2019 {\em JHEP\/} {\bf 10} 140
  (\textit{Preprint} \eprint{1905.11983})

\bibitem{Giusto:2020neo}
Giusto S, Russo R, Tyukov A and Wen C 2020 {\em Eur. Phys. J. C\/} {\bf 80} 736
  (\textit{Preprint} \eprint{2005.08560})

\bibitem{Wen:2021lio}
Wen C and Zhang S~Q 2021 {\em JHEP\/} {\bf 07} 125 (\textit{Preprint}
  \eprint{2106.03499})

\bibitem{Aprile:2021mvq}
Aprile F and Santagata M 2021 {\em Phys. Rev. D\/} {\bf 104} 126022
  (\textit{Preprint} \eprint{2104.00036})

\bibitem{Alday:2021odx}
Alday L~F, Behan C, Ferrero P and Zhou X 2021 {\em JHEP\/} {\bf 06} 020
  (\textit{Preprint} \eprint{2103.15830})

\bibitem{Drummond:2022dxd}
Drummond J~M, Glew R and Santagata M 2022  (\textit{Preprint}
  \eprint{2202.09837})

\bibitem{1706.00756}
Giombi S, Roiban R and Tseytlin A~A 2017 {\em Nucl. Phys. B\/} {\bf 922}
  499--527 (\textit{Preprint} \eprint{1706.00756})

\bibitem{1802.05201}
Giombi S and Komatsu S 2018 {\em JHEP\/} {\bf 05} 109 [Erratum: JHEP 11, 123
  (2018)] (\textit{Preprint} \eprint{1802.05201})

\bibitem{1806.01862}
Liendo P, Meneghelli C and Mitev V 2018 {\em JHEP\/} {\bf 10} 077
  (\textit{Preprint} \eprint{1806.01862})

\bibitem{1811.02369}
Giombi S and Komatsu S 2019 {\em J. Phys. A\/} {\bf 52} 125401
  (\textit{Preprint} \eprint{1811.02369})

\bibitem{1812.04593}
Kiryu N and Komatsu S 2019 {\em JHEP\/} {\bf 02} 090 (\textit{Preprint}
  \eprint{1812.04593})

\bibitem{2001.11039}
Grabner D, Gromov N and Julius J 2020 {\em JHEP\/} {\bf 07} 042
  (\textit{Preprint} \eprint{2001.11039})

\bibitem{2103.10440}
Ferrero P and Meneghelli C 2021 {\em Phys. Rev. D\/} {\bf 104} L081703
  (\textit{Preprint} \eprint{2103.10440})

\bibitem{2107.08510}
Cavagli\`a A, Gromov N, Julius J and Preti M 2022 {\em Phys. Rev. D\/} {\bf
  105} L021902 (\textit{Preprint} \eprint{2107.08510})

\bibitem{2108.13432}
Barrat J, Gimenez-Grau A and Liendo P 2021  (\textit{Preprint}
  \eprint{2108.13432})

\bibitem{2110.13126}
Giombi S, Komatsu S and Offertaler B 2022 {\em JHEP\/} {\bf 03} 020
  (\textit{Preprint} \eprint{2110.13126})

\bibitem{2112.10780}
Barrat J, Liendo P, Peveri G and Plefka J 2021  (\textit{Preprint}
  \eprint{2112.10780})

\bibitem{2202.07627}
Giombi S, Komatsu S and Offertaler B 2022  (\textit{Preprint}
  \eprint{2202.07627})

\bibitem{Cavaglia:2022qpg}
Cavagli\`a A, Gromov N, Julius J and Preti M 2022  (\textit{Preprint}
  \eprint{2203.09556})

\bibitem{Aprile:2021pwd}
Aprile F and Heslop P 2021  (\textit{Preprint} \eprint{2112.12169})

\bibitem{Dolan:2004mu}
Dolan F~A, Gallot L and Sokatchev E 2004 {\em JHEP\/} {\bf 09} 056
  (\textit{Preprint} \eprint{hep-th/0405180})

\bibitem{Dolan:2003hv}
Dolan F~A and Osborn H 2004 {\em Nucl. Phys. B\/} {\bf 678} 491--507
  (\textit{Preprint} \eprint{hep-th/0309180})

\bibitem{Isachenkov:2017qgn}
Isachenkov M and Schomerus V 2018 {\em JHEP\/} {\bf 07} 180 (\textit{Preprint}
  \eprint{1711.06609})

\end{thebibliography}

\end{document}